\documentclass[12pt]{article}
\usepackage{amsthm,amsmath,amssymb,bbm}
\usepackage{natbib}
\usepackage{multirow}
\usepackage[pdftex]{graphicx}
\usepackage{caption}
\usepackage{subcaption}
\usepackage{makecell}
\usepackage{booktabs}
\usepackage{array}
\usepackage{tabularx}
\usepackage{tabulary}
\usepackage{url}
\usepackage{algorithm}
\usepackage{algorithmic}
\usepackage{bm}
\usepackage{wrapfig}
\usepackage{lipsum}
\usepackage{mathrsfs}
\usepackage{dsfont}
\usepackage{titling}
\usepackage{paralist}
\usepackage{accents}
\usepackage{csvsimple}
\usepackage{enumerate}
\usepackage[flushleft]{threeparttable}
\usepackage{hyperref}
\usepackage{amsthm,amsmath,amssymb,bbm}
\usepackage{natbib}
\usepackage{multirow}
\usepackage[pdftex]{graphicx}
\usepackage{makecell}
\usepackage{booktabs}
\usepackage{array}
\usepackage{tabularx}
\usepackage{tabulary}
\usepackage{caption}
\usepackage{url}
\usepackage{algorithm}
\usepackage{algorithmic}
\usepackage{bm}
\usepackage{wrapfig}
\usepackage{lipsum}
\usepackage{mathrsfs}
\usepackage{dsfont}
\usepackage{titling}
\usepackage{paralist}
\usepackage{accents}
\usepackage{csvsimple}
\usepackage{enumerate}
\usepackage{url} 
\usepackage[flushleft]{threeparttable}
\usepackage{hyperref}

\newcolumntype{L}{>{\centering\arraybackslash}m{3.1cm}}

\def\T{{ \mathrm{\scriptscriptstyle T} }}

\def\##1\#{\begin{align}#1\end{align}}
\def\$#1\${\begin{align*}#1\end{align*}}

\def\sn{\sum_{i=1}^n}

\newcommand{\SQ}{\mathcal{S}}
\newcommand{\EE}{E}
\newcommand{\RR}{\mathbb{R}}
\newcommand{\cB}{\mathcal{B}}
\newcommand{\cD}{\mathcal{D}}
\newcommand{\cE}{\mathcal{E}}
\newcommand{\cG}{\mathcal{G}}
\newcommand{\cQ}{\mathcal{Q}}
\newcommand{\cL}{\mathcal{L}}
\newcommand{\cS}{\mathcal{S}}
\newcommand{\cN}{\mathcal{N}}
\newcommand{\cR}{\mathcal{R}}

\newcommand{\cO}{\mathcal{O}}
\newcommand{\wt}{\widetilde}

\newcommand{\argmin}{\mathop{\mathrm{argmin}}}
\newcommand{\var}{\text{var}}
\newcommand{\BB}{\mathbb{B}}

\newcommand{\CC}{{\rm C}}

\numberwithin{equation}{section}

\newcommand{\PP}{P}
\usepackage {xcolor}
\newtheorem{theorem}{Theorem}[section]
\newtheorem{condition}{Condition}[section]
\newtheorem{proposition}{Proposition}[section]
\newtheorem{corollary}{Corollary}[section]
\newtheorem{lemma}{Lemma}[section]
\ifx\QED\undefined
\def\QED{~\rule[-1pt]{5pt}{5pt}\par\medskip}
\fi

\newcommand{\blind}{1}

\usepackage{times}
\usepackage{bm}
\usepackage{natbib}

\def\T{{ \mathrm{\scriptscriptstyle T} }}

\def\##1\#{\begin{align}#1\end{align}}
\def\$#1\${\begin{align*}#1\end{align*}}
\def\sn{\sum_{i=1}^n}

\numberwithin{equation}{section}

\usepackage{xcolor}
\allowdisplaybreaks

\numberwithin{equation}{section}

\usepackage{setspace}
\begin{document}

\def\spacingset#1{\renewcommand{\baselinestretch}%
{#1}\small\normalsize} \spacingset{1}


\if1\blind
{
  \title{\bf High-Dimensional Expected Shortfall Regression}
  \author{Shushu Zhang$^\dagger$, Xuming He$^\ddag$, Kean Ming Tan$^\dagger$, and Wen-Xin Zhou$^*$\\ 
  University of Michigan$^{\dagger}$,  Washington University in St. Louis$^\ddag$,\\ and  University of Illinois Chicago$^{*}$} 
  \maketitle
} \fi

\if0\blind
{
 \bigskip
  \bigskip
  \bigskip
  \begin{center}
    {\Large\bf High-Dimensional Expected Shortfall Regression}
\end{center}
  \medskip
} \fi

\bigskip
\begin{abstract}
Expected shortfall is defined as the average over the tail below (or above) a certain quantile of a probability distribution. Expected shortfall regression provides powerful tools for learning the relationship between a response variable and a set of covariates while exploring the heterogeneous effects of the covariates. In the health disparity research, for example,   the lower/upper tail of the conditional distribution of a health-related outcome, given high-dimensional covariates, is often of importance. Under sparse models, we propose the lasso-penalized expected shortfall regression and establish non-asymptotic error bounds, depending explicitly on the sample size, dimension, and sparsity, for the proposed estimator. To perform statistical inference on a covariate of interest, we propose a debiased estimator and establish its asymptotic normality, from which asymptotically valid tests can be constructed. We illustrate the finite sample performance of the proposed method through numerical studies and a data application on health disparity.
\end{abstract}

\noindent%
\vfill

\newpage
\spacingset{1.9} 
	\section{Introduction}
	\label{sec:1}
\begin{figure}[!t]
		\centering
		\includegraphics[scale=0.225]{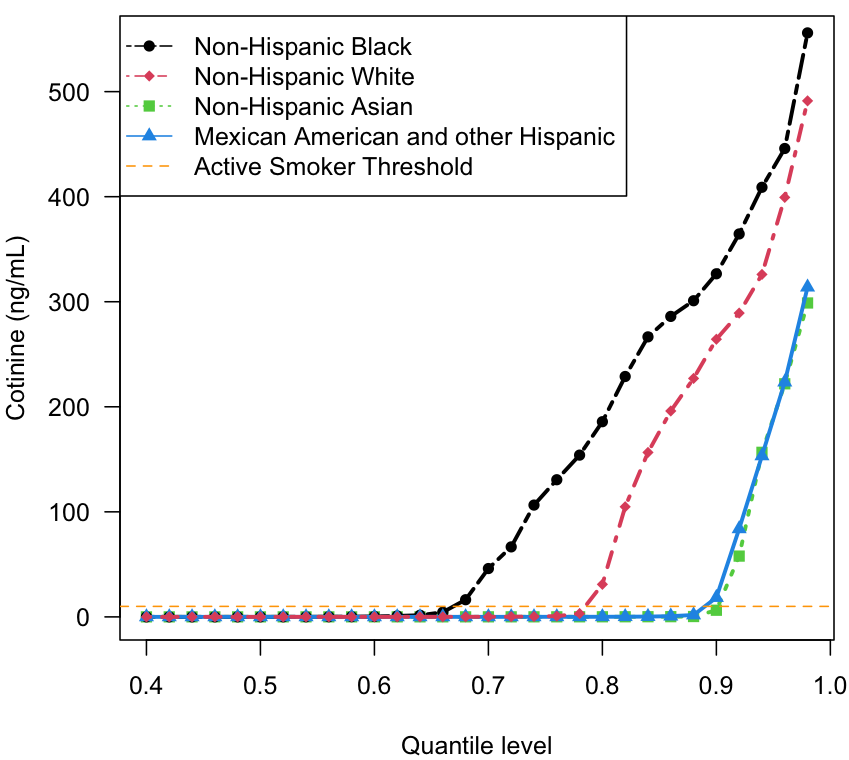}
		\caption{	\label{fig:cotinine} Cotinine levels for four different ethnic groups across the 0.4 quantile to 0.99 quantile. }
	\end{figure}

Recent years have witnessed a proliferation of heterogeneous data with a large number of covariates in various applications, such as genomics \citep{lee2018two} and bioinformatics \citep{cai2021individual}, due to the development of computing facilities and automated data collection technologies such as sensors. As an example,   we consider the health disparity research that has drawn increasing attention from policymakers and regulatory bodies \citep{CDC2005}.  We aim to investigate whether there are disparities in cotinine levels among different ethnic groups given a large number of covariates, where cotinine levels measure exposure to nicotine \citep{Benowitz1996}. 
    In particular, we focus on subpopulations with high cotinine levels who tend to have higher risks for lung cancer \citep{boffetta2006serum,timofeeva2011genetic} and other diseases \citep{mccann1992nicotine}.  Figure~\ref{fig:cotinine} is a plot of cotinine levels for four different ethnic groups at various quantile levels based on the data retrieved from the National Health and Nutrition Examination Survey\footnote{
    Centers for Disease Control and Prevention (CDC). National Center for Health Statistics (NCHS). National Health and Nutrition Examination Survey (NHANES) Data. Hyattsville, MD: U.S. Department of Health and Human Services, Centers for Disease Control and Prevention, 2017-2020, \url{https://wwwn.cdc.gov/nchs/nhanes/continuousnhanes/default.aspx?Cycle=2017-2020}.}. 
    From Figure~\ref{fig:cotinine}, we observe that most of the populations have low cotinine values that are below the active smoker threshold, resulting in no disparity among the ethnic groups at the 0.7 quantile or lower, and the disparity appears larger at the higher quantiles. While quantile regression, first introduced by \cite{KB1978}, is often used to model such data heterogeneity in the tail region,  it concerns a particular quantile level of interest but not about the further tail beyond the quantile.

    To summarize the accumulative information in the tail of an outcome distribution, we consider the expected shortfall (ES), also known as the conditional value-at-risk or superquantile \citep{Rockafellar.etal:2000,A2002}. 
     The ES measure has been used across many disciplines such as finance \citep{rockafellar2002conditional,jorion2003financial,B2019}, operations research \citep{rockafellar2008risk,ROCKAFELLAR2014140,rockafellar2013superquantiles}, and clinical trials \citep{he2010detection}. In particular, the Basel Committee (\url{https://www.bis.org/bcbs/publ/d457.htm})
     confirms the replacement of the quantile with the ES as the standard risk measure in banking and insurance, one reason being that the ES is a coherent risk measure while the quantile is not \citep{artzner1999coherent}.

For notational convenience, let $Z$ be a real-valued random variable. In the regression setting, $Z$ can be interpreted as the response $Y \in \mathbb{R}$ given the covariates $X \in \mathbb{R}^p$.  Let $F_Z: \RR \to [0, 1]$ and $\cQ_{\cdot}(Z): (0, 1) \to \RR$ be its cumulative distribution function and quantile function, respectively, i.e., $F_Z(z) = \PP( Z \leq z)$ for $z\in \RR$  and  $\cQ_{\tau}(Z) = \inf \{ z \in \RR : F_Z(z) \geq \tau \}$ for $\tau  \in (0 ,1)$. The ES represents the average of the tail below or above a given quantile level. Formally, the lower $\tau$th ES of $Z$ is defined as 
 \$
	\cS_\tau(Z) = \frac{1}{ \tau } \int_0^\tau   \cQ_{u}( Z) \,{\rm d} u ,  \ \  \tau  \in (0, 1).  
	\$
 For a continuous random variable $Z$, the lower $\tau$th ES is equivalently given by $\cS_\tau(Z)  = E \{ Z | Z \leq  \cQ_{\tau}(Z) \}$. The upper $\tau$th ES is symmetrically defined as the tail average over the interval $[\tau, 1)$.

To estimate the ES in the regression setting, one challenge is the lack of elicitability \citep{G2011}, i.e., there does not exist a loss function such that the ES is the minimizer of the expected loss. To overcome this challenge, several approaches have been proposed for estimating the (conditional) ES, such as the weighted sum of the estimated quantiles \citep{scaillet2004nonparametric,van2000asymptotic}, 
a function of the estimated conditional cumulative distribution function \citep{chetverikov2022weighted}, or the minimizer of an implicit loss function \citep{rockafellar2013superquantiles,ROCKAFELLAR2014140}. However, the loss function cannot be easily generalized to the regression framework \citep{li2022contributions}. \cite{FZ2016} show that the quantile and the ES are jointly elicitable, and construct a class of joint loss functions. Inspired by this, \cite{DB2019} and \cite{PZC2019} propose a joint regression framework for the conditional quantile and ES in the presence of covariates.
From a different perspective, \cite{B2020} proposes a two-step method for the ES regression and establishes its asymptotic properties under the fixed-$p$ regime. \cite{HTZ2022} investigate this two-step method under the increasing-$p$ regime in which $p=O(n^a)$ for some $a\in (0, 1)$, and further propose a robust two-step estimator that accommodates heavy-tailed data.

To our knowledge, existing inferential methods for the ES regression have been developed mainly for low- to moderate-dimensional covariates. In a data-rich environment, large-$p$ problems arise due to the availability of many features as in genomic studies or the use of multi-way interactions among a moderate amount of covariates as in econometric applications.
Our goal is to develop new estimation and inference methods for the ES regression in high dimensions when $p$ is comparable to or larger than the sample size $n$. Our work is motivated by the aforementioned health disparity study in which a large number of covariates, such as various demographic variables, hospital utilization, access to care, and other survey-type questions, are present. In the health disparity study, it is of particular interest to perform statistical inference on the ethnic groups after adjusting for a large number of explanatory covariates.

High-dimensional regression has been extensively studied under linear and generalized linear models \citep{B2011,W2019}.  One of the most widely used assumptions for high-dimensional regression is sparsity, i.e.,  only a small number of predictors among a large pool are needed in the regression model.  To fit sparse models, various sparsity-inducing penalty functions have been introduced, such as the lasso/$\ell_1$-penalty \citep{T1996}, the elastic net \citep{Z2005} as a combination of the $\ell_1$- and $\ell_2$-penalties, and other non-convex penalty functions \citep{FL2001,Z2010}. Although the rates of convergence of those estimators are well-understood under mild conditions, performing statistical inference in the high-dimensional setting is challenging due to the lack of a tractable limiting distribution of the penalized estimator when $p\gg n$. To overcome this, various debiased/de-sparsified lasso estimators have been proposed \citep{ZZ2014,G2014,NL2017,JM2018}, and proven to follow normal distributions asymptotically under certain growth conditions.

In this article, we propose a regularized two-step method for fitting the ES regression in high-dimensional settings. To perform statistical inference on the impact of a specific covariate (e.g., ethnic group) on the tail average of an outcome (e.g., cotinine levels) given many control variables, we further propose a debiased estimator and establish its asymptotic normality. While our proposal is inspired by the existing debiased procedures (see, for instance, \citealp{NL2017,JM2018}), its theoretical analyses require more delicate investigations due to the presence of high-dimensional nuisance parameters in the conditional quantile specification. Theoretically, we first establish non-asymptotic bounds for the proposed estimator under the $\ell_1$- and $\ell_2$-norms. We then establish the asymptotic normality of the debiased estimator and explicitly characterize its asymptotic variance. We further propose to estimate the asymptotic variance using the refitted cross-validation approach with provable consistency \citep{FGH2012}. Applying the proposed inferential method to a National Health and Nutrition Examination Survey dataset, we gain insights into the disparities in cotinine levels among different ethnic groups after adjusting for many covariates.

	\section{Preliminaries}
	\label{sec:2.1}
	
	Assume that we observe a random sample $\{ (Y_i,  X_i) \}_{i=1}^n$ of size $n$ from $(Y,X) \in \RR \times \RR^p$. We consider the joint regression framework for modeling the conditional quantile and ES of $Y_i$ given $X_i$. In particular, at a given quantile level $\tau  \in (0 , 1)$, assume that
	\#
	 \cQ_\tau( Y_i |  X_i ) = X_i^\T \beta^*(\tau)  , \qquad  \SQ_\tau( Y_i |    X_i ) = X_i^\T \theta^*(\tau) , \qquad i = 1, \ldots, n,\label{joint.reg.model}
	\# 
	where $\beta^*(\tau), \theta^*(\tau) \in \RR^p$ are the true underlying quantile and ES regression parameters. For simplicity, we will suppress the dependency on $\tau$ and write $\beta^*$ and $\theta^*$ throughout the paper. Moreover, the first component of each $X_i$ is one so that the first coordinates of $\beta$ and $\theta$ indicate intercept terms for the quantile and ES regression models, respectively. 
 
 \cite{FZ2016} show that the quantile and the ES are jointly elicitable, and propose a class of strictly consistent joint loss functions for the unconditional quantile and ES estimation. In the regression framework,  let 
 	\#
	S(\beta, \theta; Y, X)  & = \{  \tau  -  \mathbbm{1}( Y \leq X^\T \beta )  \}  \{ G_1(Y) -  G_1(X^\T \beta) \} \label{def:rho}  \\
	&~~~~ + \frac{ G_2( X^\T \theta) }{\tau }  \big\{  (   Y - X^\T \beta ) \mathbbm{1} (Y\leq X^\T \beta)  +  \tau X^\T \beta   -    \tau    X^\T  \theta     \big\}  - \cG_2(X^\T \theta) , \nonumber
	\#
	where $G_1$ and $G_2$ are user-specified functions that satisfy certain conditions. In particular, $G_1$ is increasing and integrable, $G_2=\cG_2'$ is the first order derivative of $\cG_2$, and $G_2$ and $\cG_2$ are strictly positive. Under the model   assumption~\eqref{joint.reg.model},   $(\beta^*, \theta^*)$ is the unique minimizer of $E \{ S(\beta, \theta; Y, X)  \}$;  see Corollary~5.5 in \cite{FZ2016}. Theoretically,  \cite{DB2019} establish the consistency and asymptotic normality for the joint $M$-estimator, defined as $(\wt \beta , \wt \theta) \in \argmin_{ (\beta, \theta) \in \,  \Theta } \sn S(\beta, \theta; Y_i, X_i)$, as $n\to \infty$ with dimension $p$ fixed, where $\Theta \subseteq \RR^{2p}$ is the parameter space, assumed to be compact, convex, and have a non-empty interior. Since the loss function is non-differentiable and non-convex for any feasible choice of the functions $G_1$ and $G_2$ \citep{FZ2016},  it is challenging to obtain the global minimum when $p$ is large.

The ES depends on the quantile but not vice versa. Therefore, the estimation of $\beta^*$ and $\theta^*$ can be disentangled from the joint loss function in~\eqref{def:rho}.  Suppose that the estimation and inference on $\theta^*$ are of primary interest, then $\beta^*$ serves as a nuisance parameter. Motivated by the idea of using Neyman-orthogonal scores to reduce sensitivity with respect to nuisance parameters \citep{N1979, CCDDHNR2018}, \cite{B2020} proposes a two-step procedure that bypasses the non-convex optimization problem. The first step computes the standard quantile regression estimate $\hat \beta$ by minimizing  $\sn \rho_\tau( Y_i - X_i^\T \beta)$ over $\beta$, where $\rho_\tau(u) = \{ \tau - \mathbbm{1} (u<0) \} u$ is the quantile loss function. 
The second step estimates $\theta^*$ through the least squares regression of the adjusted response $Z(\hat \beta)$ on $\tau X$, that is, 
	\#
	\hat \theta  \in \argmin_{\theta \in \RR^p} \frac{1}{n} \sn \{ Z_i(\hat \beta) - \tau X_i^\T \theta \}^2 ,\label{sq.est_low}
	\#
	where
	\#
	Z_i (\beta ) =    (Y_i - X_i^\T \beta   ) \mathbbm{1}(Y_i\leq X_i^\T \beta)    +  \tau X_i^\T \beta, \quad  i=1,\ldots,n. \label{def:Zi}
	\#

Let $ \psi_0(\beta, \theta; X)  =  \tau X [E \{ Z(\beta) | X \} - \tau X^\T \theta]$ be the score function of the expected loss in~\eqref{sq.est_low}, where $Z(\beta)=(Y - X^\T \beta) \mathbbm{1}(Y \leq X^\T \beta)    +  \tau X^\T \beta$. The rationale behind the estimation procedure in~\eqref{sq.est_low} is twofold. First, $\theta^*$ is the unique minimizer of the expected loss function $E[\{Z(\beta^*)-\tau X^T \theta^*\}^2]$ using the property that $\psi_0(\beta^* , \theta^* ; X) = 0$ under~\eqref{joint.reg.model}. 
Second, the quantile regression estimation error is first-order negligible to the ES regression estimation and therefore does not affect the asymptotic distribution of $\hat \theta$ \citep{B2020}, which can be explained by the following orthogonality property.  Let $F_{Y | X}(\cdot)$ be the conditional distribution function of $Y$ given $X$. Provided $F_{Y | X}(\cdot)$ is continuous, it can be shown that for any $\theta \in \RR^p$,
	\#
	\partial_\beta  \psi_0(\beta, \theta; X)  \big|_{\beta = \beta^* } =  \tau^2 X X^\T \{  \tau - F_{Y | X}(X^\T \beta^* )  \} =  0,     \label{neyman.cond}
	\#
 where $\partial_{\beta} \psi_0(\beta, \theta; X)$ denotes the partial derivative of $\psi_0(\beta, \theta; X)$ with respect to $\beta$. We refer to \eqref{neyman.cond} as the Neyman orthogonality.

\section{Methodology}
\label{sec:2}

\subsection{Proposed method}
\label{sec:2.2}
    
We develop a new framework for estimating the ES regression coefficients and performing statistical inference on a particular coefficient of interest, under the joint model \eqref{joint.reg.model} in high dimensions.  We focus on the high-dimensional regime in which $p$ may be greater than $n$, 
and both regression coefficients $\beta^*$ and $\theta^*$ are assumed to be sparse, that is, only a few coordinates of $\beta^*$ and $\theta^*$ are non-zero.
The rationale of the proposed framework will be given in Section~\ref{subsec:onepara}. 

Let $\theta^*_j$ be the $j$th coordinate of $\theta^*$.
For a given $1\le j \le p$ and a pre-determined constant $c_0$, 
our goal is to test the hypothesis 
\begin{equation}
\label{eq:hypothesisnull}
H_0: \theta^*_j = c_0 \qquad \mathrm{versus} \qquad H_1 : \theta^*_j \neq c_0.  
\end{equation}
We start with a two-step $\ell_1$-penalized approach to estimate the ES regression coefficients. 
In the first step, we estimate the quantile regression coefficients by a lasso-penalized estimator given by
	\#
	\hat \beta_{\lambda_q}(\tau)  \in \argmin_{\beta \in \RR^p} \bigg\{ \frac{1}{n} \sn \rho_\tau( Y_i - X_i^{\T} \beta) + \lambda_q    \| \beta \|_1 \bigg\}, \label{qr.est}
	\#
	where $\lambda_q>0$ is a sparsity tuning parameter and $\| \cdot \|_1$ is the $\ell_1$-norm.  Recall from~\eqref{def:Zi} that $Z_i(\beta) =(Y_i - X_i^\T \beta   ) \mathbbm{1}(Y_i\leq X_i^\T \beta)    +  \tau X_i^\T \beta$. 
    Given the above quantile regression estimator $\hat \beta_{\lambda_q}(\tau)$, we propose the $\ell_1$-penalized orthogonal-score least squares estimator 
	\#
	\hat \theta_{\lambda_e}(\tau) \in \argmin_{\theta \in \RR^p}  \bigg\{ \frac{1}{2n} \sn  \big(   Z_i   ( \hat \beta_{\lambda_q}(\tau)   )  - \tau X_i^{\T} \theta \big)^2  + \tau \lambda_e  \| \theta \|_1  \bigg\}  , \label{lasso.ES}
	\#
	where $\lambda_e>0$ is a second sparsity tuning parameter. For notational simplicity, we suppress the dependency of $\tau$, $\lambda_q$, and $\lambda_e$ for the quantile and ES regression estimators in the sequel.

To test the hypothesis  ~\eqref{eq:hypothesisnull}, a naive approach is to construct a test statistic based on $\hat \theta_j$ obtained from~\eqref{lasso.ES}. 
However, even with $\hat \beta_{\lambda_q}$ replaced by the true coefficient $\beta^*$,   $\hat \theta_j$ does not have a tractable limiting distribution due to the $\ell_1$-penalty \citep{G2014}.
To address this challenge, we construct a new score function and propose a debiased estimator for performing valid inference on $\theta^*_j$.  

We introduce some additional notation that will be used throughout the paper. 
Let $X_{i,j}$ be the $j$th covariate for the $i$th observation and let $X_{i,-j} \in \RR^{p-1}$ be a subvector of $X_{i}$ obtained by removing the $j$th entry.  
Similarly, let $\theta_{-j} \in \RR^{p-1}$ be a subvector of $\theta$ with the $j$th element removed. Let $\Sigma = (\sigma_{jk})_{1\leq j, k \leq p} = E(X X^{\T})$, and $\Sigma_{-j}$ be a submatrix of $\Sigma$ with $j$th row and column removed. Let 
\#
\hat \gamma \in \argmin_{\gamma \in \RR^{p-1}}  \frac{1}{2n } \sn (X_{i,j} - X_{i,-j}^{\T} \gamma)^2 + \lambda_m \| \gamma \|_1 \label{decorr.est}
\#
be an estimator obtained by projecting $X_{i,j}$ onto $X_{i,-j}$.  
We define $S_n(\theta_j, \theta_{-j} , \beta ,   \gamma ) = n^{-1} \sn  \{ Z_i(\beta) - \tau X_{i,j} \theta_j - \tau X_{i,-j}^{\T} \theta_{-j} \} ( X_{i,j} - X_{i,-j}^{\T} \gamma)$ as the triply-orthogonal score function. To test the hypothesis \eqref{eq:hypothesisnull}, we propose the decorrelated score test that rejects the null hypothesis with significance level $\alpha$ if 
\begin{equation}\label{reject}
    | n^{1/2} S_n( c_0 ,  \hat \theta_{-j} (c_0) , \hat  \beta ,  \hat \gamma ) / \hat \sigma_{{\rm s}} | > \Phi^{-1} ( 1-\alpha/2),
\end{equation}
where $\hat \theta_{-j} (c_0)$ is obtained by minimizing \eqref{lasso.ES} with $\theta_j$ fixed at $c_0$, to be defined formally in~\eqref{initial.estimators}, and $\hat \sigma_{{\rm s}}$ is an estimator of the variance term to be defined in Section~\ref{sec:2.4}. 

To construct a confidence interval for $\theta^*_j$, we propose the   following debiased estimator 
 \begin{equation}\label{debiased.estimator}
      \begin{aligned}
     \wt \theta_j &= \hat \theta_j -  \frac{S_n(\hat \theta_j,\hat \theta_{-j}, \hat \beta, \hat \gamma)} {\partial_{b} S_n(b,\hat \theta_{-j}, \hat \beta, \hat \gamma) |_{b = \hat \theta_j}}\\
 &=\hat \theta_j +  \frac{\sn  \{ Z_i(\hat \beta) - \tau X_{i,j} \hat \theta_j - \tau X_{i,-j}^{\T} \hat \theta_{-j} \} ( X_{i,j} - X_{i,-j}^{\T} \hat \gamma)} {\tau \sn   X_{i,j} ( X_{i,j} - X_{i,-j}^{\T} \hat \gamma)}.  
 \end{aligned}
 \end{equation}
Then, a Wald-type 100$(1-\alpha)$\% confidence interval for $\theta_j^*$ can be constructed as
	\#
	\wt \theta_j \pm \Phi^{-1}(1-\alpha/2) \frac{  \hat \sigma_{{\rm s}} }{  n^{1/2} \tau \hat \sigma_\omega^2}, \label{wald.ci}
	\#
where both $\hat{\sigma}_{\omega}^2$ and $\hat{\sigma}_{s}^2$ are constructed using a refitted cross-validation procedure described in Section~\ref{sec:2.4}.
		
	\subsection{Motivations for the proposed method}
	\label{subsec:onepara}
	In this section, we provide the motivations of the proposed inferential framework in Section~\ref{sec:2.2}. 
 Let $l_n(\beta, \theta_j, \theta_{-j})=(2\tau n)^{-1} \sn \{ Z_i(\beta) - \tau X_{i,j}^\T \theta_j - \tau X_{i,-j}^\T \theta_{-j} \}^2,$ and  let 
 \#
	 	\hat \theta_{-j}(c_0)  \in \argmin_{ \theta_{-j} \in \RR^{p-1}  }  \frac{1}{2 n}\sn \{ Z_i(\hat \beta)  - \tau X_{i,j} c_0 -  \tau X_{i,-j}^{\T}  \theta_{-j}\}^2 + \tau \lambda_e \|   \theta_{-j} \|_1 ,
 \label{initial.estimators}
	\#
 which provides an estimate of $\theta^*$ with $\hat{\theta}_j$ constrained to equal  $c_0$. 
 As mentioned in Section~\ref{sec:2.2}, inference using $\hat \theta_j$ is challenging because the score function $n^{1/2} \partial_{\theta_j} l_n( \hat \beta, \theta_j, \hat \theta_{-j}(c_0) )|_{\theta_j=c_0}$ does not have a tractable limiting distribution. 
 
Following the ideas behind the debiased procedures for penalized linear regression methods, we construct a triply-orthogonal score function \begin{equation}\label{decorr.score}
     \begin{aligned}
         S_n(\theta_j,\theta_{-j},\beta,\gamma) & = \partial_{\theta_j} l_n(\beta,\theta_j,\theta_{-j}) - \gamma^\T \partial_{\theta_{-j}} l_n(\beta, \theta_j,\theta_{-j})\\
         & = \frac{1}{n} \sn  \{ Z_i(\beta) - \tau X_{i,j} \theta_j - \tau X_{i,-j}^{\T} \theta_{-j} \} ( X_{i,j} - X_{i,-j}^{\T} \gamma)
     \end{aligned}
 \end{equation}
 that is orthogonal to the nuisance score function $\partial_{\theta_{-j}} l_n(\beta, \theta_j, \theta_{-j})$ with $\gamma \in \RR^{p-1}$. 
 For convenience in the presentation, we introduce the following model that linearly projects the covariate of interest $X_{i,j}$ on the remaining covariates $X_{i, -j} \in \RR^{p-1}$, i.e.,
	\#
	X_{i,j} = X_{i,-j}^{\T} \gamma^* + \omega _i ,    \quad  E( \omega_i  X_{i,-j} )= 0,   \quad  i = 1,\ldots, n , \label{auxiliary.model}
	\# 
 where $\gamma^* = \argmin_{\gamma \in \RR^{p-1}} E (X_{i,j} - X_{i,-j}^{\T} \gamma)^2 = \Sigma_{-j}^{-1} E ( X_{i,j} X_{i,-j} )$ under the condition that $\Sigma_{-j } = E (X_{i,-j} X_{i,-j}^{\T}) \in \RR^{(p-1)\times(p-1)}$ is non-singular.  Moreover, we write $\sigma_{\omega}^2 = E(\omega_i^2) = \sigma_{jj} - (\gamma^*)^\T \Sigma_{-j} \gamma^*$.

Under \eqref{joint.reg.model} and \eqref{auxiliary.model}, the following orthogonality conditions 
 \begin{equation}\label{orthogonal.moments}
 \begin{aligned}
     \partial_{{\beta}} E \{S_n(\theta_j , \theta_{-j}  , {\beta} ,  \gamma  ) \} \big|_{ \beta   = \beta^*}  = 0  & \mbox{ for any}~  \theta_j, \theta_{-j} , \gamma  , \\  
		\partial_{\theta_{-j}} E  \{S_n(\theta_j , \theta_{-j} , {\beta}  ,  \gamma^* ) \}   = 0 &\mbox{ for any}~ \theta_j, \theta_{-j} ,  \beta  ,  \\  
		\partial_{\gamma} E \{S_n(\theta_j^*, \theta_{-j}^* , {\beta}^* ,  \gamma )  \}   = 0 & \mbox{ for any}~  \gamma,
\end{aligned}
 \end{equation}
play a key role in controlling the statistical errors from estimating the nuisance parameters $\beta^* \in  \RR^{p}$  and $\theta^*_{-j}, \gamma^* \in \RR^{p-1}$, which may converge at a slower than $n^{1/2}$ rate due to high-dimensionality.
As a result, the estimation errors of the three penalized estimators $\hat \beta$ and $\hat \theta_{-j}(c_0)$ and $\hat \gamma$ are first-order negligible to the estimation of $\theta_j$ so that $n^{1/2} S_n(c_0 ,  \hat \theta_{-j} (c_0) , \hat  \beta ,  \hat \gamma)$ and $ n^{1/2} S_n(c_0 ,    \theta^*_{-j},  \beta^* ,  \gamma^*)$ have the same asymptotic distribution under $H_0$ in~\eqref{eq:hypothesisnull}.

	To test the hypothesis \eqref{eq:hypothesisnull}, we first obtain estimators of $\beta^*$, $\theta_{-j}^*$ and $\gamma^*$ by solving \eqref{qr.est}, \eqref{initial.estimators}, and \eqref{decorr.est}, respectively. 
Given the estimators $\{\hat  \beta ,\hat \theta_{-j} (c_0) ,   \hat \gamma\}$, we will show in Section~\ref{sec:theory} that under certain moment and scaling conditions, 
	 $n^{1/2} S_n(c_0 ,  \hat \theta_{-j} (c_0) , \hat  \beta ,  \hat \gamma)  / \sigma_{{\rm s}}$ converges in distribution to $ \mathcal{N}(0, 1)$ under $H_0$, where
  $\sigma_{{\rm s}}^2 = E\{ \omega_i^2 \var(\varepsilon_{i,-}  | X) \}$ with $\omega_i = X_{i,j} - X_{i,-j}^{\T} \gamma^*$,  $\varepsilon_{i,-} = \min(\varepsilon_i, 0)$, and $\varepsilon_i = Y_i- X_i^{\T} \beta^*$. 
 Let $\hat \sigma_{{\rm s}}$ be a generic estimator of $\sigma_{{\rm s}}$.  
 Then, we reject $H_0$ in~\eqref{eq:hypothesisnull} if \eqref{reject} holds with significance level $\alpha$.  
 
	To construct a confidence interval for $\theta_j^*$, we propose a debiased estimator by using a one-step approximation to the triply-orthogonal score function. We show that the proposed debiased estimator is asymptotically normal for which normal-based confidence intervals can be constructed using the estimated asymptotic variance. 
	 To begin with, let $\hat \beta$, $\hat \theta$, and $\hat \gamma$ be defined as in \eqref{qr.est}, \eqref{lasso.ES}, and \eqref{decorr.est}, respectively.  Although $\hat \theta_j$ is a consistent estimate of $\theta^*_j$ under certain conditions, we will show in Section~\ref{sec:theory} that $\hat \theta_j$ is not root-$n$ consistent. Motivated by the classical one-step construction \citep{B1975} to the triply-orthogonal score function in~\eqref{decorr.score}, which aims at improving an initial estimator that is already consistent but not efficient, we propose the debiased estimator as defined in~\eqref{debiased.estimator}. We will show in Section~\ref{sec:theory} that $n^{1/2} \tau (\wt \theta_j - \theta^*_j )$ converges in distribution to $ \mathcal{N}(0, \, \sigma_{{\rm s}}^2 /  \sigma^4_\omega )$, which enables the construction of Wald-type confidence intervals in~\eqref{wald.ci}.


	\subsection{Variance estimation based on refitted cross-validation}
	\label{sec:2.4}
	
	The construction of confidence intervals based on either the decorrelated score test or the Wald-type statistic depends on the estimation of the asymptotic variance, or equivalently, $\sigma_{{\rm s}}^2$ and $\sigma_\omega^2$ as defined in Section~\ref{subsec:onepara}.
	The variance estimation in high-dimensional linear models is an inherently difficult task because false discoveries occur on the lasso path due to spurious correlations,  which leads to a downward bias for the lasso residual sum of squares estimator \citep{FGH2012}. 
We propose a refitted cross-validation method to estimate $\sigma_{{\rm s}}^2$ and $\sigma_\omega^2$.



 To begin with, we randomly split the dataset into two approximately equal-sized parts,  indexed by $S_1$ and $S_2$ satisfying $|S_1 | - |S_2| \in \{0,1\}$ and $S_1 \cup S_2 = \{1, \ldots, n\}$.  Use the first half $S_1$ to compute $\hat \beta^{{\rm cv}}$,  $\hat \theta^{{\rm cv}}$, and  $\hat \gamma^{{\rm cv}}$ by solving \eqref{qr.est}, \eqref{lasso.ES}, and \eqref{decorr.est},  respectively, where the tuning parameters $\lambda_q$, $\lambda_e$, and $\lambda_m$ are determined by cross-validations. Let $\hat S_q$, $\hat S_e$, and $\hat S_m$ be the sets of selected variables with cardinalities $\hat s_q$, $\hat s_e$, and $\hat s_m$, respectively. With the $\ell_1$-penalties that encourage sparsities for $\hat \beta^{{\rm cv}}$,  $\hat \theta^{{\rm cv}}$, and  $\hat \gamma^{{\rm cv}}$, we implicitly assume that $|S_2 | > \hat s_m +\hat s_q +\hat s_e$. Then we use the second half $S_2$ to compute 
\$
	\begin{cases}
		\hat \beta^{(2)} \in \argmin_{ \beta \in \RR^{\hat s_q} }  \sum_{i\in S_2} \rho_\tau(Y_i - X_{i, \hat S_q}^{\T}  \beta)  , \\
		\hat \theta^{(2)}  \in \argmin_{\theta \in \RR^{\hat{s}_e}  }  \sum_{i\in S_2} ( \hat Z_i^{(2)} - \tau X_{i, \hat S_e}^{\T}  \theta)^2 , \\
		\hat \gamma^{(2)} \in \argmin_{\gamma \in \RR^{\hat s_m}}  \sum_{i\in S_2} (X_{i,j} - X_{i, \hat S_m}^{\T} \gamma)^2,
	\end{cases}
	\$
	where $\hat Z_i^{(2)} = (Y_i - X_{i,\hat S_q}^{\T} \hat \beta^{(2)} ) \mathbbm{1}(Y_i \leq X_{i,\hat S_q}^{\T} \hat \beta^{(2)} ) + \tau X_{i,\hat S_q}^{\T} \hat \beta^{(2)} $.  We define intermediate variance estimators of $\sigma_s^2$ and $\sigma_{\omega}^2$ as
	\# \label{eq:asympvariances}
	\hat \sigma_{{\rm s}, 1}^2 =  \frac{1}{|S_{2}|-\hat s_m -\hat s_q -\hat s_e  } \sum_{i \in S_{2}}   (\hat \omega^{(2)}_i)^2 \big( \hat Z_i^{(2)}  - \tau X_{i,  \hat S_e }^{\T} \hat \theta^{(2)} \big)^2   , \quad \hat \sigma^2_{\omega, 1} = \frac{1}{| S_2 | - \hat s_m } \sum_{i\in S_2}  (\hat \omega^{(2)}_i)^2 ,
	\#
	where $\hat \omega^{(2)}_i = X_{i, j}  - X_{i, \hat S_m}^{\T} \hat \gamma^{(2)}$.  Next,  we switch the roles of $S_1$ and $S_2$ and repeat the above process to obtain  the variance estimates $\hat \sigma_{{\rm s}, 2}^2$ and  $ \hat \sigma^2_{\omega, 2}$.  The final variance estimators of $\sigma_s^2$ and $\sigma_{\omega}^2$ are given by
    \# \label{eq:asympvariances.est}
	\hat{\sigma}_{{\rm s}}^2 = (\hat{\sigma}_{{\rm s}, 1}^{2 } + \hat{\sigma}_{{\rm s}, 2}^{2 })/2 ~~\mbox{ and }~~
	\hat{\sigma}_{\omega}^2 = (\hat{\sigma}_{\omega, 1}^{2  } + \hat{\sigma}_{\omega, 2}^{2 })/2.
    \#

	\section{Theoretical analysis}
	\label{sec:theory}
	We first provide the convergence rates for the two-step $\ell_1$-penalized ES regression estimator through non-asymptotic high probability bounds. Then, we establish the asymptotic normality of the triply-orthogonal score function in~\eqref{decorr.score} under $H_0$ and the proposed debiased estimator $\wt \theta_j$ defined in~\eqref{debiased.estimator}. Finally, we demonstrate the consistency of the refitted cross-validated estimator to the asymptotic variance as described in Section~\ref{sec:2.4}.
	  The proof of all the theorems and propositions in this section can be found in Section S 1 of the online Supplementary Materials.

 We analyze the proposed estimators in the high-dimensional setting in which $p$ may be larger than $n$. For any $p$-dimensional vector $u =(u_1,\ldots, u_p)^\T \in \mathbb{R}^p$, we use $\| u \|_0 = \sum_{j=1}^p \mathbbm{1} (u_j\ne 0)$ to denote the number of nonzero elements. Let $\|u\|_2 = (\sum_{j=1}^p u_j^2)^{1/2}$ and let $\| u \|_{\infty} = \max_{i=1,\ldots,p} |u_i|$ be the $\ell_2$-norm and $\ell_{\infty}$-norm of $u$, respectively.  
Throughout our analysis, we assume that $\beta^*$, $\theta^*$, and $\gamma^*$ in \eqref{joint.reg.model} and \eqref{auxiliary.model} are sparse with sparsity levels $s =  \max (\| \beta^*\|_0, \| \theta^* \|_0)$ and $s_0  = \|\gamma^*\|_0$. 
In the following, we define additional notation that will be used throughout the paper. Let $\sigma_X^2 = \max_{1\leq j\leq p} \sigma_{jj}$,  and $\mathbb{S}^{p-1} = \{u \in \RR^p : \|u\|_2=1\}$. For any subset $\cS \subseteq \{1,\ldots,p\}$, let $\delta_{\cS}$ 
be a subvector of $\delta$ indexed by $\cS$.
Let  $\mathbb{C}(\cS ) = \big\{ \delta \in \RR^p : \| \delta_{\cS^{{\rm c}}} \|_1 \leq 3 \| \delta_{\cS} \|_1 \big\}$ be an $\ell_1$-cone.
 
 Recall that $\varepsilon = Y - X^{\T} \beta^*$ denotes the residual in the linear quantile model \eqref{joint.reg.model}, satisfying $\PP(\varepsilon\leq 0| X) =\tau$,
 and  $\varepsilon_-  = \min(\varepsilon, 0)$ is its negative part.
 We further impose the following regularity conditions.
 
\begin{condition} \label{cond:density}
The conditional cumulative distribution function $F_{\varepsilon | X}(\cdot)$ of $\varepsilon$ given $X$ is continuously differentiable and satisfies $| F_{\varepsilon | X}(t) -F_{\varepsilon | X}(0) |  \leq  f_u |t|$ for all $t\in \RR$, where $f_u $ is a positive constant.   Moreover, $\varepsilon_-$ satisfies a conditional Bernstein-type condition (e.g., Theorem 2.10 in \citealp{BLM2013}), that is, $E(\varepsilon_-^2| X) \leq \sigma_\varepsilon^2$ and $E( |\varepsilon_-|^k | X) \leq  k! \sigma_\varepsilon^2 b_\varepsilon^{k-2} /2$ for any integer $k\geq 3$ and some constants  $b_\varepsilon \geq \sigma_\varepsilon>0$. 
	\end{condition}

	\begin{condition} \label{cond:covariate}
 The quantile and ES regression coefficients $\beta^*$ and $\theta^*$ are sparse with $s =  \max (\| \beta^*\|_0, \| \theta^* \|_0)\in [1,\min(p, n)]$. There exist constants $b_X,m_4 \geq 1$ such that 
		(i) $\| X\|_\infty \leq b_X$ (almost surely)  and 
		(ii) $\sup_{u \in \mathbb{S}^{p-1}} E \{(X^{\T} u)^4\}/  [ E \{(X^{\T} u)^2\}   ]^2  \leq m_4$. In addition, let $\Sigma = E(X X^\T)$ and  define the following restricted minimum eigenvalue
		\#
		\underline{\phi}^2 =  \inf_{\cS \subseteq \{1,\ldots,p\}: | \cS |\leq s } ~\inf_{\delta \in \mathbb{C}(  \cS ) } \frac{\delta^{\T} \Sigma \delta }{\| \delta_{\cS} \|_2^2 } >0 . \label{re.cond}
		\# 
	\end{condition}
	
\noindent Condition~\ref{cond:density} requires the negative part of $\varepsilon$ to satisfy conditional Bernstein-type conditions, which are weaker than sub-Gaussian conditions that are often imposed in high-dimensional statistics \citep{W2019}. Condition~\ref{cond:covariate} is mainly used to guarantee that certain population and empirical quantities, e.g., the objective or the gradient functions, are uniformly close to each other in a compact region. The boundedness condition on $X$ can be replaced by a sub-Gaussian assumption for the same results to hold but at the cost of a stronger sample size requirement.  In this case,  we have $\| X \|_\infty \lesssim  \sqrt{\log (p)}$ with high probability.

Given a positive definite matrix $\Sigma \in \RR^{p \times p}$ and $r>0$, write $\BB_1(r) =  \{u \in \RR^p: \|u\|_1 \leq r \}$ and  $\BB_{\Sigma}(r) = \{u \in \RR^p: \|u\|_\Sigma \leq r\}$, where $\|u\|_\Sigma=\|\Sigma^{1/2}u\|_2$. Let $m_3 = \sup_{u \in \mathbb{S}^{p-1}} E (|X^{\T} u|^3) /\{ E (X^{\T} u)^2 \}^{3/2}$. 
In the following theorem, we establish upper bounds of the estimation error for the proposed ES estimator obtained from~\eqref{lasso.ES} under the $\ell_{\Sigma}$-  and $\ell_1$-norms.

	\begin{theorem} 
 \label{thm:lasso-ES}
		Assume Conditions~\ref{cond:density} and \ref{cond:covariate} hold. For any $r_0, r_1 >0$ and $t, u >0$, let 
	$\lambda_e \geq  \max\{   (5.5 \sigma_X \sigma_\varepsilon + 7  \sigma_X b_X    r_1 )  \sqrt{t/n} , \,  f_u m_3  \underline{\phi} s^{-1/2} r_0^2 \}$.	 
	 Then,  the $\ell_1$-penalized ES estimator $\hat \theta$ in~\eqref{lasso.ES} satisfies, with probability at least $1- e^{-t + \log(2p+2p^2)} - e^{-u}$ conditioned on $\{\hat \beta \in \beta^* + \BB_\Sigma(r_0) \cap \BB_1(r_1) \}$, the error bounds
		\#
		\tau  \| \hat \theta  - \theta^* \|_\Sigma   \leq 4   \underline{\phi}^{-1} s^{1/2} \lambda_e 
		~~\mbox{ and }~~  \tau  \| \hat \theta  - \theta^* \|_1 \leq  20\underline{\phi}^{-2} s \lambda_e  ,  \label{ES.est.bound}
		\#
		 provided that $n \geq C \max [  m_4   \{\sigma_X^2 \underline{\phi}^{-2} s \log(2p) + u \},   \,   (b_X/\sigma_X)^2 \{\log (2p)  + t\} ]$ for some sufficiently large constant $C>1$.
	\end{theorem}

Note that the ES estimation error bounds in Theorem~\ref{thm:lasso-ES} are conditioned on the $\ell_1$-penalized quantile regression estimation error bounds, namely, $\|\hat \beta - \beta^*\|_{\Sigma} \leq r_0$ and $\|\hat \beta - \beta^*\|_1 \leq r_1$. 
 To derive the upper bounds in~\eqref{ES.est.bound} explicitly with respect to $s$, $p$, and $n$, Proposition~\ref{prop:qr} below provides the convergence rates of $\hat \beta$ under $\ell_{\Sigma}$-  and $\ell_1$-norms. These results recover those from \cite{BC2011} and \cite{WH2021} under comparable conditions but are more transparent with explicit dependency on different model parameters.

	\begin{condition} \label{cond:density2}
		The conditional density function of $\varepsilon$ given $X$, denoted by $f_{\varepsilon | X}(\cdot)$, exists and is continuous on its support. Moreover, there exist constants $f_l , l_0 > 0$ such that $ f_{\varepsilon | X}(0) \geq  f_l$ and $|f_{\varepsilon | X}(t) - f_{\varepsilon | X}(0) | \leq l_0 |t|$ for all $t\in \RR$ almost surely over $X$.
	\end{condition}
	
	\begin{proposition} \label{prop:qr}
		Assume Conditions~\ref{cond:covariate} and~\ref{cond:density2} hold. For any $t>0$, let
\#
		\lambda_q \geq  2 \big\{ \sigma_X \sqrt{   2 \tau(1-\tau) t/n  } + \bar \tau b_X t/(3n)  \big\} \label{qr.lambda.constraint}
		\#
  where $\bar \tau = \max(\tau, 1-\tau)$.
  Then,  the $\ell_1$-penalized quantile regression estimator $\hat \beta =\hat{\beta}_{\lambda_q}$ defined in \eqref{qr.est} 
		satisfies the error bounds
		\$
		\| \hat \beta - \beta^* \|_\Sigma \leq \frac{2}{f_l}    s^{1/2} r(n,   t)  ~~\mbox{ and }~~
		\| \hat \beta - \beta^* \|_1 \leq  \frac{8}{\underline{\phi}  f_l }  s \cdot r(n, t)  
		\$
		with probability at least $1-  e^{-t+\log(4p)}$ as long as $s^{1/2} r(n,  t)  < 3 f_l^2/(4  m_3 l_0 )$, where $r(n,   t) =  \underline{\phi}^{-1}  \{   32 \bar \tau\sigma_X     \sqrt{2t /n}   + 32 \bar \tau b_X   t/(3n)   + 2 \lambda_q \}$.
	\end{proposition}

As a direct consequence of Theorem~\ref{thm:lasso-ES} and Proposition~\ref{prop:qr}, 
we present explicit error bounds depending on $(n, p, s)$ in  Corollary~\ref{cor:ES_bound}.

\begin{corollary}\label{cor:ES_bound}
Assume Conditions~\ref{cond:density}--\ref{cond:density2} hold. Under the sample size requirement 
$n \geq (C s)^2 \log(p)$ where $C>0$ is a constant,
with penalization parameters $\lambda_q$ and $\lambda_e$ satisfying $C_1 \sqrt{\log(p)/n} \leq \lambda_q \leq C_2 \sqrt{\log(p)/n}$, 
$C_3 \sqrt{\log(p)/n} \leq \lambda_e \leq C_4 \sqrt{\log(p)/n}
$
where 
$C_1 = 4 \sigma_X \sqrt{   \tau(1-\tau)} + 4 \bar \tau b_X/(3 C)$ and $C_3 = \max\{5.5 \sigma_X \sigma_\epsilon + 6  \sigma_X b_X C_6/C, f_u m_3 \underline{\phi}/C \}$,
and $C_2$ and $C_4$ are some sufficiently large constants, 
the $\ell_1$-penalized ES estimator $\hat \theta$ satisfies the bounds 
$\tau \| \hat \theta - \theta^* \|_\Sigma \lesssim \sqrt{s\log(p)/n}$ and $\tau \| \hat \theta - \theta^* \|_1 \lesssim  s \sqrt{\log(p)/n} $
 with probability at least $1-9/p$.
\end{corollary}

From Theorem~\ref{thm:lasso-ES} and Proposition~\ref{prop:qr}, we observe that the quantile regression estimation error only affects the final convergence rates of the ES estimation through higher-order terms under the scaling condition $n\gtrsim s^2 \log(p)$. Specifically, the terms $s^{-1/2} r_0^2 \asymp s^{1/2} \log(p) /n$ and $r_1 \sqrt{\log(p)/n} \asymp s \log(p) /n$ are dominated by the leading term $\sqrt{\log(p)/n}$ in $\lambda_e$. This coincides with the orthogonality condition in~\eqref{neyman.cond}. Next,  we establish the asymptotic normality for both the triply-orthogonal score in~\eqref{decorr.score} under the null hypothesis \eqref{eq:hypothesisnull} and the debiased estimator $\wt \theta_j$ in~\eqref{debiased.estimator} under  models~\eqref{joint.reg.model} and~\eqref{auxiliary.model}. These asymptotic results further require the following condition on the estimation error $\omega_i$ from regressing $X_{i, j}$ on $X_{i,-j}$. 
	
\begin{condition} \label{cond:moment.U}
The projection coefficient vector $\gamma^* = \argmin_{\gamma \in \RR^{p-1} } E(X_{i,j} - X_{i,-j}^\T \gamma)^2$ is $s_0$-sparse with $1\leq s_0<\min(p, n)$.
The projection residual $\omega_i=X_{i,j} - X_{i, -j}^\T \gamma^*$ satisfies $|\omega_i |\leq b_X$.
Moreover, assume that Condition~\ref{cond:covariate} applies also to $X_{i,-j}$ and $\Sigma_{-j}=E(X_{i,-j}X_{i,-j}^\T)$.
\end{condition}

To establish convergence rates for the lasso regression estimator from~\eqref{decorr.est} in Proposition~\ref{prop:gamma.rate} under $\Sigma_{-j}$- and $\ell_1$-norms, we require an additional restricted eigenvalue condition similar to Condition~\ref{cond:covariate} but imposed on $X_{i,-j}$. This condition is implied by Condition~\ref{cond:covariate} by taking $\delta_j=0$ in~\eqref{re.cond}. For consistency, we assume Condition~\ref{cond:covariate} for the error bounds in Proposition~\ref{prop:gamma.rate}.

	\begin{proposition} \label{prop:gamma.rate}
		Assume Conditions~\ref{cond:covariate} and~\ref{cond:moment.U} hold. For any $t, u>0$, the lasso estimator $\hat \gamma$ in \eqref{decorr.est}  with $ \lambda_m \geq 2 b^2_X \sqrt{2t /n}$ satisfies the bounds  $\| \hat \gamma - \gamma^* \|_{\Sigma_{-j}} \leq 3 \underline{\phi}^{-1} s_0^{1/2} \lambda_m$ and $\| \hat \gamma - \gamma^* \|_1 \leq 12  \underline{\phi}^{-2} s_0  \lambda_m$ with probability at least $1-2e^{-t+\log (p-1)} - e^{-u}$, provided $n \geq C \max [  m_4   \{\sigma_X^2 \underline{\phi}^{-2} s_0 \log(2p-2) + u \},   \,   (b_X/\sigma_X)^2 \{\log (2p-2)  + t\} ]$ for some sufficiently large constant $C>1$. 
	\end{proposition}

 With the error bounds for the quantile, ES, and lasso regression estimation in Theorem~\ref{thm:lasso-ES}, Proposition~\ref{prop:qr}, and Proposition~\ref{prop:gamma.rate}, respectively, next we establish the asymptotic normality for the triply-orthogonal score function under $H_0$ in the following theorem. 
  
	\begin{theorem}  \label{thm:score.clt}
  Assume Conditions~\ref{cond:density}--\ref{cond:moment.U} hold. Under the null hypothesis \eqref{eq:hypothesisnull}, let $\hat \beta$, $\hat \theta_{-j}(c_0)$, and $\hat \gamma$ be the estimators given in~\eqref{qr.est},~\eqref{initial.estimators}, and~\eqref{decorr.est}, respectively. 
Then,  with regularization parameters $\lambda_q, \lambda_e, \lambda_m$ all in the order of $\sqrt{\log(p)/n }$,  $n^{1/2}  S_n(  c_0 ,   \hat \theta_{-j}(c_0), \hat \beta,  \hat \gamma )   \to  \mathcal{N}(0,  \sigma_{{\rm s}}^2 )$ in distribution as $n, p \to \infty$, subject to $\max(s, s_0) \log (p) = o(n^{1/2})$. 
	\end{theorem}

 Based on the asymptotic normality for the triply-orthogonal score function $S_n$ in Theorem~\ref{thm:score.clt}, and that the partial derivative of $S_n$ with respect to $\theta_j$ converges to some constant in probability, we show that the debiased estimator defined in~\eqref{debiased.estimator} is asymptotic normal in Theorem~\ref{thm:debias.clt}.
	
\begin{theorem} \label{thm:debias.clt}
Assume Conditions~\ref{cond:density}--\ref{cond:moment.U} hold with $\sigma_\omega^2 = E(\omega_i^2)$ bounded away from zero.
  Then, the debiased estimator $\wt \theta_j$ defined in \eqref{debiased.estimator} satisfies $n^{1/2} \tau( \wt \theta_j - \theta^*_j) \to  \cN( 0 ,   \sigma_{{\rm s}}^2/\sigma_\omega^4 )$ in distribution as $n, p \to \infty$, subject to $\max(s, s_0) \log (p) = o(n^{1/2})$.
	\end{theorem}

To conclude this section, we establish the consistency of the two variance estimators $\hat{\sigma}_{\omega}^2$ and $\hat{\sigma}_{{\rm s}}^2$, obtained through refitted cross-validation. 
The result involves more delicate investigations than \cite{FGH2012} since $\hat{\sigma}_{{\rm s}}^2$ involves a summation of products of two residual squares, along with three sources of bias in $\hat \beta$, $\hat \theta$, and $\hat \gamma$. 
Let $S_q^*$, $S_e^*$, and $S_m^*$ be the support of $\beta^*$, $\theta^*$, and $\gamma^*$, respectively. 
The verification of the consistency here further requires the following sure screening condition, which is the weakest model consistency property for sparse regression in the high-dimensional setting \citep{FGH2012}.



\begin{condition}\label{cond:sure_screen}
The support of the lasso estimators from the first stage, namely, $\hat S_q$, $\hat S_e$, and $\hat S_m$ and their corresponding cardinalities $\hat s_q$, $\hat s_e$, and $\hat s_m$, as defined in Section~\ref{sec:2.4}, satisfy $P(S_q^* \subseteq \hat S_q) \to 1, P(S_e^* \subseteq \hat S_e) \to 1$, and $P(S_m^* \subseteq \hat S_m) \to 1$, as $n,p \to \infty$,  with $\hat s_q \leq b_n$, $\hat s_e \leq b_n$, and $\hat s_m \leq b_n$, where $b_n$ is some sequence satisfying $b_n^2 \log(p)=o(n)$. 
\end{condition}

 \begin{theorem} \label{thm:var.est}
    In addition to Conditions~\ref{cond:density}--\ref{cond:sure_screen}, assume that $\| \gamma^* \|_1$ is bounded and $   \underline{\lambda}  \le \lambda_{\min} (\Sigma) \le \lambda_{\max} (\Sigma)\le \bar{\lambda}$ for some positive constants $\bar{\lambda} \geq \underline{\lambda}$. With penalization parameters $\lambda_q, \lambda_e \asymp  \sqrt{\log(p)/n}$ and the sample size satisfying $\max(s^2,s_0^2) \log(p)=o(n)$,
        the refitted cross-validated variance estimator $\hat{\sigma}_{{\rm s}}^2/\hat{\sigma}_{{\omega}}^4$ as defined in~\eqref{eq:asympvariances.est} is consistent.
 \end{theorem}

	\section{Numerical studies}
	\label{sec:numerical}
		We conduct numerical studies to examine the empirical performance of the proposed estimator for both estimation and inference in Sections~\ref{subsec:numerical_est} and~\ref{subsec:Inf}, respectively. 
	\subsection{Estimation on a linear heteroscedastic model}
	\label{subsec:numerical_est}
The proposed two-step method involves computing an $\ell_1$-penalized quantile regression estimator and a lasso-type estimator with the adjusted response variable defined in~\eqref{def:Zi}. 
For computational purposes, we employ the \texttt{R} package \texttt{conquer} with the default tuning parameters to obtain an $\ell_1$-penalized smoothed quantile regression estimator  \citep{M2022}, which is shown to be 
 statistically first-order equivalent to the $\ell_1$-penalized quantile estimator in~\eqref{qr.est} \citep{TWZ2021}.
 Given a quantile regression estimator  $\hat \beta$, we compute  $\hat \theta$ in \eqref{lasso.ES} via the \texttt{glmnet} package. Both sparsity tuning parameters $\lambda_q$ and $\lambda_e$ are selected using ten-fold cross-validation.

We compare the proposed method with the oracle ES ``estimator'' obtained by regressing $\{ Z_i(\beta^*) \}_{i=1}^n$ on $\{ \tau X_{i, S_e^* } \}_{i=1}^n$. 
To reveal the non-negligible bias induced by the $\ell_1$-penalty, we refit the ES regression on the support of $\hat{\theta}$. For each method, we calculate the relative $\ell_2$-error $ \|\hat{\theta}-\theta^*\|_2/\|\theta^*\|_2$, the true positive rate and false positive rate, defined as the proportion of correctly identified non-zeros and· falsely identified non-zeros, respectively.  

 We generate data from a linear heteroscedastic model
	\#
	\label{response}
	y_i = X_i^{\T} \zeta^*+ (X_i^{\T}\eta^* )\xi_i, \quad i=1,\ldots,n,
	\#
 where $\xi_1,\ldots, \xi_n \sim  \mathcal{N}(0, 1)$ are independent, and $X_i  = (X_{i1}, \ldots, X_{ip})^\T \in\mathbb{R}^p$ are independent across $i=1,\ldots,n$ and follows one of the following three distributions: 
 (1) $X_{ij} = |z_{ij}|$ and $z_i = (z_{i1},\ldots,z_{ip})^{\T} \sim \mathcal{N}(0,\mathbb{I}_p)$, 
 (2) $X_{ij} = |z_{ij}|$ and $z_i = (z_{i1},\ldots,z_{ip})^{\T} \sim N\{0,\Sigma= (\sigma_{jk})_{1\leq j, k\leq p}\}$  with $\sigma_{jk} = 0.8^{|j-k|}$, and (3) $X_{ij}$'s are  independent and identically distributed from $\mathrm{Uniform}(0,1.5)$. Here, $\zeta^* = (\zeta_1^*,\ldots,\zeta_s^*,0,\ldots,0)^{\T}\in \RR^p$ with $\zeta_j^*=2$ for $j=1,\ldots,\lceil s/2 \rceil$ and $\zeta_j^*=1$ for $j=\lceil s/2  \rceil + 1,\ldots,s$, and $\eta^* = (\eta_1^*,\ldots,\eta^*_{\lceil s/2   \rceil},0,\ldots,0)\in \RR^p$ with $\eta_j^* = 1/3$ for  $j=1,\ldots,\lceil s/2   \rceil$. Under the above model, the true conditional quantile and ES coefficients are $\beta^* = \zeta^* + \eta^* Q_{\tau}(\xi)$, and $\theta^* = \zeta^* + \eta^* \cS_{\tau}(\xi)$.

In our numerical experiments, we set $\tau \in (0.05,0.1,0.2)$, $s=10$, $p=1000$, and $n\in (\lceil 30 s/\tau  \rceil,\lceil  50 s / \tau  \rceil)$.  The simulation results, averaged over 500 replications, are reported in Table~\ref{tab:est}. The true positive rates of the cross-validated two-step estimator are all ones due to the fact that the minimum signal strength, defined as $\min_{j : \theta^*_j \neq 0} | \theta^*_j|$, is sufficiently large in our settings. 
In summary, we find that the proposed $\ell_1$-penalized ES regression estimator performs slightly worse than the oracle estimator, which is not available in practice, across various settings.  
Moreover, we observe that the estimation errors decrease as we increase the sample size. 
The refitted estimator has a similar estimation error to the oracle estimator for the true positive coefficients, suggesting that the higher estimation error in our proposed $\ell_1$-penalized estimator is due to bias induced by the $\ell_1$-penalty. However, the refitted estimator is not recommended due to the high estimation error for the false positives. More numerical studies 
are reported in Section S 3 
of the online Supplementary Materials.

\begin{table}[h!]
	\fontsize{8}{9}\selectfont
	\caption{
 Numerical comparisons of the two-step, refitted two-step, and oracle two-step methods under the linear heteroscedastic model~\eqref{response} with $(s, p) =(10, 1000)$, $n\in \{\lceil 30 s/\tau \rceil, \lceil 50 s/\tau \rceil\}$ and $\tau \in \{0.05, 0.1, 0.2\}$. Estimation errors under the relative $\ell_2$-norm,  the false positive rate, averaged over 500 replications, are reported. The true positive rates are all one. The largest standard error is 0.003 for Error (FP) and 0.001 for Error (P) and FPR in the table. }
 \begin{center}
		\begin{tabular}{| c | l | c c c | c c c|}
			\hline
   \multicolumn{2}{|c|}{} & \multicolumn{3}{c|}{ $(n = \lceil 30 s/\tau \rceil, p = 1000)$} & \multicolumn{3}{c}{ $(n = \lceil 50 s/\tau \rceil, p =1000)$} \vline\\ \hline
			\multicolumn{8}{|c|}{Covariates $X_i= |z_i|$ where $z_i \sim  \mathcal{N}(0,\mathbb{I}_p)$}\\ \hline
			$\tau$ & Methods & Error (P) & Error (FP) & FPR & Error (P) & Error (FP) & FPR \\
			\hline
   0.05 & two-step &     0.132 &         0.060  &  0.044 &     0.105 &        0.047  &  0.043 \\
     & two-step+refitted &     0.064 &        0.269  &  0.044 &     0.049 &        0.216  &  0.043 \\
     & two-step oracle &     0.064 &          0 &  NA &     0.049 &          0 &  NA \\
     \hline
0.10 & two-step &     0.141 &        0.062  &  0.043 &      0.11 &        0.051  &  0.045 \\
     & two-step+refitted &     0.069 &        0.276  &  0.043 &     0.052 &        0.228  &  0.045 \\
     & two-step oracle &     0.067 &          0 &  NA &     0.051 &          0 &  NA \\
     \hline
0.20 & two-step &     0.148 &        0.072  &  0.047 &     0.118 &        0.054  &  0.044 \\
     & two-step+refitted &     0.076 &        0.294  &  0.047 &     0.058 &        0.235  &  0.044 \\
     & two-step oracle &     0.072 &          0 &  NA &     0.054 &          0 &  NA \\
			\hline
			\multicolumn{8}{|c|}{Covariates $X_i= |z_i|$ where $z_i \sim N\{0,\Sigma= (0.8^{|j-k|})_{1\leq j, k\leq p}\}$}\\ \hline
			$\tau$ & Methods & Error (P) & Error (FP) & FPR & Error (P) & Error (FP)& FPR \\
			\hline
0.05 & two-step &     0.107 &        0.037  &  0.018 &     0.085 &        0.029  &  0.019 \\
     & two-step+refitted &     0.095 &        0.186  &  0.018 &     0.074 &        0.149  &  0.019 \\
     & two-step oracle &       0.1 &          0 &  NA &     0.078 &          0 &  NA \\
     \hline
0.10 & two-step &     0.112 &         0.04  &  0.019 &     0.088 &        0.031  &  0.017 \\
     & two-step+refitted &       0.1 &        0.197  &  0.019 &     0.078 &        0.154  &  0.017 \\
     & two-step oracle &     0.105 &          0 &  NA &     0.082 &          0 &  NA \\
     \hline
0.20 & two-step &     0.121 &        0.042  &  0.018 &     0.096 &        0.033  &  0.018 \\
     & two-step+refitted &     0.109 &        0.203  &  0.018 &     0.085 &        0.166  &  0.018 \\
     & two-step oracle &     0.112 &          0 &  NA &     0.087 &          0 &  NA \\
			\hline
				\multicolumn{8}{|c|}{Covariates $X_i$ have independent $\mathrm{Uniform}(0,1.5)$ coordinates}\\ \hline
				$\tau$ & Methods & Error (P) & Error (FP)& FPR & Error (P) & Error (FP)& FPR \\
				\hline
0.05 & two-step &     0.168 &        0.077  &  0.044 &     0.134 &        0.061  &  0.044 \\
     & two-step+refitted &     0.076 &        0.354  &  0.044 &     0.061 &        0.281  &  0.044 \\
     & two-step oracle &     0.075 &          0 &  NA &     0.059 &          0 &  NA \\
     \hline
0.10 & two-step &     0.177 &        0.084  &  0.046 &     0.139 &        0.064  &  0.043 \\
     & two-step+refitted &     0.082 &         0.37  &  0.046 &     0.063 &        0.292  &  0.043 \\
     & two-step oracle &     0.077 &          0 &  NA &     0.061 &          0 &  NA \\
     \hline
0.20 & two-step &     0.188 &        0.094  &  0.048 &     0.149 &        0.072  &  0.046 \\
     & two-step+refitted &     0.093 &        0.385  &  0.048 &     0.069 &         0.31  &  0.046 \\
     & two-step oracle &     0.085 &          0 &  NA &     0.065 &          0 &  NA \\
			\hline	
		\end{tabular}
  \end{center}
 \label{tab:est}
 \begin{tablenotes}
 \small
 \item
Error (P): relative $\ell_2$-error on the support, i.e., $ \|\hat{\theta}_{S_e^*}-\theta^*_{S_e^*}\|_2/\|\theta^*\|_2$. Error (FP): relative $\ell_2$-error on the false positives, i.e., $ \|\hat{\theta}_{(S_e^*)^c}\|_2/\|\theta^*\|_2$. Here $S_e^* = (1,\ldots,s)$ and $(S_e^*)^c=(s+1,\ldots,p)$. FPR: false positive rate. 
\end{tablenotes}
\end{table}

\subsection{Inference on debiased estimator}\label{subsec:Inf}
To perform statistical inference on $\theta_j^*$ in a high-dimensional model, we compute the debiased estimator $\tilde{\theta}_j$ in~\eqref{debiased.estimator} and its associated 95\% confidence interval. To this end, we use the \texttt{glmnet} package to obtain the projection estimator $\hat \gamma$ \eqref{decorr.est} with $\lambda_m$ chosen by ten-fold cross-validation with the ``one standard error rule" (see Section~7.10 in \citealp{hastie2009elements}), that is, we choose the largest lambda whose error is no more than one standard error above the minimum mean cross-validated error. Given $\hat \beta$, $\hat \theta$, and $\hat \gamma$, we compute the triply-orthogonal score function in~\eqref{decorr.score} and the debiased estimator in~\eqref{debiased.estimator}. We then estimate the asymptotic variance of the debiased estimator using refitted cross-validation as described in Section~\ref{sec:2.4}. The corresponding 95\% confidence interval is given in~\eqref{wald.ci}.

Without loss of generality, we consider testing the hypotheses $H_0: \theta_2^* = 0$ versus $H_1: \theta_2^* \ne 0$. The data-generating process is the same as in Section~\ref{subsec:numerical_est}.	In addition to the proposed debiased estimator of $\theta_2^*$, we also report the $\ell_1$-penalized estimator without debiasing, the oracle estimator, and the bootstrap estimator defined as the average of 100 $\ell_1$-penalized estimators constructed from the bootstrap samples. The bootstrap estimator here is a standard and practical competitor to our proposed debiased estimator. The bias of each estimator under comparison, the squared error, and the coverage probability of the proposed confidence interval, averaged over 500 replications, are summarized in Table~\ref{table:inference}. We see that the $\ell_1$-penalized estimators, with and without bootstrapping, have non-negligible bias. In contrast, the debiased estimator has substantially lower bias and mean squared error, and the average coverage probabilities of the corresponding confidence intervals are close to the nominal level across different $\tau$-values and designs.
	\begin{table}[h!]
		 \fontsize{8}{9}\selectfont
		 \centering
		 \caption{Estimation and inference on $\theta_2^*$ under the linear heteroscedastic model~\eqref{response} with $(s, p) = (10, 1000)$,  $n=\lceil 50 s/\tau \rceil$ and $\tau \in \{ 0.05, 0.1, 0.2 \}$. The biases, squared errors of the four methods, and coverage probabilities of the 95\% confidence interval, averaged over 500 replications, are reported. The largest standard error is 0.005 for the bias and mean squared error  and 0.01 for the coverage probability in the table.
}		
  \begin{center}
			\begin{tabular}{| c |  c | cccc | c |}
				\hline
    \multicolumn{2}{|c|}{} &\multicolumn{4}{c|}{Bias ($\times 10^{-2}$)/MSE ($\times 10^{-2}$)}&\multicolumn{1}{c|} {Inference}\\\hline
    \multicolumn{7}{|c|}{Covariates $X_i= |z_i|$ where $z_i \sim  \mathcal{N}(0,\mathbb{I}_p)$}\\ \hline
				\multirow{2}{*}{$\tau$}&\multirow{2}{*}{$\theta_2^*$} &\multirow{2}{*}{two-step} &\multirow{2}{*}{\makecell[c]{debiased\\two-step}}  &\multirow{2}{*}{\makecell[c]{two-step\\oracle}}&\multirow{2}{*}{\makecell[c]{two-step\\bootstrap}} &\multirow{2}{*}{\makecell[c]{coverage\\ probability}}\\	
				&&&&&&\\
				\hline
0.05  &        1.312 &  -10.4/1.5 &       0.3/0.4 &      -0.1/0.4 &    -5.8/0.7 &    0.95 \\
0.10  &        1.415 &  -11.4/1.8 &       0.3/0.5 &       0.1/0.6 &    -6.7/1.0 &    0.94 \\
0.20  &        1.533 &  -12.3/2.1 &       0.9/0.6 &       0.6/0.6 &    -7.5/1.1 &    0.95 \\
    \hline
 \multicolumn{7}{|c|}{Covariates $X_i= |z_i|$ where $z_i \sim  \mathcal{N}(0,\Sigma= (0.8^{|j-k|})_{1\leq j, k\leq p})$}\\ \hline
				\multirow{2}{*}{$\tau$}&\multirow{2}{*}{$\theta_2^*$} &\multirow{2}{*}{two-step} &\multirow{2}{*}{\makecell[c]{debiased\\two-step}}  &\multirow{2}{*}{\makecell[c]{two-step\\oracle}}&\multirow{2}{*}{\makecell[c]{two-step\\bootstrap}} &\multirow{2}{*}{\makecell[c]{coverage\\ probability}}\\	
				&&&&&&\\
    \hline
    0.05  &        1.312 &   -8.1/1.4 &       1.3/0.7 &       0.0/0.8 &    -4.5/0.9 &    0.94 \\
0.10  &        1.415 &   -8.8/1.6 &       1.8/0.8 &       0.1/0.9 &    -5.1/1.1 &    0.97 \\
0.20  &        1.533 &   -9.5/2.0 &       2.8/1.1 &       0.8/1.2 &    -5.6/1.4 &    0.95 \\
    \hline
			\end{tabular}
		\end{center}
  \label{table:inference}
   \begin{tablenotes}
\small
\item
\noindent
Bias: $M^{-1}\sum_{m=1}^M \{\hat \theta^{(m)} - \theta_2^*\}$, where $ \hat \theta^{(m)}$ denotes a generic ES regression estimator. MSE: $M^{-1}\sum_{m=1}^M  \{ \hat \theta^{(m)} - \theta_2^* \}^2$. Here $M$ is the number of replications, which is set as 500.
\end{tablenotes}
	\end{table}

\section{Data application on health disparity}
\label{sec:data}

Disparity research has drawn increasing attention, especially from regulatory bodies such as the National Institutes of Health and the Centers for Disease Control and Prevention \citep{CDC2005,LFA2016}. Health disparity is defined as the differences in health outcomes between advantaged and disadvantaged groups, often referred to as majority and minority groups, respectively \citep{Braveman2006}.

Our goal is to investigate whether there are disparities in high cotinine levels among different ethnic groups, after adjusting for a large number of observed covariates that potentially contribute to the marginal disparities observed from Figure~\ref{fig:cotinine}. Cotinine is a metabolite of nicotine, indicating recent exposure to nicotine, either through consuming products containing nicotine or being exposed to second-hand tobacco smoke \citep{Benowitz1996}. To investigate the health disparity in cotinine, we analyze the National Health and Nutrition Examination Survey data from 2017 to 2020\footnote{\url{https://wwwn.cdc.gov/nchs/nhanes/continuousnhanes/default.aspx?Cycle=2017-2020}.}. 
We use the cotinine variable from the serum dataset as the response variable. The covariates are obtained by merging 22 related health determinants datasets via the respondent sequence number, including but not limited to demographic variables, health insurance, hospital utilization and access to care, smoking behaviors, drug use, medical conditions, and living conditions such as secondhand smoke exposure and food security. After merging the 22 health determinant datasets and the serum dataset that contains the response, we remove individuals with missing response variable, i.e., cotinine level. 
 We drop continuous variables with more than 10\% missingness, then drop individuals with missingness in those variables. Furthermore, we convert all of the categorical variables into dummy variables, including level NA if any. 
 The post-processed dataset contains $n=2143$ observations and $p=473$ covariates.
  
\begin{figure}[!t]
		\centering
  \begin{subfigure}[b]{0.5\textwidth}
         \centering
         \includegraphics[height=7.5cm]{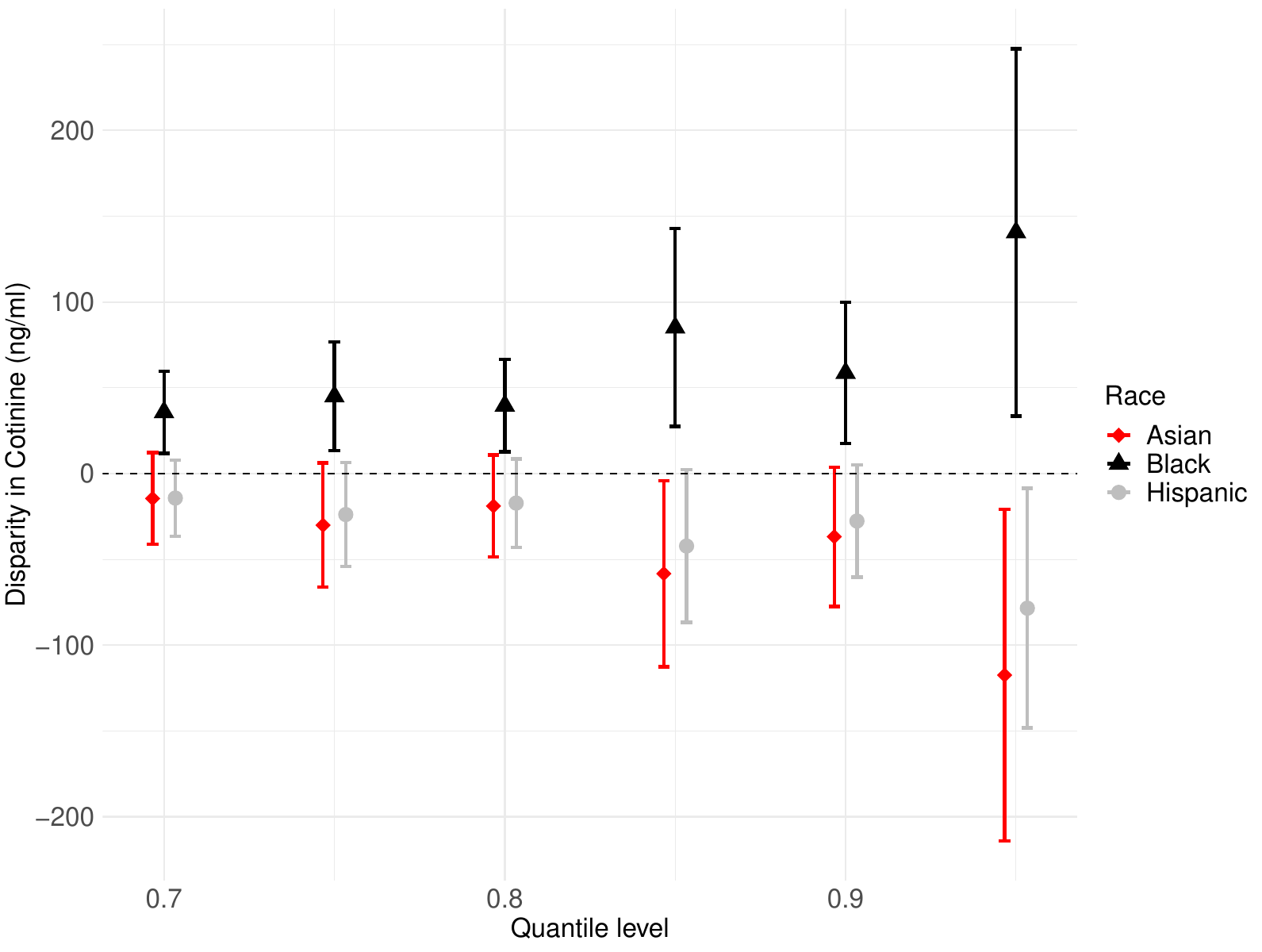}
         \caption{ES regression}
         \label{fig:ci_ES}
     \end{subfigure}
     \hfill
     \begin{subfigure}[b]{0.4\textwidth}
         \centering
         \includegraphics[height=7.5cm]{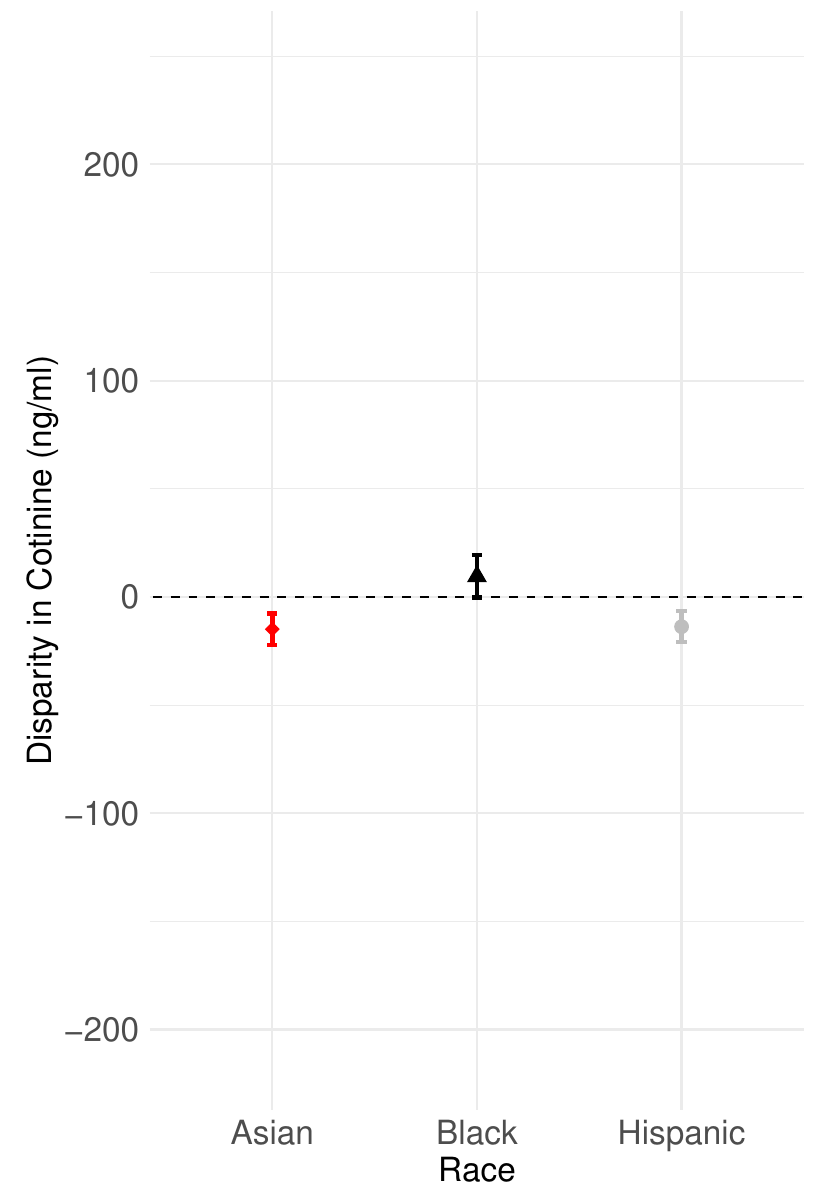}
         \caption{Mean regression}
         \label{fig:ci_Lasso}
     \end{subfigure}
		\caption{The 95\% confidence intervals for the estimated coefficients of cotinine levels (ng/ml) on the race dummy variables by (a) upper ES regressions with $\tau \in \{0.7,0.75,0.8,0.85,0.9,0.95\}$ using the proposed debiased ES methods and (b) mean regression using desparsified lasso, conditioned on 473 covariates.	\label{fig:ci} }
	\end{figure}

We fit the $\ell_1$-penalized ES regression to assess whether there are indeed health disparities in terms of cotinine levels among the different ethnic groups by focusing on the upper tail of the conditional distribution of cotinine levels. 
Unless otherwise stated, the covariates are standardized before fitting all of the involved penalized estimators.
The primary covariate of interest is race, induced by three dummy variables on Non-Hispanic Asian, Non-Hispanic Black, Mexican American and Other Hispanic, using Non-Hispanic White as the baseline. Thus, the estimated coefficients for the three dummy variables of race represent the conditional disparity of the cotinine between the majority and the minority. The debiased estimator for the parameter of interest is then computed for statistical inference. While the proposed method defined in the previous sections mainly focuses on the ES regression for the lower tail of the conditional distribution, it can be easily modified to accommodate the upper tail. Specifically, we fit the proposed method at the $(1-\tau)$th level with a negative cotinine response, and flip the sign of the resulting estimators and confidence intervals to obtain an estimate for the regression coefficients of interest. 
We compare our proposed method with desparsified lasso that focuses on the conditional mean, implemented using the \texttt{desla} package in \texttt{R} \citep{G2014}. The results for the ES regression at various quantile levels and lasso regression are summarized in Figure~\ref{fig:ci}. 

\begin{figure}[!t]
		\centering
		\includegraphics[scale=0.6]{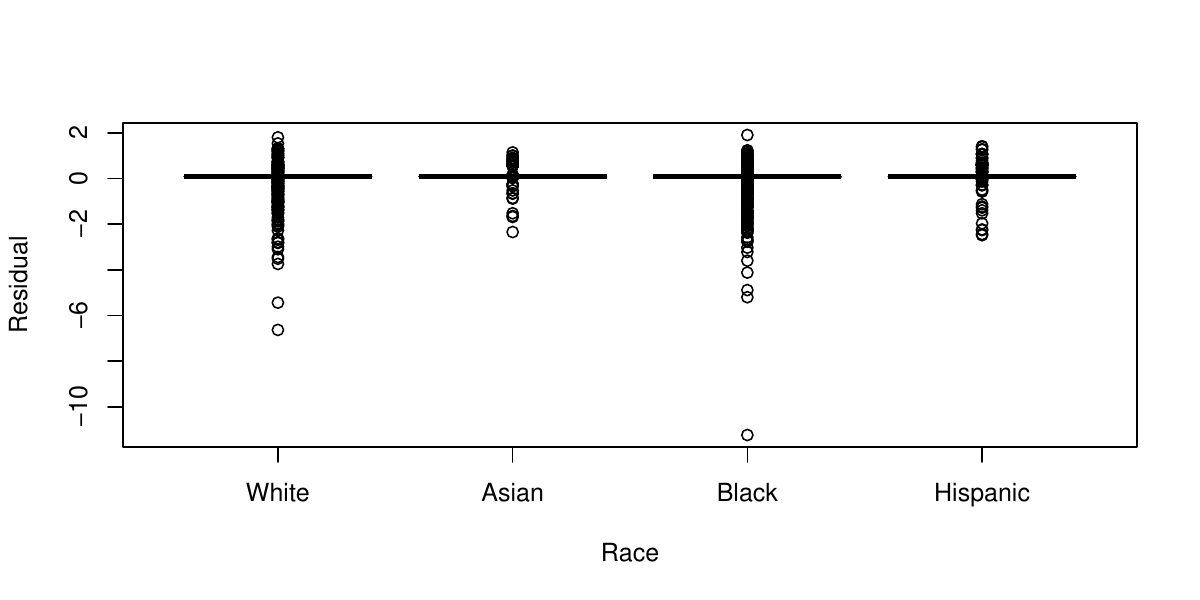}
		\caption{	\label{fig:resid_lasso} Boxplot of the residuals from the mean lasso regression of cotinine levels (ng/ml) on 473 covariates for four different races.}
	\end{figure}

 In Figure~\ref{fig:ci}, 
 the 95\% confidence intervals for the proposed debiased ES estimators that measure the disparity between Black and White remain consistently positive across all quantile levels, even after covariates adjustment. This suggests that Black has a significantly higher cotinine value than White in the subpopulations with high cotinine levels. For the disparity of Asian and Hispanic compared with White, the ES estimators are only significant at $0.95$ quantile level. Compared with the proposed method that characterizes the tail, the desparsified lasso method that focuses on the mean does not provide valid inference, since the homogeneity assumption on the error term in lasso modeling is violated according to the residual plot for the lasso regression in Figure~\ref{fig:resid_lasso}. Therefore, we cannot conclude from the seemingly significant results for the disparity of Asian and Hispanic compared with White in Figure~\ref{fig:ci_Lasso} that Asian and Hispanic have significantly lower cotinine 
 values than White.  
  
 Given the findings from the ES regression analysis in the upper tail, one naturally asks what the underlying factors, possibly beyond the observed covariates, are for such disparities and what interventions might be taken to address the disparities. We do not yet have the answers to those questions, but this example shows that the proposed method can play a valuable role in collaborative health disparity research.

\section{Discussions on tuning parameter selection}
\label{sec:discuss}

Throughout the numerical studies in Section~\ref{sec:numerical}, we used cross-validation to select the regularization parameters $\lambda_q$ and $\lambda_e$, guided by their theoretical order. Specifically, we set $\lambda = c \sqrt{\log(p)/ n}$ and adjusted $c$ via cross-validation, either from a grid of constants (for quantile regression) or from a grid of constants multiplied by a standard deviation estimator using the fitted residuals (for ES regression). Since cross-validation aims for prediction accuracy, it tends to select a smaller tuning parameter, resulting in a model that typically contains false discoveries and is larger in size. Over the past two decades, many studies have shown that when nonconvex penalization is combined with certain generalized or modified Bayesian information criteria (BIC), the underlying model can be recovered with probability approaching one, under additional conditions such as the minimum signal strength condition. Notable works in this direction include \cite{CC2008}, \cite{WZ2011}, \cite{KKC2012}, \cite{FT2013}, \cite{WKL2013}, among others.

Extending \eqref{qr.est}, we begin with a more general penalized quantile regression estimator $\hat \beta_{\lambda} \in \argmin_{\beta = (\beta_1, \ldots, \beta_p)^\T } \{ (1/n) \sn \rho_\tau(Y_i - X_i^\T \beta) + \sum_{j=1}^p P_\lambda(|\beta_j|) \}$, where $P_\lambda(\cdot)$ denotes a folded concave penalty \citep{FXZ2014}, encompassing penalties such as lasso, SCAD \citep{FL2001} and MCP \citep{Z2010}. 
Motivated by \cite{WKL2013}, we consider a modified high-dimensional BIC, defined as 
high-dimensional BIC (HBIC), defined as 
$$
{\rm HBIC}_q (\lambda) = \log \bigg(  \frac{1}{n} \sn \rho_\tau (Y_i - X_i^\T \hat \beta_{\lambda} ) \bigg) + \| \hat \beta_\lambda \|_0 \frac{C_n \log(p)}{n} ,
$$
where $C_n\geq 1$ is a sequence of integers that slowly diverge with $n$. For some prespecified integer $K_n > s$, we choose the data-driven penalty level as $\hat \lambda_q = \argmin_{\lambda >0: \| \hat \beta_\lambda \|_0 \leq K_n } {\rm HBIC}_q (\lambda) $. In the second step, define surrogate response variables $\hat Z_i = Z_i(\hat \beta_{\hat \lambda_q} )$, $i=1,\ldots, n$, and the corresponding HBIC for ES as
$$
{\rm HBIC}_e (\lambda) = \log \bigg(  \frac{1}{2 n} \sn (\hat Z_i - \tau X_i^\T \hat \theta_\lambda )^2 \bigg) + \| \hat \theta_\lambda \|_0 \frac{D_n \log(p)}{n} ,
$$
where $\hat \theta_\lambda \in \argmin_{\theta\in\RR^p}   \{(2n)^{-1} \sn (\hat Z_i - \tau X_i^\T \hat \theta_\lambda )^2 + \sum_{j=1}^p P_\lambda(|\theta_j|) \}$, and $D_n\geq 1$ is also a sequence of integers similar to $C_n$. As a result, we select the second penalty level $\lambda_e$ in \eqref{lasso.ES} as $\hat{\lambda}_e = \argmin_{\lambda > 0 : | \hat{\theta}_\lambda |_0 \leq K_n } {\rm HBIC}_e(\lambda)$. 
Due to space constraints, we examine the numerical performance of this approach in Section S 3.1 of the online Supplementary Materials.

To establish the selection consistency of the HBIC procedure, namely, $P(\hat S_e = S_e) \rightarrow 1$ as $n \rightarrow \infty$ where $\hat{S}_e = \text{supp}(\hat{\theta}_{\hat{\lambda}_e})$ and $S_e = \text{supp}(\theta^*)$, $\ell_1$-penalty is insufficient as it tends to over-penalize large coefficients. Therefore, we need to develop a self-contained theory for the nonconvex penalized ES regression, particularly the oracle properties that concern support recovery. A key step is to show that for $\lambda_e$ in a certain range, the nonconvex penalized estimator converges to an oracle, defined as $ \hat \theta^{{\rm o}} = \argmin_{\theta \in \RR^{p} : \theta_{S_e^{{\rm c}}} = 0 }  \sn (\hat Z_i - \tau X_i^\T \theta  )^2$, under a minimum signal strength condition. \cite{GZ2020} proves the selection consistency of the HBIC for penalized composite quantile regression based on such oracle properties and under the minimum signal strength condition that $\min_{j\in {\rm supp}(\beta^*)} |\beta^*_j| \gtrsim \sqrt{s \log (p)/n}$. One question is whether the $\sqrt{s}$-factor can be removed. Additionally, to establish oracle properties similar to those in \cite{TWZ2021}, one needs sharper bounds on $\|  \hat \theta^{{\rm o}}  - \theta^* \|_2$, $\|  \hat \theta^{{\rm o}} - \theta^* \|_\infty$ and  $ \| (1/n) \sn (\hat Z_i - \tau X_i^\T \hat \theta^{{\rm o}} ) X_i \|_\infty$. To this end, does the $\ell_1$-penalized quantile regression estimator suffice, or do we need a nonconvex penalized estimator with an improved convergence rate? 
We defer a rigorous investigation of nonconvex penalized ES regression 
to future research. While there exist completely tuning-free approaches for certain steps of the proposed ES regression, a completely tuning-free approach for the proposed approach with multiple steps is unknown to us, which we also leave as future work.

	\section*{Acknowledgement and disclosure statement}
XH acknowledges NSF Awards DMS-2345035 and DMS-1951980. KT acknowledges NSF CAREER DMS-2238428 and NSF DMS-2113346. WZ acknowledges NSF DMS-2401268. 
	The authors thank Dr. Grace Hong from the National Institutes of Health for her valuable discussions on health disparity research. They also extend their gratitude to the anonymous Associate Editor and referees for their constructive comments, which contributed to improvements in the paper. The authors report there are no competing interests to declare.

\renewcommand*{\thesection}{S \arabic{section}}
\renewcommand*{\thesubsection}{\thesection.\arabic{subsection}}

\def\spacingset#1{\renewcommand{\baselinestretch}%
{#1}\small\normalsize} \spacingset{1}

\setcounter{equation}{0}
\setcounter{section}{0}

\section{Proofs of Main Results}
\label{sec:main_proof}

\subsection{Preliminaries}
\label{sec:pre}

We first introduce some notation that will be used throughout the supplementary materials. We use $\| \cdot \|_{\max}$ to denote maximum norm in $\RR^{p \times p}$, i.e., for any $U = (u_{jk})_{j,k=1,\ldots,p}$, we denote $\| U \|_{\max} = \max_{j,k=1,\ldots,p}|u_{jk}|$. Define the quantile and expected shortfall regression residuals as  
\#
  \varepsilon_i = Y_i - X_i^\T \beta^* , \quad  e_i(\beta) = Z_i(\beta) - \tau X_i^\T \theta^*    \quad ( i = 1,\ldots, n) . \label{def:residual}
\#
respectively, where $Z_i(\beta)=(Y_i - X_i^\T \beta)\mathbbm{1}(Y_i \leq X_i^\T \beta) + \tau X_i^\T \beta$. In particular, we write $Z_i^* = Z_i(\beta^*)$, $e_i = e_i(\beta^*)$, and $\hat e_i = \hat Z_i - \tau X_i^\T \hat \theta$ where  $\hat Z_i = Z_i(\hat \beta)$ for some estimator $\hat \beta$ and $\hat \theta$ of $\beta^*$ and $\theta^*$ respectively.  Let $E_{X_i}$ be the conditional expectation given $X_i$. 
Define $W_i = \Sigma^{-1/2} X_i$ recalling $\Sigma = E(X X^{\T})$. Let $\hat \Sigma =  (1/n) \sn X_i X_i^\T$. Similarly, define $W_{i,-j} = \Sigma_{-j}^{-1/2} X_{i,-j}$ and $\hat \Sigma_{-j} = (1/n) \sn X_{i,-j} X_{i,-j}^\T$.  The standardized covariates $W_i$ and $W_{i,-j}$ are such that $E(W_i W_i^\T) = {\rm I}_p$ and $E(W_{i,-j} W_{i,-j}^\T) = {\rm I}_{p-1}$.
A key observation is that, under the joint model (2.1), we have
\#
E_{X_i}(e_i) &= E_{X_i}\{(Y_i - X_i^\T \beta^*)\mathbbm{1}(Y_i \leq X_i^\T \beta^*)\} + \tau X_i^\T (\beta^*-\theta^*)\nonumber \\
&= E_{X_i}\{(Y_i - X_i^\T \beta^*) |Y_i \leq X_i^\T \beta^*\} \PP(Y_i \leq X_i^\T \beta^*) + \tau X_i^\T (\beta^*-\theta^*)\nonumber \\
&= (X_i^\T \theta^* - X_i^\T \beta^*) \tau + \tau X_i^\T (\beta^*-\theta^*)\nonumber \\
&=0. \label{eq:e=0}
\#
Recall from main text, we define $\mathbb{C}(\cS) = \big\{ \delta \in \RR^p : \| \delta_{\cS^{{\rm c}}} \|_1 \leq 3 \| \delta_{\cS} \|_1 \big\}$ as an $\ell_1$-cone.
For every $l_1>0$, define the cone-like set 
\#
	\CC(l_1)  = \big\{ \delta \in \RR^p: \| \delta \|_1 \leq l_1 \| \delta \|_\Sigma \big\}. \label{def:cone}
\#
For any subset $\cS \subseteq \{1,\ldots,p\}$ satisfying $| \cS |\leq s$,  under Condition 4.2 it holds that $\mathbb{C}(\cS) \subseteq  \CC(l_1)$ with $l_1 =4 \underline{\phi}^{-1} |\cS|^{1/2}$.

In the following, we present all the technical lemmas that will be used to prove the main results.   Proofs of the lemmas are postponed to Section~\ref{sec:proof_lemma}.

\begin{lemma}   \label{lem:first-order.error}
Assume Conditions 4.1 and  4.2 hold. For any $r_1>0$ and $t>0$,
\begin{align}\label{eq::first-order.error}
   &  \sup_{ \beta \in   \beta^* + \BB_1( r_1)  }  \bigg\| \frac{1}{n} \sn (1 -E)   \{  e_i(\beta)   - e_i  \} X_i   \bigg\|_\infty    \leq  4  b_X r_1   \bigg(  \sigma_X  \sqrt{\frac{2t}{n}} +  b_X\frac{ t}{3n} \bigg)
\end{align}
holds with probability at least $1- 2 p^2 e^{-t}$. When replacing $X_i$ with $X_{i,-j}$ in~\eqref{eq::first-order.error}, the bound holds with probability at least $1-2p(p-1) e^{-t}$. Similarly, Condition 4.4 ensures that with probability at least $1-2p e^{-t}$, 
\$
\sup_{ \beta \in   \beta^* + \BB_1( r_1)  }  \bigg| \frac{1}{n} \sn (1 - E) \{  e_i(\beta)   - e_i  \} \omega_i  \bigg| \leq   4 b_X   r_1   \bigg(  \sigma_X  \sqrt{\frac{2t}{n}}  +   b_X\frac{ t}{3n} \bigg).
\$
\end{lemma}

\begin{lemma} \label{lem:approximate.neyman}
Assume Conditions 4.1 and 4.2 hold. For  any $r_0 >0$,  we have
\#
	 \sup_{\beta \in \beta^* + \BB_\Sigma(r_0) }\| \EE \{ e_i(\beta)  W_{i,-j} \}   \|_2 \leq
   \sup_{\beta \in \beta^* + \BB_\Sigma(r_0) } \big\|  E  \{  e_i(\beta) W_i  \}   \big\|_2   \leq   \frac{1}{2}  f_u  m_3   r_0^2    , \label{mean.grad.ubd}
\#
where $e_i(\beta)$ is defined in \eqref{def:residual}, and $W_i = \Sigma^{-1/2} X_i$.  In addition, if Condition 4.4 holds,  then $ \sup_{\beta \in \beta^* + \BB_\Sigma(r_0) } | E  \{  e_i(\beta) \omega_i    \}  | \leq   f_u  b_X r_0^2/2 $.
\end{lemma}

\begin{lemma}  \label{lem:score.bound}
Assume Conditions 4.1 and  4.2 hold. For any $t > 0$, it holds with probability at least $1-2p e^{-t}$ that
\#\label{eq::bound_e_X}
\bigg\| \frac{1}{n} \sn   e_i   X_i   \bigg\|_\infty  \leq    \sigma_\varepsilon \sigma_X  \sqrt{\frac{2t}{n}} + 2 b_\varepsilon b_X \frac{t}{n}.
\#
When replacing $X_i$ with $X_{i,-j}$ in~\eqref{eq::bound_e_X}, the bound holds with probability at least $1-2(p-1) e^{-t}$. Similarly, with probability at least $1-2(p-1) e^{-t}$,
\$
\bigg\| \frac{1}{n} \sn  \omega_i   X_{i, -j}   \bigg\|_\infty  \leq  b^2_X   \sqrt{\frac{2 t}{n}}.
\$

\end{lemma}

\begin{lemma} \label{lem:RE}
Assume Condition~ 4.2.  Then, for any $l_1 >0$ and $t \geq 0$,
\$
	 \inf_{\delta \in \CC(l_1) }  \frac{\delta^\T \hat \Sigma \delta}{ \| \delta \|_\Sigma^2} \geq \frac{3}{4} - 10 m_4^{1/2}   l_1 \bigg\{  \sigma_X \sqrt{\frac{2\log(2p)}{n}} + b_X\frac{\log(2p)}{3 n} \bigg\} -   \sqrt{\frac{2m_4 t}{n}} -  \frac{4.4 t}{n} 
\$
holds with probability at least $1-e^{-t}$.
Similarly, 
\$
	 \inf_{\delta \in \CC(l_1) }  \frac{\delta^\T \hat \Sigma_{-j} \delta}{ \| \delta \|_{\Sigma_{-j}}^2} \geq \frac{3}{4} - 10 m_4^{1/2}   l_1 \bigg\{  \sigma_X \sqrt{\frac{2\log(2p-2)}{n}} + b_X\frac{\log(2p-2)}{3 n} \bigg\} -   \sqrt{\frac{2m_4 t}{n}} -  \frac{4.4 t}{n} 
\$
holds with probability at least $1-e^{-t}$.
\end{lemma}

 Turning to the empirical quantile loss $\hat{\mathcal{Q}}(\beta) = n^{-1} \sum \rho_\tau(Y_i - X_i^\T \beta)$,   define the loss difference $\hat{\mathcal{D}}(\delta) =\hat{\mathcal{Q}}(\beta^* + \delta ) - \hat{\mathcal{Q}}(\beta^*)$ and its population counterpart $\mathcal{D}(\delta ) = E \{\hat{\mathcal{D}}(\delta)\}$ for $\delta \in \RR^p$.
 
 	\begin{lemma} \label{lem:qr.diff.bound}
 	Assume Condition~ 4.2.  For any $r_0 , l_1 >0$ and $t>0$,
 	\#
 	\sup_{\delta \in \CC(l_1) \cap \BB_\Sigma(r_0) }  \{ \cD(\delta) - \hat \cD(\delta)   \} \leq  4 \bar \tau   l_1 r_0 \bigg(  \sigma_X \sqrt{\frac{2t}{n}} +  b_X \frac{t}{3n} \bigg)
 	\#
 	holds with probability at least $1-2 p e^{-t}$, where $\bar \tau = \max(\tau, 1-\tau)$.
 \end{lemma}

 \begin{lemma} \label{lem:qr.score.bound}
 	Assume Conditions 4.2 and 4.3.  Then, with probability at least $1-2p e^{-t}$,
 	\#
 	\bigg\| \frac{1}{n} \sn \{ \mathbbm{1}(\varepsilon_i \leq 0) - \tau  \} X_i \bigg\|_\infty  \leq 
  \sigma_X \sqrt{ \tau(1-\tau)\frac{2 t}{n}} + \bar \tau b_X \frac{t}{3n},
 	\#
 	where $\bar \tau = \max(\tau, 1-\tau)$.
 \end{lemma}

 The following result is a variant of Lemma~C.1 in \cite{SZF2020} obtained by replacing the gradient with any subgradient.
 
 \begin{lemma} \label{lem:convexity}
 	Let $f:\RR^p\to \RR$ be a convex function, and define the corresponding symmetrized Bregman divergence 
 	$\cB_f(\beta_1, \beta_2) =  (w_{\beta_2}  - w_{\beta_1} )^\T( \beta_2 - \beta_1)$ for $\beta_1, \beta_2 \in \RR^p$ and subgradients $ w_{\beta_1} \in \partial f(\beta_1),  w_{\beta_2} \in \partial f(\beta_2)$. Then, for any $\beta, \delta \in \RR^p$ and $\lambda\in [0, 1]$,  $\cB_f( \beta_\lambda , \beta) \leq \lambda \cdot \cB_f(\beta_1 , \beta )$, where $\beta_\lambda =\beta + \lambda\delta $ and $\beta_1 = \beta + \delta$.
 \end{lemma}
 
 The following three lemmas provide useful tools from the empirical process theory. We present them here for readers' convenience.

  \begin{lemma}[Bernstein's Inequality---Theorem 2.10 in \cite{BLM2013}]
  \label{lem:Berstein}
 If $X_1,\ldots,X_n$ are independent random variables. Assume that there exist $v>0$ such that $\sn E(X_i^2) \leq v$, and $\sn E\{(X_i)^k_{+}\} \leq k! v c^{k-2}/2$ for all integers $k \geq 3$ where $x_{+} = \max(x,0)$. Write $S = \sn X_i - \EE (X_i)$. For any $t > 0$, 
 \begin{equation}\label{eq:bernstein}
 	\PP( |S| \geq \sqrt{2vt} + ct ) \leq \exp(-t).
 \end{equation} 
In particular,   if $X_1,\ldots,X_n$ satisfy $|X_i| \leq b$ almost surely for all $i \leq n$,  \eqref{eq:bernstein} holds with $c=b/3$.
  \end{lemma}
 
 \begin{lemma}[Rademacher symmetrization---Lemma 6.3 in \cite{LT1991}]
 \label{lem:Rademacher}
 	Let $\epsilon_1,\ldots, \epsilon_n$ be independent Rademacher random variables. Let $\mathcal{F}$ be a collection of function $f: \mathcal{X} \to \RR$ and $X_1,\ldots,X_n$ be independent. For any convex function $\phi: \RR^{+} \to \RR^{+}$,
 	\#
  \EE \phi\bigg\{ \sup_{f\in \mathcal{F}} \bigg|\sn f(X_i) - \EE f(X_i)  \bigg|\bigg\}  \leq  \EE \phi\bigg\{2 \sup_{f\in \mathcal{F}} \bigg|\sn \epsilon_i f(X_i)\bigg|\bigg\} 
 	\#
 \end{lemma}
 
 \begin{lemma}[Ledoux-Talagrand contraction---Theorem 4.12 in \cite{LT1991}]
 \label{lem:Ledoux}
 	Let $\epsilon_1,\ldots, \epsilon_n$ be independent Rademacher random variables, and $\phi: \RR^{+} \to \RR^{+}$ be a convex and increasing function.  Assume that functions $f_i: \RR \to \RR$ are Lipschitz continuous with $f_i(0)=0$, i.e. $|f_i(a)-f_i(b)|\leq L |a-b|$.  Then, for any $T\subset\RR^n$, 
 	\#
 	\EE \phi\bigg\{\frac{1}{2} \sup_{t \in T} \bigg|\sn \epsilon_i f_i(t_i)\bigg|\bigg\} \leq \EE \phi\bigg(L \sup_{t\in T} \bigg|\sn \epsilon_i t_i \bigg| \bigg) 
 	\#
 \end{lemma}


 \subsection{Proof of Theorem~4.1}\label{pf:thm:lasso-ES}
 
 Let $\hat \cL(\theta) = (2n)^{-1} \sn (\hat Z_i - \tau X_i^\T \theta)^2$ be the empirical loss function, where $\hat Z_i = Z_i(\hat \beta)$ and $\hat \beta$ is an $\ell_1$-penalized quantile regression estimator of $\beta^*$. 
 By standard optimality conditions for a convex program, for any optimum $\hat \theta \in \argmin_\theta \{ \hat \cL(\theta) + \tau \lambda_e \| \theta \|_1 \}$, there exists a subgradient
 vector $\hat g \in  \partial \| \hat \theta\|_1$ satisfying $ \hat g^\T \hat \theta = \| \hat \theta \|_1$, $\|\hat g\|_\infty \leq 1$ and $\nabla \hat \cL(\hat \theta) +\tau \lambda_e \hat g = 0$. On the other hand, observe that $ \{\nabla \hat \cL(\hat \theta)  -  \nabla \hat \cL(\theta^*)\}^\T (\hat \theta - \theta^*) = \tau^2 (\hat \theta-\theta^*)^\T\, \hat \Sigma\, (\hat \theta - \theta^*)$. Putting these two observations together, we obtain
 \begin{equation} \label{opt.cond1}
 	\begin{aligned}
 		\tau^2 (\hat \theta-\theta^*)^\T \, \hat \Sigma \, (\hat \theta - \theta^*) & = - \tau \lambda_e \hat g^\T (\hat \theta - \theta^*)    -  \nabla \hat \cL(\theta^*)^\T (\hat \theta - \theta^*)\\
 		& \leq \tau \lambda_e ( \| \theta^* \|_1 - \| \hat \theta \|_1 )  -  \nabla \hat \cL(\theta^*)^\T (\hat \theta - \theta^*).
 	\end{aligned}
 \end{equation}

 \noindent
 {\sc Step 1 (Deterministic analysis).}  Let $\hat \delta = \hat \theta - \theta^*$ denote the error vector, and write $\cS = {\rm supp}(\theta^*)$. Then we have $\| \theta^* \|_1 - \| \hat \theta \|_1 = \| \theta^*_{\cS} \|_1 - \| \hat \delta_{\cS^{{\rm c}}} + (\hat \delta + \theta^*)_{\cS} \|_1 \leq \| \hat \delta_{\cS} \|_1 - \| \hat \delta_{\cS^{{\rm c}}} \|_1$.  To bound $ -  \nabla \hat \cL(\theta^*)^\T (\hat \theta - \theta^*) $, consider the decomposition
 \#
 -\nabla \hat \cL(\theta^*) &  = \frac{\tau}{n} \sn    e_i(\hat \beta) X_i  =  \frac{\tau}{n} \sn    e_i X_i  + \frac{\tau}{n} \sn   (1- \EE) \{  e_i(\hat \beta) - e_i \} X_i  
 + \tau \EE \{ e_i(\beta) X_i \} \big|_{\beta = \hat \beta},
 \#
 since $\EE(e_i X_i) = \EE\{\EE_{X_i}(e_i)X_i\} = 0$ by \eqref{eq:e=0}. 
 Conditioned on the event $\{ \hat \beta \in \beta^* + \BB_\Sigma(r_0) \cap \BB_1(r_1)  \}$, it follows that
 \begin{equation}
 	\begin{aligned}
 		&   -  \nabla \hat \cL(\theta^*)^\T (\hat \theta - \theta^*)\\
 		& \leq \bigg\{  \bigg\| \frac{1}{n} \sn e_i X_i \bigg\|_\infty +   \sup_{\beta \in \beta^* + \BB_1(r_1) }  \bigg\| \frac{1}{n} \sn   (1- \EE) \{  e_i( \beta) - e_i \} X_i   \bigg\|_\infty  \bigg\} \cdot \tau \|\hat \delta \|_1   \\ 
 		& ~~~~~~ +  \sup_{\beta \in \beta^* + \BB_\Sigma(r_0) } \big\| \EE   \{ e_i(\beta) W_i \} \big\|_2  \cdot \tau \| \hat \delta \|_\Sigma \\
 		& =   \big( E_1 +E_2   )   \cdot  \tau \|\hat \delta \|_1  + \frac{1}{2} f_u  m_3   r_0^2 \cdot \tau \|\hat \delta \|_\Sigma , \label{opt.cond2}
 	\end{aligned}
 \end{equation}
 where $E_1 = \| n^{-1} \sn e_i X_i \|_\infty$, $E_2 =   \sup_{\beta \in \beta^* + \BB_1(r_1) }   \|  n^{-1} \sn   (1- \EE) \{  e_i( \beta) - e_i \} X_i   \|_\infty$, $W_i = \Sigma^{-1/2} X_i$, and
 the last step follows from Lemma~\ref{lem:approximate.neyman}.
 Conditioned on the event $\{ \lambda_e \geq 2 (E_1 + E_2) \}$,  it follows from \eqref{opt.cond1} and \eqref{opt.cond2} that $ \tau^2 (\hat \theta-\theta^*)^\T \,  \hat \Sigma \,(\hat \theta - \theta^*) \leq  2^{-1}\tau \lambda_e \big( 3 \| \hat \delta_{\cS} \|_1 - \| \hat \delta_{\cS^{{\rm c}}} \|_1 \big) +  2^{-1}\tau f_u  m_3 r_0^2 \| \hat \delta \|_\Sigma$, which implies the cone  property $\| \hat \delta_{\cS^{{\rm c}}} \|_1    \leq 3  \| \hat \delta_{\cS} \|_1 + f_u  m_3 \lambda_e^{-1} r_0^2 \| \hat \delta \|_\Sigma $. Under the restricted eigenvalue condition (4.1) and if $\lambda_e \geq f_u  m_3 \underline{\phi} s^{-1/2} r_0^2$, we have 
 \#
 \| \hat \delta \|_1 \leq 4 \| \hat \delta_{\cS} \|_1 + \underline{\phi}^{-1} s^{1/2} \| \hat \delta \|_\Sigma \leq 4s^{1/2} \| \hat \delta_{\cS} \|_2 + \underline{\phi}^{-1} s^{1/2} \| \hat \delta \|_\Sigma \leq  5 \underline{\phi}^{-1} s^{1/2}  \| \hat \delta \|_\Sigma, \label{l1-sigma-norm}
 \# 
 and hence $\hat \delta \in \CC(l_1)$ with $l_1 = 5 \underline{\phi}^{-1} s^{1/2}$.

 To bound the left-hand side of \eqref{opt.cond1} from below,  define the event 
 \$
 \cE_{{\rm re}}( \kappa  ) =  \big\{    \delta^\T  \hat \Sigma  \delta \geq  \kappa \| \delta \|_\Sigma^2   \mbox{ for all } \delta \in \CC(l_1) \big\} , \ \ \kappa \in (0, 1).
 \$
 Conditioned on $\cE_{{\rm re}}(c ) $ for some  constant $c \in  (0, 1)$,   we conclude from  the  above upper and lower bounds that
 \$
 c \tau^2 \| \hat \delta \|_\Sigma^2  &  \leq  \frac{3}{2}\tau \lambda_e  \| \hat \delta_{\cS} \|_1   +  \frac{1}{2} f_u  m_3   r_0^2 \cdot \tau \|\hat \delta \|_\Sigma   \leq  \bigg(  \frac{3}{2 \underline{\phi} }  s^{1/2} \lambda_e + \frac{1}{2 \underline{\phi}} s^{1/2} \lambda_e  \bigg)  \cdot \tau \| \hat \delta \|_\Sigma.
 \$
 Canceling out a factor of $\tau \| \hat \delta \|_\Sigma$ from both sides yields
 \#
 \tau  \| \hat \theta - \theta^* \|_\Sigma \leq  \frac{2  }{ c   \underline{\phi} } s^{1/2} \lambda_e  . \label{deterministic.error.bound}
 \#
 
 \noindent
 {\sc Step 2 (Probabilistic analysis).}   The error bound \eqref{deterministic.error.bound} is deterministic given the stated conditioning.  Probabilistic claims enter in certifying that the ``good" events
 \$
 \cE_{{\rm re}}(c) ~~\mbox{ and }~~ \big\{ \lambda_e \geq 2(E_1 +E_2 )   \big\}
 \$
 hold with high probability with a specified choice of $c$ and $\lambda_e$. For the former, we  may take $c=1/2$ so that, by Lemma~\ref{lem:RE}, $\PP\{ \cE_{{\rm re}}(1/2) \} \geq 1-e^{-u}$ as long as $n \geq \max[ C m_4  \{\underline{\phi}^{-2} \sigma_X^2 s\log (2p) + u\} , (b_X/\sigma_X)^2 \log (2p)]$ for some sufficiently large constant $C$.  Turning to $E_1$ and $E_2$, it follows from Lemma~\ref{lem:first-order.error} and Lemma~\ref{lem:score.bound} that with probability at least $1-2p(p+1) e^{-t}$,
 \$
 E_1 \leq  \sigma_\varepsilon \sigma_X \sqrt{\frac{2t}{n}} + 2 b_\varepsilon b_X \frac{t}{n}~~\mbox{ and }~~E_2\leq 4 b_X  r_1  \bigg(   \sigma_X  \sqrt{\frac{2t}{n}} + b_X  \frac{t}{3 n} \bigg).
 \$
As long as  
 $n \geq C  (\frac{b_\varepsilon b_X}{\sigma_\varepsilon \sigma_X})^2 t$ for some absolute constant $C>1$, then with the penalization parameter $\lambda_e$ satisfying 
 $$
 \lambda_e \geq  (4+\sqrt{2}) \sigma_\varepsilon \sigma_X \sqrt{\frac{t}{n}} + (4\sqrt{2}+4/3) b_X \sigma_X r_1 \sqrt{\frac{t}{n}} ,
 $$
 we have $\PP \{ \lambda_e \geq 2(E_1 + E_2 ) \} \geq 1 - 2p(p+1)e^{-t}$.  Combining this with \eqref{deterministic.error.bound} gives the upper bound for $\tau  \| \hat \theta - \theta^* \|_\Sigma$. Combining this with \eqref{l1-sigma-norm} gives the upper bound for $\tau  \| \hat \theta - \theta^* \|_1$. \qed

 \subsection{Proof of Proposition~4.1}\label{pf:prop:qr}

 Define the function 
 \$
 \hat \cR(\delta) = \hat \cQ(\beta^* + \delta) - \hat \cQ(\beta^*) + \lambda_q \big( \| \beta^* + \delta \|_1 - \| \beta^* \|_1 \big) , \ \ \delta \in \RR^p, 
 \$
 satisfying $\hat \cR(0)=0$ and $\hat \cR(\hat \delta) \leq 0$ by the optimality of $\hat \beta$, where $\hat \delta = \hat \beta - \beta^*$. Moreover, write $w_\beta = n^{-1} \sn   \{ \mathbbm{1}(Y_i \leq X_i^\T \beta ) - \tau  \} X_i$ for $\beta \in \RR^p$ so that $w_\beta \in \partial \hat \cQ(\beta)$ is a subgradient of $\hat \cQ$ at any $\beta$.  
 Denote the support of $\beta^*$ by $\cS= {\rm supp}(\beta^*)$ satisfying $|\cS|\leq s$. Then,  applying Proposition~9.13 and (9.50) in \cite{W2019} with $\cL_n = \hat \cQ$ and $\Phi(\cdot) = \| \cdot \|_1$, we obtain that conditioned on the event $\{ \lambda_q \geq 2 \| w_{\beta^*} \|_\infty \}$,  the error $\hat \delta = \hat \beta - \beta^*$ belongs to the cone set $\mathbb{C}(\cS)$,  and for any $\delta \in  \mathbb{C}(\cS)$,
 \begin{equation}  \label{qr.remainder.lbd1}
 	\begin{aligned}
 		\hat \cR(\delta) & \geq  \hat \cQ(\beta^* + \delta) - \hat \cQ(\beta^*) + \lambda_q \big(  \| \delta_{\cS^{{\rm c}}} \|_1 - \| \delta_{\cS} \|_1 \big) \\
 		& \geq  \hat \cQ(\beta^* + \delta) - \hat \cQ(\beta^*) - \underline{\phi}^{-1} s^{1/2} \lambda_q \| \delta \|_\Sigma, 
 	\end{aligned}
 \end{equation}
 where the second inequality follows from the RE condition (4.1).  
 
 Next, let $\cQ(\beta)=\EE \{\hat \cQ(\beta)\}$ be the population quantile loss, satisfying $\nabla \cQ(\beta^*) = 0$ and $\nabla^2 \cQ(\beta) = \EE [ f_{\varepsilon_i | X_i}\{ X_i^\T (\beta - \beta^*) \} X_i X_i^\T ]$. Under Condition 4.3,  it holds for any $\delta \in \RR^p$ and $t \in [0, 1]$ that
 \$
 \delta^\T  \nabla^2 \cQ(\beta^* + t \delta ) \delta & = \EE \big\{ f_{\varepsilon_i | X_i}(t X_i^\T \delta ) (X_i^\T \delta)^2 \big\} \\
 & =  \EE \big\{ f_{\varepsilon_i | X_i}(0) (X_i^\T \delta)^2 \big\} +  \EE  \big\{ f_{\varepsilon_i | X_i}(t X_i^\T \delta ) - f_{\varepsilon_i | X_i}(0)\big\} \big (X_i^\T \delta )^2  \\\
 & \geq  f_l  \| \delta \|_\Sigma^2 - l_0 t \cdot \EE \big(| X_i^\T \delta |^3\big) \\
 & \geq  f_l  \| \delta \|_\Sigma^2 - l_0 t \sup_{u \in \mathbb{S}^{p-1}}\frac{\EE \big(| X_i^\T u |^3\big)}{(u^\T \Sigma u)^{3/2}}  ( \delta^\T \Sigma \delta)^{3/2} \\
 &\geq  f_l  \| \delta \|_\Sigma^2 - l_0 m_3 t \cdot \| \delta \|_\Sigma^3.
 \$
 This together with the
 fundamental theorem of calculus implies
 \begin{equation*}
     \begin{aligned}
         \cQ(\beta^* + \delta ) - \cQ(\beta^*)  &    =     \nabla  \cQ(\beta^*)^\T \delta + \int_0^1 \{\nabla \cQ(\beta^* + t \delta  ) - \nabla \cQ(\beta^*)\}^\T \delta  {\rm d} t \\
 & =  0 +   \int_0^1  \int_0^1  t \delta^\T  \nabla^2 \cQ(\beta^* + t u \delta) \delta  {\rm d} u {\rm d} t  \\
 & \geq \frac{1}{2} f_l \,\| \delta  \|_\Sigma^2 - \frac{1}{6} l_0 m_3 \| \delta \|_\Sigma^3.
     \end{aligned}
 \end{equation*}
 For some $r_0, l_1  >0$ to be determined, it follows from Lemma~\ref{lem:qr.diff.bound} that, with probability at least $1-2 p e^{-t}$ for any $t\geq 0$,
 \$
 \sup_{\delta \in \CC(l_1) \cap \BB_\Sigma(r_0) }    \big\{   \cQ(\beta^* + \delta ) - \cQ(\beta^*)  -  \hat \cQ(\beta^* + \delta) + \hat \cQ(\beta^*) \big\} \leq r_0 \cdot \gamma(n, t),
 \$
 where $\gamma(n,   t)  = 4 \max(\tau, 1-\tau) l_1 \{  \sigma_X \sqrt{2t/n}   +  b_X t /(3n) \}$. 
 The previous two inequalities and \eqref{qr.remainder.lbd1} show that with probability at least $1-e^{-t}$,
 \#
 \hat \cR(\delta ) \geq  \frac{r_0}{2}  \Big\{   f_l  r_0 - \frac{1}{3}l_0 m_3 r_0^2 - 2 \gamma(n,   t) - 2\underline{\phi}^{-1} s^{1/2} \lambda_q \Big\}
 \#
 holds for any $\delta \in \BB_\Sigma(r_0) \cap \CC(l_1)$. We thus choose $r_0 = 2  f_l ^{-1}\{ 2 \gamma(n,   t) + 2\underline{\phi}^{-1} s^{1/2} \lambda_q\}$ and let $(n, \lambda_q)$ satisfy $2 \gamma(n,  t) + 2\underline{\phi}^{-1} s^{1/2} \lambda_q < 3  f_l ^2(4 l_0 m_3)^{-1}$. Then, with high probability $\hat \cR(\delta) >0$ for all $\delta \in \BB_\Sigma(r_0) \cap \CC(l_1)$. Recall that $\hat \cR(\hat \delta) \leq 0$ and $\hat \delta \in \mathbb{C}(\cS) \subseteq \CC(l_1)$ with $l_1 = 4\underline{\phi}^{-1} s^{1/2}$ conditioned on $\{ \lambda_q \geq 2 \| w_{\beta^* } \|_\infty \}$.  Consequently, we conclude from Lemma~9.21 in \cite{W2019} and the convexity of $\hat \cQ(\cdot)$ that $\hat \delta \in \BB_\Sigma(r_0)$ with probability at least $1-e^{-t}$  conditioned on $\{ \lambda_q \geq 2 \| w_{\beta^* } \|_\infty \}$.
 Combining this with Lemma~\ref{lem:qr.score.bound} 
 establishes the claim. \qed

 \subsection{Proof of Proposition~4.2}\label{pf:prop:gamma.rate}
 
 The proof is based on the same arguments used to derive Theorem~4.1. With slight abuse of notation, we use $\mathbb{C}(S_0)$ and ${\rm C}(l)$ to denote the cones $\{ \delta \in \RR^{p-1}: \| \delta_{S_0^{{\rm c}}} \|_1 \leq 3 \| \delta_{S_0} \|_1 \}$ and $\{ \delta \in \RR^{p-1} : \| \delta \|_1 \leq l_1 \|\delta \|_{\Sigma_{-j}} \}$, respectively, where $S_0= {\rm supp}(\gamma^*) $. Conditioned on the event $\mathcal{E}_1 := \{ \lambda_m \geq 2 \| n^{-1} \sn \omega_i X_{i, -j}  \|_\infty \}$,  the error $\hat \gamma - \gamma^*$ belongs to the cone set $\mathbb{C}( S_0 )$ and hence also to $\CC(l_1)$ with $l_1 = 4 \underline{\phi}^{-1} s_0^{1/2}$. Conditioned further on $\mathcal{E}_2(c) :=\{ \inf_{\delta \in \CC(l_1) } \delta^\T \, \hat \Sigma_{-j} \delta /  \| \delta \|_{\Sigma_{-j}}^2  \geq c \}$ for some $c \in (0, 1)$, it follows  that $c \| \hat \gamma - \gamma^* \|_{\Sigma_{-j}}^2 \leq (3/2) \lambda_m \| ( \hat \gamma - \gamma^*  )_{S_0} \|_1 \leq 3(2 \underline{\phi})^{-1} s_0^{1/2} \lambda_m \| \hat \gamma - \gamma^* \|_{\Sigma_{-j}}$,  implying $\| \hat \gamma - \gamma^* \|_{\Sigma_{-j}} \leq 3 (2 c \underline{\phi})^{-1} s_0^{1/2} \lambda_m$ and $\| \hat \gamma - \gamma^* \|_1\leq 6 c^{-1} \underline{\phi}^{-2} s_0 \lambda_m$.

The rest of the proof follows the same argument as in Step 2 of Theorem~4.1, using Lemmas~\ref{lem:score.bound} and \ref{lem:RE} to certify that the event $\mathcal{E}_1\cap \mathcal{E}_2(1/2)$ holds with probability at least $1-2(p-1)e^{-t}-e^{-u}$ under the claimed sample size requirement. \qed

\subsection{Proof of Corollary~4.1}\label{pf:cor:ES_bound}
For the $\ell_1$-penalized quantile regression, by taking $t = \log(p^2)$ in Proposition~4.1, we have that, with sample size $n \geq C^2 s^2 \log(p)$, and the penalization parameter $\lambda_q$ satisfies $$ C_1 \sqrt{\log(p)/n  } \leq \lambda_q \leq C_2 \sqrt{\log(p)/n  },$$ 
where $C_1 = 4 \sigma_X \sqrt{   \tau(1-\tau)} + 4 \bar \tau b_X/(3 C)$ and  $C_2$ is some sufficiently large constant,
\$
		\| \hat \beta - \beta^* \|_\Sigma \leq
  C_5(f_l, \underline{\phi}, \sigma_X, b_X ) \sqrt{\frac{s \log (p)}{n}}  ~~\mbox{ and }~~
		\| \hat \beta - \beta^* \|_1 \leq  C_6(f_l, \underline{\phi}, \sigma_X, b_X )s\sqrt{\frac{ \log (p)}{n}}
		\$
		with probability at least $1- 4/p$, 
  where $C_5(f_l, \underline{\phi}, \sigma_X, b_X ) = 2f_l^{-1}\underline{\phi}^{-1} \{   64 \bar \tau\sigma_X    + 64 \bar \tau b_X (3C)   + 2 C_2 \}$ and $C_6(f_l, \underline{\phi}, \sigma_X, b_X ) = 8\underline{\phi}^{-2}  f_l^{-1}  \{   64 \bar \tau\sigma_X    + 64 \bar \tau b_X (3C)   + 2 C_2 \} $, as long as $s^{1/2} r\{n,  \log(p^2)\} < 3 f_l^2/(4  m_3 l_0 )$. 
  
  For the $\ell_1$-penalized expected shortfall regression, taking $r_0 = C_5 \sqrt{s \log(p)/n}$, $r_1 = C_6 s \sqrt{\log(p)/n}$, $t=\log(p^3)$, and $u = \log(p)$ in Theorem~4.1, we have that,  with sample size $n \geq C^2 s^2 \log(p)$, and the penalization parameter $\lambda_e$ satisfies
  $$
  C_3 \sqrt{\log(p)/n}\leq \lambda_e \leq C_4 \sqrt{\log(p)/n},
  $$
  where $C_3 = \max\{5.5 \sigma_X \sigma_\epsilon + 6  \sigma_X b_X C_6/C, f_u m_3 \underline{\phi}/C \}$, and $C_4$ is a some sufficiently large constant, with probability at least $1-5/p$, 
  \$
  \tau \| \hat \theta - \theta^* \|_\Sigma \leq 4 C_4 \underline{\phi}^{-1} s^{1/2}  \sqrt{\log(p)/n}
  ~~\mbox{ and }~~
  \tau  \| \hat \theta  - \theta^* \|_1 \leq  20 C_4 \underline{\phi}^{-2} s \sqrt{\log(p)/n}.
  \$
The statement holds since $r_0 = C_5 \sqrt{s \log(p)/n}$ and $r_1 = C_6 s \sqrt{\log(p)/n}$ hold with probability at least $1-4/p$ under certain conditions. \qed

\subsection{Proof of Theorem~4.2}\label{pf:thm:score.clt}

We first derive the asymptotic normality for the triply-orthogonal score function in general cases, which will be used in proving Theorem~4.3. Then, the conclusion hold similarly under $H_0$. To begin with,  we decompose $S_n(\theta^*_j, \hat \theta_{-j} ,    \hat \beta,  \hat \gamma)- S_n(\theta^*_j,   \theta_{-j}^* ,     \beta^*,  \gamma^*)$ as 
 \begin{equation*}
 	\begin{aligned}
 		& S_n(\theta^*_j , \hat \theta_{-j} ,    \hat \beta,  \hat \gamma) - S_n( \theta^*_j ,  \theta_{-j}^* ,     \beta^*,  \gamma^*)  \\
 		& =  S_n(\theta^*_j ,  \hat \theta_{-j} ,    \hat \beta,  \hat \gamma) -  S_n(\theta^*_j , \hat \theta_{-j} ,    \hat \beta,  \gamma^*) +  S_n(\theta^*_j ,  \hat \theta_{-j} ,    \hat \beta,  \gamma^*) - S_n(\theta^*_j ,  \theta_{-j}^* ,    \hat  \beta,  \gamma^*)   \\
 		& ~~~~~+   S_n(\theta^*_j ,  \theta_{-j}^* ,    \hat   \beta,  \gamma^*) - S_n(\theta^*_j ,  \theta_{-j}^* ,     \beta^*,  \gamma^*)    \\
 		& = - \frac{1}{n} \sn  \{ Z_i(\hat \beta) -  \tau X_{i,j} \theta^*_j -  \tau X^\T_{i,-j} \hat \theta_{-j}  \} X^\T_{i,-j}( \hat \gamma - \gamma^* )    \\
 		&~~~~~ -  \frac{\tau}{n} \sn  \omega_i X^\T_{i,-j}(\hat \theta_{-j} - \theta_{-j}^* ) + \frac{1}{n} \sn \{ Z_i(\hat \beta) - Z_i (\beta^* ) \} \omega_i \\
 		& = - \frac{1}{n} \sn e_i(\hat \beta) X^\T_{i,-j}( \hat \gamma - \gamma^* )  + \frac{\tau}{n} \sn (\hat \theta _{-j}- \theta_{-j}^*)^\T X_{i,-j} X^\T_{i,-j} (\hat \gamma - \gamma^*)   \\
 		& ~~~~~ -  \frac{\tau}{n} \sn  \omega_i X^\T_{i,-j}(\hat \theta_{-j} - \theta_{-j}^* ) + \frac{1}{n} \sn \{ e_i(\hat \beta) - e_i \} \omega_i  ,
 	\end{aligned}
 \end{equation*}
 where $e_i(\beta) = Z_i(\beta) - \tau X_i^\T \theta^*$, and $e_i = e_i(\beta^*)$. 
 Conditioned on the event  $\{ \|  \hat \beta - \beta^*  \|_{\Sigma} \leq r_0   \} \cap  \{ \| \hat \beta - \beta^* \|_1 \leq r_1   \}$, it follows that
 \begin{equation*}
 	\begin{aligned}
 		& | S_n(\theta^*_j , \hat \theta_{-j}  ,    \hat \beta,  \hat \gamma) - S_n( \theta^*_j ,  \theta_{-j} ^* ,     \beta^*,  \gamma^*) | \\ 
 		& = \bigg|-\frac{1}{n} \sn (1-\EE)   [\{e_i(\hat \beta)-e_i\} X_{i,-j}^\T  ] ( \hat \gamma - \gamma^* ) - \frac{1}{n} \sn e_i X^\T_{i,-j}( \hat \gamma - \gamma^* )\\
 		&~~~~~ - \EE\{e_i(\hat \beta) W^\T_{i,-j}\} \Sigma^{1/2}_{-j} ( \hat \gamma - \gamma^* ) + \tau (\hat \theta_{-j}  - \theta_{-j} ^*)^\T (\hat \Sigma_{-j} - \Sigma_{-j}) ( \hat \gamma - \gamma^* ) \\
 		&~~~~~+  \tau (\hat \theta_{-j}  - \theta_{-j} ^*)^\T  \Sigma_{-j} ( \hat \gamma - \gamma^* )  - \frac{\tau}{n} \sn  \omega_i X^\T_{i,-j}(\hat \theta_{-j} - \theta_{-j}^* ) \\
 		&~~~~~ + \frac{1}{n} \sn  (1-\EE)  [\{ e_i(\hat \beta) - e_i  \}  \omega_i ] + \EE \{ e_i(\hat \beta) \omega_i \}\bigg| \\
 		& \leq    \sup_{\beta \in \beta^* + \BB_1(r_1) } \bigg\| \frac{1}{n} \sn  (1-\EE ) [\{ e_i(\beta) - e_i \} X_{i,-j} ] \bigg\|_\infty \cdot  \| \hat \gamma - \gamma^* \|_1  \\ 
 		&~~~~~+   \bigg\| \frac{1}{n} \sn e_i X_{i,-j} \bigg\|_\infty  \cdot  \| \hat \gamma - \gamma^* \|_1 +  \sup_{\beta \in \beta^* + \BB_{\Sigma}(r_0) }  \| \EE \{ e_i (\beta) W_{i,-j} \} \|_2 \cdot  \| \hat \gamma - \gamma^* \|_{\Sigma_{-j}}  \\
 		&~~~~~+     \| \hat \Sigma_{-j}  - \Sigma_{-j}     \|_{\max} \cdot \tau \|  \hat \theta_{-j}  - \theta_{-j} ^* \|_1 \cdot \| \hat \gamma - \gamma^* \|_1   +  \tau \|  \hat \theta_{-j} - \theta_{-j}^* \|_{\Sigma_{-j}}  \cdot \| \hat \gamma - \gamma^* \|_{\Sigma_{-j}}  \\
 		&~~~~~  +    \bigg\| \frac{1}{n} \sn \omega_i X_{i,-j} \bigg\|_\infty  \cdot \tau \| \hat \theta_{-j}  - \theta_{-j} ^* \|_1   +    \sup_{\beta \in \beta^* + \BB_1(r_1) }  \bigg| \frac{1}{n} \sn  (1-\EE) [\{ e_i(\beta) - e_i  \}  \omega_i ] \bigg|   \\
 		&~~~~~  +  \sup_{\beta \in \beta^* + \BB_{\Sigma}(r_0) }   | \EE \{ e_i(\beta) \omega_i \} |,
 	\end{aligned}
 \end{equation*}
 where $W_{i,-j} = \Sigma_{-j}^{-1/2} X_{i,-j}$, and $\hat \Sigma_{-j} = n^{-1} \sn X_{i,-j} X_{i,-j}^\T$.  Since $\EE(e_i | X_i) = 0$, the first equality follows from $\EE(e_iX_{i,-j}) = \EE \{\EE (e_i X_{i,-j}|X_i)\} = \EE \{X_{i,-j} \EE (e_i |X_i)\} = 0 $ and similarly $\EE(e_i\omega_i)=0$.

 For the bias terms $\| \EE \{ e_i (\beta) W_{i,-j} \} \|_2$ and $ | \EE \{ e_i(\beta) \omega_i \} |$,  Lemma~\ref{lem:approximate.neyman} states that
 $$
 \sup_{\beta \in \beta^* + \BB_\Sigma(r_0) }  \| \EE \{ e_i (\beta) W_{i,-j} \} \|_2 \leq \frac{1}{2} f_u   m_3 r_0^2 ~~\mbox{ and }~~\sup_{\beta \in \beta^* + \BB_\Sigma(r_0) } | \EE \{ e_i(\beta) \omega_i \} | \leq \frac{1}{2} f_u  b_X r_0^2.
 $$
 Taking $t  = 3  \log(p)$ in Lemma~\ref{lem:first-order.error} and $t  = 2  \log(p)$ in Lemma~\ref{lem:score.bound}, we obtain that with probability at least $1-6/p$,
 \$
 \sup_{\beta \in \beta^* + \BB_1(r_1) } \bigg\| \frac{1}{n} \sn  (1-\EE ) \{ e_i(\beta) - e_i  \} X_{i,-j} \bigg\|_\infty \leq C_1   b_X \sigma_X   \sqrt{\frac{\log(p)}{n}}  \cdot r_1 , \\
 \sup_{\beta \in \beta^* + \BB_1(r_1) }  \bigg| \frac{1}{n} \sn (1-\EE) [\{ e_i(\beta) - e_i \}  \omega_i  ] \bigg|\leq C_1   b_X \sigma_X  \sqrt{\frac{\log(p)}{n}}  \cdot r_1  , \\ 
 \bigg\| \frac{1}{n} \sn e_i X_{i,-j}\bigg\|_\infty\leq C_2 \sigma_\varepsilon \sigma_X \sqrt{\frac{\log(p)}{n}} ~~\mbox{ and }~~
 \bigg\| \frac{1}{n} \sn \omega_i X_{i,-j} \bigg\|_\infty \leq C_2  b^2_X   \sqrt{\frac{\log(p)}{n}} ,
 \$
 provided  $n\geq  (\frac{b_X b_\varepsilon}{\sigma_X \sigma_\varepsilon})^2 \log(p)$.  
 
 Turning to $\| \hat \Sigma_{-j}  - \Sigma_{-j}     \|_{\max} $,  note that for every pair $\{(l,k): 1\leq l\leq k\leq p, l \neq j,k \neq j\}$,  $\EE \{(X_{i,l} X_{i,k})^2\} \leq  \{ \EE (X_{i,l}^4) \}^{1/2} \{ \EE (X_{i,k}^4) \}^{1/2} \leq m_4^{1/2} \EE(X_{i,l}^2) m_4^{1/2} \EE(X_{i,k}^2) = m_4\sigma_{ll} \sigma_{kk} \leq m_4 \sigma_X^4$ and $| X_{i,l} X_{i,k} |\leq b_X^2$ almost surely from Condition~ 4.2.  Consequently, by Bernstein's inequality, i.e. taking $v = n m_4 \sigma_X^4$, $c = \frac{b_X^2}{3}$, and $t=3\log(p)$  in Lemma~\ref{lem:Berstein},  
 \$
 \PP \bigg\{ \bigg| \frac{1}{n} \sn (X_{i,l} X_{i,k} - \sigma_{lk}) \bigg| \geq   m_4^{1/2} \sigma_X^2 \sqrt{\frac{6 \log (p)}{n}} + b_X^2 \frac{\log (p)}{n} \bigg\} \leq  \frac{2}{p^3} .
 \$
 Then, by the union bound over $\{(l,k): 1\leq l\leq k\leq p, l \neq j,k \neq j\}$, 
 \$
 \PP \bigg( 	\| \hat \Sigma_{-j}  - \Sigma_{-j}     \|_{\max}  \geq  m_4^{1/2} \sigma_X^2 \sqrt{\frac{6 \log (p)}{n}}  + b_X^2 \frac{\log (p) }{n} \bigg) \leq \frac{2}{p}.
 \$
 Putting together the pieces yields that with probability at least $1-8/p$ conditioned on $\{ \|  \hat \beta - \beta^*  \|_{\Sigma} \leq r_0  \} \cap  \{ \| \hat \beta - \beta^* \|_1 \leq r_1   \}$,
 \begin{equation}
 	\begin{aligned}
 		& | S_n(\theta_j^* , \hat \theta_{-j} ,    \hat \beta,  \hat \gamma) - S_n( \theta^*_j ,  \theta^*_{-j} ,      \beta^*,  \gamma^*) |    \\
 		& \leq C_2  b_X \sigma_X \sqrt{\frac{\log (p)}{n}} \cdot r_1 +  f_u  b_X  \cdot r_0^2  + ( C_1 b_X \cdot r_1  + C_2 \sigma_\varepsilon  ) \sigma_X  \sqrt{\frac{\log (p)}{n}} \cdot \| \hat \gamma - \gamma^* \|_1    \label{score.error.ubd1}  \\
 		&~~~~~ + \bigg( \frac{1}{2} f_u  m_3 \cdot r_0^2 + \tau \| \hat \theta_{-j} - \theta_{-j}^* \|_{\Sigma_{-j}} \bigg) \cdot \| \hat \gamma - \gamma^* \|_{\Sigma_{-j}}  \\
 		&~~~~~ + \bigg( m_4^{1/2} \sigma_X^2  \sqrt{\frac{6\log (p)}{n}} + b_X^2 \frac{\log (p)}{n} \bigg)  \cdot \tau \|  \hat \theta_{-j} - \theta_{-j}^* \|_1 \cdot \| \hat \gamma - \gamma^* \|_1 \\
 		&~~~~~ + C_2 b^2_X  \sqrt{\frac{\log (p)}{n}} \tau \|  \hat \theta_{-j} - \theta_{-j}^* \|_1.
 	\end{aligned}
 \end{equation}

 For the initial  estimators $\hat \beta$,  $\hat \theta$ and $\hat \gamma$,  we conclude from taking $t = C\log (p)$ for some absolute constant $C>1$ in Theorem~4.1 and Propositions~4.1,  4.2 that with penalization parameters $(\lambda_q, \lambda_e, \lambda_m)$ all in the order of  $\sqrt{\log(p)/n}$,  the following error bounds 
 $$	
 \begin{cases}
 	\| \hat \beta - \beta^* \|_{\Sigma}  \lesssim  \sqrt{s\log(p) / n }  , ~~ \| \hat \beta - \beta^* \|_1 \lesssim  s\sqrt{\log(p)/n}   \\
 	\tau \| \hat \theta - \theta^* \|_{\Sigma} \lesssim \sqrt{s\log(p) / n } , ~~  \tau \| \hat \theta - \theta^* \|_1 \lesssim s \sqrt{\log(p)/n}  \\
 	\| \hat \gamma - \gamma^* \|_{\Sigma_{-j}} \lesssim  \sqrt{s_0\log(p) / n }  ,~~  \| \hat \gamma - \gamma^* \|_1 \lesssim s_0 \sqrt{\log(p)/n}
 \end{cases}
 $$
 hold with probability at least $1- 7/p$ , provided the sample size $n$ satisfies $n\gtrsim s^2 \log(p)$.  Substituting these into the right-hand side of \eqref{score.error.ubd1} yields that with probability at least $13/p$,
 \#
 | S_n(\theta_j^* , \hat \theta_{-j},    \hat \beta ,  \hat \gamma) - S_n( \theta^*_j ,  \theta_{-j}^* ,     \beta^*,  \gamma^*) |    
 \lesssim \frac{\max(s,s_0)\log (p)}{n} 
 \#
 provided $n\gtrsim s^2 \log(p)$. As $n , p \to \infty$, this further implies 
 \#
 n^{1/2}  \big\{ S_n(\theta_j^* , \hat \theta_{-j} ,    \hat \beta,  \hat \gamma) - S_n( \theta^*_j ,  \theta_{-j}^* ,     \beta^*,  \gamma^*) \big\} = \cO_{p} \bigg\{ \frac{\max(s,s_0) \log (p)}{ n^{1/2} } \bigg\}.   \label{score.error.ubd2}  
 \#
 
 Applying the central limit theorem to the ``oracle" score $S_n( \theta^*_j ,  \theta^*_{-j} ,     \beta^*,  \gamma^*) $, we see that
 $$
 n^{1/2}    S_n(\theta^*_j,  \theta_{-j}^* ,     \beta^*,  \gamma^*) =  \frac{1}{\sqrt{n}} \sn e_i \omega_i   \to \cN\big(0,  \, \sigma_{{\rm s}}^2 \big) ,
 $$
 in distribution, where $\sigma_{{\rm s}}^2 = \EE\{(e_i \omega_i  )^2\}$, or equivalently,  recalling $e_i = (Y_i-X_i^\T \beta^*)\mathbbm{1}(Y_i \leq X_i^\T  \beta^*)  - \tau X_i (\theta^*-\beta^*) =  \varepsilon_{i, -} - \EE(\varepsilon_{i,-} | X_i)$ with  $\varepsilon_{i,-} = \min\{ \varepsilon_i, 0\} = (Y_i - X_i^\T \beta^*) \mathbbm{1}(Y_i \leq X_i^\T  \beta^*)$,  $\sigma_{{\rm s}}^2=  \EE\{\omega_i^2 \EE(e_i ^2|X_i)\}= \EE \{ (X_{i,j} - X_{i,-j}^\T \gamma^*)^2 \var_{X_i}( \varepsilon_{i, -} ) \}$.  Provided $\max(s,s_0) \log(p) = o(n^{1/2} )$ as $n, p \to \infty$,  this together with \eqref{score.error.ubd2} proves that $n^{1/2} S_n(\theta_j^*,\hat \theta_j,\hat \beta, \hat \gamma) \to \cN\big(0,  \, \sigma_{{\rm s}}^2)$ in distribution. 
 
 Under $H_0: \theta_j^* = c_0$, we substitute the estimation of $\theta_{-j}$ with $\hat \theta_{-j}(c_0)$ from~(3.8) in the triply-orthogonal score function. We proves the claim by the same arguments with the same error bounds $\tau \| \hat \theta_{-j}(c_0) - \theta^* \|_{\Sigma} \lesssim  \{n^{-1} s \log (p) \}^{1/2}$ and $\tau \| \hat \theta_{-j}(c_0)- \theta^* \|_1 \lesssim s  \{n^{-1} \log (p) \}^{1/2}$.  \qed

 \subsection{Proof of Theorem~4.3}\label{pf:thm:debias.clt}
 
 Let $\hat S_n(\theta_j) =  S_n(\theta_j ,   \hat \theta_{-j}, \hat \beta,  \hat \gamma)$. 
 For any $\theta_j\in \RR$,  note that
 \$
 \partial_{\theta_j}  \hat S_n(\theta_j) = -\frac{\tau}{n} \sn X_{i,j} (X_{i,j} - X_{i,-j}^\T \hat \gamma  ),
 \$
 and that
 \$ \hat S_n(\theta_j)   - \hat S_n(\theta_j^*)     = - \frac{ \tau}{n} \sn X_{i,j} (X_{i,j} -X_{i,-j}^\T \hat \gamma) (\theta_j - \theta^*_j).
 \$
 It then follows that
 \$
 \wt \theta_j = \hat \theta_j - 	\hat S_n(\hat \theta_j) /  \partial_{\theta_j}  \hat S_n(\hat \theta_j) =  \hat \theta_j - (\hat \theta_j - \theta^*_j) -  \hat S_n(\theta^*_j) /   \partial_{\theta_j}  \hat S_n(\theta^*_j) = \theta^*_j -  \hat S_n(\theta^*_j) /   \partial_{\theta_j}  \hat S_n(\theta^*_j).
 \$
 Applying first Theorem~4.2 we see that,  in the asymptotic regime $\max(s_0, s) \log(p) = o(n^{1/2})$,
 \#
 n^{1/2}   \hat S_n(\theta^*_j) = n^{1/2}   S_n(\theta^*_j, \hat \theta_{-j}, \hat \beta, \hat \gamma) \to \cN\big( 0, \, \sigma_{{\rm s}}^2 \big)  . \label{score.clt}
 \#
 in distribution. 
 Note further that  $\partial_{\theta_j}  \hat S_n(\theta^*_j) = -\frac{\tau}{n} \sn X_{i,j} \omega_i  + \frac{\tau}{n} \sn X_{i,j} X_{i,-j}^\T( \hat \gamma -  \gamma^*  )$, and under model (3.10),  
 $$
 \EE(X_{i,j} \omega_i) = \EE\{ (X_{i,-j}^\T \gamma^* + \omega_i) \omega_i \} = \EE (\omega_i^2)=\sigma_{\omega}^2. 
 $$
 Therefore, $n^{-1}\sn X_{i,j} \omega_i   \to \sigma_{\omega}^2$ in probability, and 
 \$
 &  \bigg| \frac{1}{n} \sn X_{i,j} X_{i,-j}^\T( \hat \gamma -  \gamma^*  ) \bigg|  \\
 & = \bigg| \frac{1}{n} \sn \{ (\gamma^{*} )^\T X_{i,-j} X_{i,-j}^\T + \omega_i X_{i,-j}^\T \} ( \hat \gamma -  \gamma^*  ) \bigg|\\
 & = \bigg| (\gamma^{*} )^\T (1-E) \frac{1}{n} \sn ( X_{i,-j} X_{i,-j}^\T) ( \hat \gamma -  \gamma^* ) + (\gamma^{*} )^\T \EE \bigg( \frac{1}{n} \sn  X_{i,-j} X_{i,-j}^\T \bigg) ( \hat \gamma -  \gamma^* ) \\
 &~~~~+ \frac{1}{n} \sn \omega_i X_{i,-j}^\T ( \hat \gamma -  \gamma^*  ) \bigg|\\
 &  \leq   \| \hat \Sigma_{-j}  - \Sigma_{-j}     \|_{\max}  \| \gamma^* \|_1  \| \hat \gamma - \gamma^* \|_1 +       \| \gamma^* \|_{\Sigma_{-j}} \|  \hat \gamma -  \gamma^* \|_{\Sigma_{-j}} + \bigg\| \frac{1}{n} \sn \omega_i X_{i,-j} \bigg\|_\infty  \| \hat \gamma - \gamma^* \|_1.
 \$
 Recall from the proof of Theorem~4.2 that
 \$
 \| \hat \Sigma_{-j}  - \Sigma_{-j}     \|_{\max}  = \cO_{p} \bigg( \sqrt{\frac{\log (p)}{n}} \, \bigg) , \quad  \bigg\| \frac{1}{n} \sn \omega_i X_{i,-j} \bigg\|_\infty = \cO_{p}  \bigg( \sqrt{\frac{\log (p)}{n}} \, \bigg) , \\
 \| \hat \gamma - \gamma^* \|_{\Sigma_{-j}} = \cO_{p} \bigg(  \sqrt{\frac{s_0 \log (p)}{n}} \, \bigg) ~~\mbox{ and }~~ \| \hat \gamma - \gamma^* \|_1 = \cO_{p} \bigg( s_0 \sqrt{\frac{\log (p)}{n}} \,  \bigg) .
 \$ 
 Moreover, by Cauchy-Schwarz inequality and Condition~ 4.2, $\|\gamma^* \|_1 \leq s_0^{1/2} \| \gamma^* \|_2 \leq \underline{\phi}^{-1} s_0^{1/2} \| \gamma^* \|_{\Sigma_{-j}} $ and $\| \gamma^* \|_{\Sigma_{-j}}^2 = (\gamma^*)^\T \Sigma_{-j} \gamma^* = (\gamma^*)^\T  \EE(X_{i,-j} X_{i,j}) = \EE(X_{i,j}^2) - \sigma_{\omega}^2$.
 Putting these bounds together yields
 \$
 \bigg| \frac{1}{n} \sn X_{i,j} X_{i,-j}^\T( \hat \gamma -  \gamma^*  ) \bigg|  = \cO_{p}  \bigg(  \sqrt{\frac{s_0 \log (p)}{n}}  +  s_0^{3/2} \frac{\log (p)}{n} \bigg) .
 \$
 Under the assumed scaling condition,  this implies $\partial_{\theta_j} \hat S_n (  \theta^*_j) \to   \tau \sigma_{\omega}^2$ in probability.  Finally, combining this with \eqref{score.clt} and Slutsky's theorem establishes the claim. \qed

\subsection{Proof of Theorem~4.4}\label{pf:thm:var.est}
We will show that $\hat{\sigma}_{{\omega}}^2 \to \sigma_{{\omega}}^2$  and  $\tilde{\sigma}_{{\rm s}}^2 \to \sigma_{{\rm s}}^2$ in probability. The conclusion follows by an application of the Slutsky's theorem. 
We begin with showing $\hat{\sigma}_{{\omega}}^2 \to \sigma_{{\omega}}^2$ in probability by applying Theorem 2 of~\cite{FGH2012}. 
For completeness, we restate Theorem 2 of ~\cite{FGH2012} under the current setting. We start by adapting the two conditions for Theorem 2 of~\cite{FGH2012}. 
\begin{condition}\label{cond::rcv1}
    The projection residuals $\omega_1,\ldots,\omega_n$ defined in~(3.10) are i.i.d. with mean zero and finite variance $\sigma_{{\omega}}^2$. 
    In addition, $\omega_i$ is independent of $X_{i,-j}$. 
\end{condition}
Let $M \subseteq \{1,\ldots,p\}$ be an index set with cardinality $|M| \le m$ and let $X_M \in \RR^{n \times |M|}$ be the submatrix of $X$ with columns indexed by $M$.  For any symmetric and positive-definite matrix $A$, let $\lambda_{\min}(A)$ and $\|A\|$ be its minimum eigenvalue and its operator norm, respectively. 
\begin{condition}\label{cond::rcv2}
    There exist  a constant $\lambda_0>0$ and a sequence $\{b_n\}$ such that $b_n^2 \log(p) = o(n)$ and $\PP\{ \min_{M:|M|\leq b_n}\lambda_{\min} (n^{-1}  X_{M}^\T X_{M}) \geq \lambda_0\} \to 1$ as $n \to \infty$. 
\end{condition}
\begin{theorem}[Theorem 2 of ~\cite{FGH2012}]
    Assume Condition~\ref{cond::rcv1} and~\ref{cond::rcv2} and $E(\omega_i^4)<\infty$. If $P(S_m^* \subseteq \hat S_m) \to 1$, as $n,p \to \infty$,  with $\hat s_m \leq b_n$, where $b_n$ is defined in Condition~\ref{cond::rcv2}, then
    \begin{equation}\label{eq::rcv_variance_asymp}
        n^{1/2} (\hat{\sigma}_{{\omega}}^2 - \sigma_{{\omega}}^2) \overset{d}{\to} N(0,E(\omega_i^4) - \sigma_{{\omega}}^4). 
    \end{equation}
\end{theorem}
As a direct consequence of~\eqref{eq::rcv_variance_asymp}, $\hat{\sigma}_{{\omega}}^2 \overset{p}{\to} \sigma_{{\omega}}^2$ since $E(\omega_i^4)\leq b_X^2 \EE(\omega_i^2) <\infty$. Therefore, it suffices to verify Condition~\ref{cond::rcv1} and~\ref{cond::rcv2} hold under Condition 4.1--4.4. 
By making a slight modification to the proof of~\cite{FGH2012}, it can be shown that the independence condition in Condition~\ref{cond::rcv1} 
can be weakened to the condition $\EE(\omega_i X_{i,-j})=0$ as in (3.10). 

Next, we show Condition~\ref{cond::rcv2}
holds.  
For any index set $M \subseteq \{1,\ldots,p\}$ satisfying $|M| = m$, we have 
\begin{equation*}
    \begin{split}
        \lambda_{\min} (n^{-1} X_{M}^\T X_{M}) &= \min_{\|z\|_2=1} z^\T (n^{-1} X_{M}^\T X_{M})  z\\
        & = \min_{\|z\|_2=1} \big[ z^\T \{n^{-1} (1-\EE) X_{M}^\T X_{M}\}  z + z^\T \EE(n^{-1} 
 X_{M}^\T X_{M})  z \big] \\
 & \geq -\| n^{-1} (1-\EE) X_{M}^\T X_{M} \| + \lambda_{\min} \{n^{-1} \EE (X_{M}^\T X_{M})\}\\
 & \geq -\| n^{-1} (1-\EE) X_{M}^\T X_{M} \| + \underline{\lambda} ,
    \end{split}
\end{equation*}
where the last inequality is due to Condition~ 4.2. 
 
We then employ an $\epsilon$-net argument to bound the term $\| n^{-1} (1-\EE) X_{M}^\T X_{M} \|$. Specifically, let $\cN$ be a $(1/4)$-net of  the unit sphere with cardinality $|\cN| \le 9^m$ such that 
$\| n^{-1} (1-\EE) X_{M}^\T X_{M} \| \le 2 \max_{u \in \mathcal{N}} |u^\T \{n^{-1} (1-\EE) X_M^\T X_M\} u| $.  It remains to control $ |u^\T \{n^{-1} (1-\EE) X_M^\T X_M\} u|$ for each direction $u\in \cN$.  
Let $X_{i,M}$ be the $i$-th row of $X_M$. Condition~ 4.2 ensures that
\begin{equation*}
    \begin{aligned}
        \EE\{(X_{i,M}^{\T} u)^4\} &\leq m_4 \{ \EE (X_{i,M}^{\T} u)^2 \}^2   = m_4 \{u^{\T} \EE (X_{i,M}^{\T} X_{i,M}) u \}^2 \leq m_4 \bar{\lambda}^2 .
    \end{aligned}
\end{equation*}
Moreover, since $|X_{i,M}^{\T} u| \leq \| X_{i,M} \|_\infty \| u \|_1 \leq b_X \sqrt{m}$, we have
\begin{equation*}
    \begin{aligned}
        \EE\{(X_{i,M}^{\T} u)^{2k}\} &= \EE\{(X_{i,M}^{\T} u)^{4} \cdot (X_{i,M}^{\T} u)^{2(k-2)}\}
         \leq m_4 \bar{\lambda}^2 (b_X^2 m)^{ k-2 }. 
    \end{aligned}
\end{equation*}
Applying the Bernstein's inequality in Lemma~\ref{lem:Berstein}, with probability at least $1-e^{-t}$,
\begin{equation*}
     |u^\T \{n^{-1} (1-\EE) X_M^\T X_M\} u|  = \bigg| \frac{1}{n}  \sn (1-\EE) (X_{i,M}^{\T}  u)^2\bigg| \leq \bar{\lambda}  \sqrt{\frac{2m_4 t}{n}}+  b_X^2 \frac{ m t}{n}.
\end{equation*}
Taking the union bound over $u \in \mathcal{N}$ yields
\begin{equation*}
    \PP\left\{2 \max_{u \in \mathcal{N}} \bigg| \frac{1}{n}  \sn (1-\EE) (X_{i,M}^{\T}  u)^2\bigg| > 2   \bar{\lambda} \sqrt{\frac{2m_4 t}{n}} + 2 b_X^2 \frac{m t}{n} \right\} < 9^m e^{-t}.
\end{equation*}
The above bound holds for any given $M$ with $m=|M|$. Taking another union bound over all subset $M$ satisfying $|M| \leq b_n$ yields
\begin{align*}
	\PP\bigg\{ \min_{M:|M|\leq b_n}\lambda_{\min} ( n^{-1} X_{M}^\T X_{M}) \leq -2   \bar{\lambda} \sqrt{\frac{2 m_4 t}{n}} - 2 b_X^2 \frac{m t}{n}  + \underline{\lambda} \bigg\}  & \leq \sum_{m=0}^{b_n} \binom{p}{m} 9^m e^{-t} \\
&\leq  \bigg( \frac{9ep}{b_n} \bigg)^{b_n}e^{-t}.
\end{align*}
Finally, setting $t =b_n \log(9ep b_n^{-1})  + \log(n)$ and under scaling condition $b_n^2 \log(p) = o(n)$, there exists some sufficiently large constant $C>0$ with $n \ge C b_n^2 \log (p)$ such that 
\begin{equation*}
	\begin{aligned}
		\min_{M:|M|\le b_n }  \lambda_{\min} (n^{-1} X_{M}^\T X_{M})
		\geq &-2 \sqrt{2m_4} \bar{\lambda} \left(\frac{\log(n) + b_n \log(9ep b_n^{-1}) }{n}\right)^{1/2} \\
		& - 2 b_n b_X^2 \frac{\log(n) + b_n \log(9ep b_n^{-1}) }{n} + \underline{\lambda} \\
		& \geq \underline{\lambda}/2,
	\end{aligned}
\end{equation*}
with probability at least $1-n^{-1}$.
Assumption 2 follows by setting $\lambda_0 = \underline{\lambda}/2$.

The proof of the consistency of $\hat{\sigma}_{{\rm s}}^2$ relies on an extension of the previous argument from \cite{FGH2012} combined with the consistency of the naive variance estimator $\tilde{\sigma}_{{\rm s}}^2 = n^{-1} \sum_{i=1}^n \hat{\omega}_i^2 (\hat Z_i - \tau X_i^\T \hat \theta)^2$.  
By the sure screening property as assumed in Condition~4.5, the refitted estimators admit the convergence rates $\sqrt{b_n/n}$ and $b_n/\sqrt{n}$ under the $\ell_2$ and $\ell_1$ norms,  respectively.  Based on this observation, the extension of  \cite{FGH2012}'s argument is a customary procedure but involves laborious calculations.  Therefore we opt to omit it here.   Turning to the naive variance estimator $\tilde{\sigma}_{{\rm s}}^2 = n^{-1} \sn \hat{\omega}_i^2 \hat e_i^2$, recall that $\hat e_i = \hat Z_i - \tau X_i^\T \hat \theta$ and $e_i =  Z_i^* - \tau X_i^\T \theta^*$, where $\hat{Z}_i = Z_i(\hat{\beta})$ and $Z_i^* = Z_i(\beta^*)$.  We consider 
\begin{equation}\label{eq:1}
	\begin{aligned}
		|\tilde{\sigma}_{{\rm s}}^2 - \sigma_{{\rm s}}^2| &= \bigg| \frac{1}{n} \sn \hat{\omega}_i^2 \hat e_i^2 - \EE(\omega_i^2 e_i^2) \bigg|\\
		&\leq \bigg| \frac{1}{n} \sn (\hat{\omega}_i^2 - \omega_i^2) \hat e_i^2 \bigg| +
		\bigg|  \frac{1}{n}  \sn \omega_i^2 (\hat e_i^2 - e_i^2) \bigg| + \bigg| \frac{1}{n}  \sn (1-\EE)  e_i^2 \omega_i^2  \bigg|.
	\end{aligned}
\end{equation}

To show that the first term $|n^{-1} \sn (\hat{\omega}_i^2 - \omega_i^2) \hat e_i^2 | = o_p(1)$, it suffices to show that $\max_{i=1,\ldots,n} |\hat{\omega}_i^2 - \omega_i^2| = o_p(1)$ and $n^{-1} \sn \hat e_i^2 = \cO_p(1)$. 
Recall from Proposition~4.2 that $\| \hat{\gamma} - \gamma^* \|_1 = o_p(1)$ under the scaling condition $s_0^2 \log(p)=o(n)$.  
Since $\|\gamma^*\|_1$ is bounded and $\|X_i\|_{\infty}\le b_X$, we have
\begin{equation*}
\begin{aligned}
 \max_{i=1,\ldots,n}  | \hat \omega_i^2 - \omega_i^2 |
    &= \max_{i=1,\ldots,n}  | (\hat{\gamma} - \gamma^*)^\T X_{i,-j} X_{i,-j}^\T (\hat{\gamma} - \gamma^*) + 2 \gamma^* X_{i,-j} X_{i,-j}^\T (\hat{\gamma} - \gamma^*) |  \\
    &\leq \| \hat{\gamma} - \gamma^* \|_1^2 \|X_{i,-j} X_{i,-j}^\T \|_{\max} + 2\| \gamma^* \|_1 \|X_{i,-j} X_{i,-j}^\T \|_{\max} \| \hat{\gamma} - \gamma^* \|_1 \\
    &= o_p(1). 
\end{aligned}
\end{equation*}
We now show that $n^{-1} \sn \hat e_i^2 = \cO_p(1)$.  
By the weak law of large numbers, $
		n^{-1} \sn e_i^2 \overset{p}{\to} \EE(e_i^2) =  \EE\{\var(\varepsilon_{i,-} | X_i)\}\leq \sigma_\varepsilon^2$.  Consequently, we have $n^{-1} \sn e_i^2 =\cO_p(1) $.
We now show that  the difference $|n^{-1} \sn  (\hat e_i^2 - e_i^2) | = o_p(1)$.  
Let $\phi(t) = t \mathbbm{1}(t \leq 0)$, $\zeta_i(\theta) = X_i^\T (\theta-\theta^*)$, and $\xi_i(\beta) = X_i^\T (\beta - \beta^*)$.
Using the fact that $\hat e_i - e_i = \phi( \varepsilon_i - \xi_i(\hat{\beta}) ) - \phi(\varepsilon_i) + \tau \xi_i(\hat{\beta}) -  \tau \zeta_i(\hat{\theta})$ and that $b^2 - a^2 = 2a(b-a) + (b-a)^2$, we obtain
\begin{equation}\label{eq:Z_diff_bound}
	\begin{aligned}
 \bigg|  \frac{1}{n} \sn (\hat e_i^2 - e_i^2)\bigg| 
    &\leq  \frac{1}{n} \sn   |\hat e_i^2 - e_i^2 | \\
    &\leq   \frac{1}{n} \sn \bigg(2 |e_i| \cdot |\phi (\varepsilon_i - \xi_i(\hat \beta) ) - \phi(\varepsilon_i) + \tau \xi_i(\hat \beta) -  \tau \zeta_i(\hat \theta)|\\
    &~~~~~ + [\phi( \varepsilon_i - \xi_i(\hat \beta) ) - \phi(\varepsilon_i) + \tau \xi_i(\hat \beta) -  \tau \zeta_i(\hat \theta)]^2 \bigg) \\
    &\leq  \frac{1}{n}  \sn  \bigg(2 |e_i| \{(1+\tau)|\xi_i(\hat \beta)| + \tau |\zeta_i(\hat \theta)|\} + \{(1+\tau)|\xi_i(\hat \beta)| + \tau |\zeta_i(\hat \theta)|\}^2 \bigg) \\
    &\leq  \frac{2}{n}   \sn |e_i| \{(1+\tau)b_X \| \hat \beta - \beta^*\|_1 + \tau b_X \| \hat \theta - \theta^*\|_1\} \\
    &~~~~~ + \{(1+\tau)b_X \| \hat \beta - \beta^*\|_1 + \tau b_X \| \hat \theta - \theta^*\|_1\}^2 \\
    & = o_p(1),
\end{aligned}
\end{equation}
where the third inequality uses the Lipschitz property of $\phi(t)$; the fourth inequality follows from Condition~ 4.2 that $\|X_i\|_{\infty}$ is bounded; and the last equality follows from Theorem~4.1 and Proposition~4.1 that $\| \hat \theta - \theta^*\|_1 = o_p(1)$ and $\| \hat \beta - \beta^*\|_1 = o_p(1)$ under the scaling condition $s^2 \log(p)=o(n)$, and the fact that $n^{-1} \sn |e_i| = \cO_p(1)$ since $n^{-1} \sn |e_i| \overset{p}{\to} \EE(|e_i|) \leq \sqrt{\EE(e_i^2)} \leq \sigma_{{\omega}}$.
Combining the fact that $n^{-1} \sn e_i^2 =\cO_p(1) $ and \eqref{eq:Z_diff_bound}, we establish that $n^{-1} \sn \hat e_i^2 = \cO_p(1)$. 
 
The second term in \eqref{eq:1} is $o_p(1)$ follows from Condition 4.4 that $\max_{i=1,\ldots,n} |\omega_i|$ is bounded and from~\eqref{eq:Z_diff_bound} that $| n^{-1} \sn (\hat e_i^2 - e_i^2)|=o_p(1)$. The third term in \eqref{eq:1} is $o_p(1)$ follows from the weak law of large number. 
Finally, combining all three  terms results in $\tilde{\sigma}_{{\rm s}}^2- \sigma_{{\rm s}}^2 = o_p(1)$, as desired.    \qed

 \section{Proof of Technical Lemmas}\label{sec:proof_lemma}

\subsection{Proof of Lemma~\ref{lem:first-order.error}}

After a change of variable $\delta = \beta - \beta^*$, we have $e_i(\beta) = \phi(\varepsilon_i - X_i^\T \delta) + \tau X_i^\T \delta +  \tau X_i^\T (\beta^* - \theta^*)$, where $\phi(t) = t\mathbbm{1}(t\leq 0)$ is a 1-Lipschitz continuous function. Then,
\$
\frac{1}{n}\sn (1-\EE)  \{  e_i(\beta) - e_i  \} X_i   & = \frac{1}{n} \sn (1-\EE) \{  \phi(\varepsilon_i - X_i^\T \delta) - \phi(\varepsilon_i) + \tau X_i^\T \delta \} X_i  .
\$
For each $l$, define the random process $\cR_l(\delta)  = n^{-1} \sn (1-\EE) [ \{  \phi(\varepsilon_i - X_i^\T \delta) - \phi(\varepsilon_i) + \tau X_i^\T \delta \} X_{i,l}]$. The key is to bound the supremum $\sup_{\delta \in \BB_1(r_1)   }  | \cR_l(\delta)  |$ for each $l$, followed by taking the union bound over $l=1,\ldots, p$.

Fix $l$, we can rewrite $\cR_l$ as $\cR_l(\delta) = n^{-1} \sn (1-\EE) \{r_l(\delta; \varepsilon_i, X_i)\}$.  Let $\partial_{u_i} r_l$ denote the derivative of $r_l(\delta; \varepsilon_l, X_l)$ with respect to $u_i = X_i^\T \delta$. By the chain rule, $\partial_{u_i} r_l(\delta; \varepsilon_l, X_l) =  \{ \tau -  \mathbbm{1}(\varepsilon_l \leq X_l^\T \delta ) \}X_{i,l}$. 
Under Condition~ 4.2, for any sample $(\varepsilon_i, X_i)$ and parameters $\delta, \delta'$, we have 
\$
| r_l(\delta; \varepsilon_i, X_i) - r_l(\delta'; \varepsilon_i, X_i) | 
= |X_i^\T \delta - X_i^\T \delta'| |\partial_{u_i} r_l(\tilde{\delta}; \varepsilon_l, X_l)|\\
\leq |X_{i,l}|   |X_i^\T \delta - X_i^\T \delta'|
\leq b_X | X_i^\T \delta - X_i^\T \delta' | .
\$
where $\tilde{\delta}$ is between $\delta$ and $\delta'$.
Next we control $|\cR_l(\delta)|$ uniformly over an $\ell_1$-ball with fixed radius. Given $r_1>0$, define the random variable 
$$
A_l( r_1) = \frac{n}{4 b_X r_1 } \sup_{\delta \in  \BB_1(r_1) } | \cR_l(\delta) | .
$$
For any $t>0$, using Chernoff's inequality gives 
\#
\PP\big\{ A_l(r_1) \geq t \big\} \leq \exp \bigg[  -\sup_{\lambda \geq 0} \big\{ \lambda t - \log \EE [e^{\lambda A_l(r_1) }] \big\} \bigg]. \label{chernoff.ineq}
\#
For the exponential moment $\EE [e^{\lambda A_l(r_1) }]$, applying first Rademacher symmetrization Lemma~\ref{lem:Rademacher},
and then the Ledoux-Talagrand contraction inequality Lemma~\ref{lem:Ledoux}, we obtain that
\$
\EE [e^{\lambda A_l(r_1) }] & \leq \EE \bigg[ \exp\bigg\{  2\lambda \sup_{\delta \in  \BB_1(r_1) } \bigg| \frac{1}{4 b_X r_1  } \sn \epsilon_i r_l(\delta; \varepsilon_i, X_i ) \bigg| \bigg\}\bigg] \\
& \leq \EE  \bigg[ \exp \bigg\{  \frac{\lambda}{r_1} \sup_{\delta \in  \BB_1(r_1) } \bigg|   \sn \epsilon_i X_i^\T \delta \bigg| \bigg\} \bigg]\\
& \leq \EE  \bigg\{ \exp  \bigg( \lambda   \bigg\| \sn \epsilon_i X_i \bigg\|_\infty  \bigg) \bigg\} ,
\$
where $\epsilon_1,\ldots, \epsilon_n$ are independent Rademacher random variables. Define $\overline X_i = (\overline X_{i,1}, \ldots,\overline X_{i,2p})^\T = (X_i^\T, -X_i^\T)^\T \in \RR^{2p}$, so that $ \|   \sn \epsilon_i X_i   \|_\infty = \max_{1\leq l\leq 2p}  \sn \epsilon_i\overline X_{i, l}$. Since $\epsilon_i$ is symmetric, $\EE\{(\epsilon_i X_{i,l})^k\}=0$ if $k$ is odd, and $\EE\{(\epsilon_i X_{i,l})^k\} = \EE (X_{i,l}^k) \leq b_X^{k-2} \sigma_{ll} \leq \frac{k!}{2} \sigma_{ll} (b_X/3)^{k-2}$ if $k \geq 4$ is even.  By Bernstein's inequality (see, e.g., Theorem~2.10 in \cite{BLM2013}), it can be shown that for any $0< \lambda < 1/c_X$ with $c_X = b_X/3$, 
$$
\log\big\{\EE  \big(e^{ \lambda \sn \epsilon_i X_{i,l} }\big) \big\} \leq \frac{n \sigma_{ll}  \lambda^2 }{2(1 - c_X \lambda  )}.
$$
This further implies
\$
\log\big\{\EE  \big(e^{\lambda A_l(r_1) }\big)\big\}    \leq   \log\bigg\{ \EE \bigg(\sum_{l=1}^{2p} e^{\lambda \sn \epsilon_i\overline X_{i, l} }\bigg) \bigg\}  \leq \log(2p) + \frac{n \sigma_X^2 \lambda^2}{2(1-c_X \lambda )} .
\$
For any $t> 0$, using the above bound and equation (2.5) from \cite{BLM2013} we obtain
\$
& \sup_{\lambda \geq 0} \bigg( \lambda t - \log  \EE  [e^{\lambda A_l(r_1) } ] \bigg)\\
& \geq - \log(2p) + \sup_{0< \lambda < 1/c_X} \bigg\{ \lambda t - \frac{n \sigma_X^2 \lambda^2}{2(1 - c_X \lambda )}  \bigg\} = - \log(2p) + \frac{n \sigma_X^2}{c_X^2}h_1( c_X t / n\sigma_X^2) ,
\$
where $h_1(u) = 1 + u - \sqrt{1+2u}$ is an increasing function from $(0, \infty)$ onto $(0, \infty)$ with inverse function $h_1^{-1}(u) = u + \sqrt{2u}$. Substituting this into \eqref{chernoff.ineq} yields
\$
\PP\big\{ A_l(r_1 ) \geq t \big\} \leq \exp\bigg\{ \log(2p) - \frac{n \sigma_X^2}{c_X^2} h_1\bigg(\frac{c_X t}{n \sigma_X^2} \bigg) \bigg\} ,
\$
or equivalently, for every $u>0$,
\$
\PP\big\{ A_l(r_1) \geq c_X u + \sigma_X (2nu)^{1/2} \big\}  \leq 2p e^{-u} .
\$
This proves the claimed bound by taking the union bound over $l=1,\ldots, p$.  If we replace $X_i$ with $X_{i,-j}$ in~\eqref{eq::first-order.error}, the same bound still holds by simply taking union bound over $l=1,\ldots,j-1, j+1,\ldots,p$. 

When $X_i$ is replaced by $\omega_i$, the same bound holds under Condition 4.4 without requiring the second union bound (over $l$).
\qed

\subsection{Proof of Lemma~\ref{lem:approximate.neyman}}

Note that
$\| \EE \{ e_i(\beta)  W_i \}   \|_2 = \sup_{u \in \mathbb{S}^{p-1}} | \EE   \{ e_i(\beta)   W_i^\T u \} |$.
Recall that the conditional CDF $F=F_{\varepsilon_i | X_i}$ of $\varepsilon_i$ given $X_i$ is continuously differentiable with $f=F'$. 
Again, let $\EE_{X_i}$ be the conditional expectation given $X_i$. For $\beta \in \RR^p$ and $u\in \mathbb{S}^{p-1}$, define $\Delta_i = \Delta_i(\beta) = X_i^\T (\beta - \beta^*)$  and  
\begin{equation*}
	\begin{aligned}
		E_i(\beta) = \EE_{X_i}   \{ e_i (\beta) \}  &= \EE_{X_i} \{(\varepsilon_i - \Delta_i) \mathbbm{1}(\varepsilon_i\leq  \Delta_i)\}+ \tau X_i^\T (\beta - \theta^*) \\
		&=   \int_{-\infty}^{\Delta_i}  (  t - \Delta_i  ) f(t) {\rm d} t + \tau X_i^\T (\beta - \theta^*)    .
	\end{aligned}
\end{equation*}
From above we know that $E_i(\beta^*) = \EE_{X_i}   ( e_i ) = 0$. Then, by the fundamental theorem of calculus we have
\$
E_i(\beta)  = E_i(\beta) - E_i(\beta^*) = \int_0^1 \nabla E_i \big\{ \beta^* + t(\beta-\beta^*) \big\}^\T (\beta - \beta^*) {\rm d} t ,
\$
where  $\nabla E_i(\beta) =   - X_i \int_{-\infty}^{\Delta_i} f(t) {\rm d} t  + \tau X_i = \{ \tau - F(\Delta_i) \} X_i$. For $t\in [0, 1]$, write $\beta_t = \beta^* + t(\beta-\beta^*)$ so that  $X_i^\T(\beta_t - \beta^*) = t \Delta_i$ and $\nabla E_i \big\{ \beta^* + t(\beta-\beta^*) \big\}^\T (\beta - \beta^*) = \{ \tau - F(t\Delta_i) \} \Delta_i$.   Condition 4.1 ensures $| \tau - F(t\Delta_i) | \leq f_u  \cdot t | \Delta_i |$ almost surely.   Putting together the pieces, 
we have 
\begin{equation}\label{eq::ei_bound}
    \EE_{X_i}   \{ e_i (\beta) \}\leq \int_0^1 f_u  t \delta_i^2 \, {\rm d} t \leq \frac{1}{2}  f_u \Delta_i^2. 
\end{equation}
We conclude that for any $\beta \in \beta^*+ \BB_\Sigma(r_0)$,
\begin{equation*}
    \begin{aligned}
        |  \EE  \{ e_i(\beta)  W_i^\T u \}    | & =  \Big|  \EE  [ \EE_{X_i}   \{ e_i (\beta) \}   W_i^\T u] \Big|  \\
& \leq  \frac{1}{2}   f_u  \EE \big(      \Delta_i^2  |  W_i^\T u | \big)    \\
& \leq  \frac{1}{2}  f_u  \EE \left[  \frac{\{W_i^\T \Sigma^{1/2} (\beta - \beta^*)\}^2}{\|\Sigma^{1/2} (\beta - \beta^*)\|_2^2} |  W_i^\T u | \right] \|\Sigma^{1/2} (\beta - \beta^*)\|_2^2 \\
&\leq \frac{1}{2}   f_u r_0^2 E(|  W_i^\T u |^3)
    \end{aligned}
\end{equation*}
Taking the supremum over $u \in \mathbb{S}^{p-1}$ yields
$\sup_{u \in \mathbb{S}^{p-1}} |  \EE  \{ e_i(\beta)  W_i^\T u \}| \leq \frac{1}{2}   f_u    m_3  r_0^2$. 
The same bounds hold for $\| \EE \{ e_i(\beta)  W_{i,-j} \}   \|_2$ since 
\begin{equation*}
\begin{aligned}
    \| \EE \{ e_i(\beta)  W_{i,-j} \}   \|_2 &= \sup_{u \in \mathbb{S}^{p-2}} |  \EE  \{ e_i(\beta)  W_{i,-j}^\T u \}| \\
    &\leq \sup_{u=(u_1,\ldots,u_{p-1}) \in \mathbb{S}^{p-2}} |  \EE  \{ e_i(\beta)  W_{i}^\T (u_1,\ldots,u_{j-1},0,u_{j},\ldots,u_{p-1})\}| \\
    &\leq \sup_{u \in \mathbb{S}^{p-1}} |  \EE  \{ e_i(\beta)  W_i^\T u \}|,
\end{aligned}
\end{equation*}
where $\tilde u \in \mathbb{S}^{p-1} $ is $u \in \mathbb{S}^{p-2}$ concatenate 0 in the $j$-th element. 

Similarly, using~\eqref{eq::ei_bound}, we conclude that for any $\beta \in \beta^*+ \BB_\Sigma(r_0)$,
\begin{equation*}
    | E  \{  e_i(\beta) \omega_i    \}  | \leq  \frac{1}{2}  f_u E(\Delta_i^2 \omega_i) \leq  \frac{1}{2} f_u r_0^2 b_X ,
\end{equation*}
as desired. \qed

\subsection{Proof of Lemma~\ref{lem:score.bound}}

Note that $\| n^{-1}  \sn   e_i   X_i  \|_\infty = \max_{1\leq l\leq p} | n^{-1} \sn e_i X_{i,l} |$.   
By Conditions 4.1 and  4.2, 
\begin{equation*}
	\begin{aligned}
	&	\EE (e_i X_{i,l})^2  \leq  \EE[X_{i,l}^2 \EE\{(e_i)^2 | X_i\}] = \EE\{X_{i,l}^2 \var(\varepsilon_{i,-} | X_i)\} \leq \sigma_\varepsilon^2 \sigma_{ll} \leq \sigma_\varepsilon^2 \sigma_X^2,\\
		&\EE   ( | e_i X_{i,l}  |^k )  \leq \EE  \{ \EE   ( e_i^2 | X_i ) X_{i,l}^2 \cdot   \EE   ( |e_i|^{k-2} | X_i)  |X_{i,l}  |^{k-2} \}
		\leq 2 k! \sigma_\varepsilon^2 \sigma_X^2  (2 b_\varepsilon b_X)^{k-2}
	\end{aligned}
\end{equation*}
 for any integer $k\geq 3$, where the last inequality follows from the fact that $\EE (|e_i|^{k} | X_i) \leq 2^{k-1} \{\EE(|\varepsilon_{i,  -}|^{k} | X_i) + \EE(|\varepsilon_{i,  -}|^{k}) \} \leq 2^{k} \EE(|\varepsilon_{i,  -}|^{k}) \leq  2 k! \sigma_\varepsilon^2 (2 b_\varepsilon)^{k-2} $. Then, taking $v = n \sigma_\varepsilon^2 \sigma_{X}^2$, $c = 2 b_\varepsilon b_X$ and $t=u$  in Lemma~\ref{lem:Berstein},  we obtain that for any $u\geq 0$,
\$
\bigg| \frac{1}{n} \sn e_i X_{i,l} \bigg| \leq  \sigma_\varepsilon \sigma_X \sqrt{\frac{2u}{n}} +2 b_\varepsilon b_X \frac{u}{n}
\$
holds with probability at least $1-2e^{-u}$. The claimed bounds follow by taking the union bound over $l=1,\ldots, p$ and $l=1,\ldots,j-1, j+1,\ldots,p$, respectively. 

Tuning to $\| n^{-1} \sn \omega_i X_{i,-j} \|_\infty$, for each $l\neq j$, applying Hoeffding's inequality yields
$$
 \bigg| \frac{1}{n}\sn \omega_i X_{i,l} \bigg| \leq   b_X^2 \sqrt{\frac{ 2t}{n}}
$$
with probability at least $1-2e^{-t}$.
Taking the union bound over $l=1, \ldots, j-1, j+1, \ldots, p$ proves the claim.
\qed

\subsection{Proof of Lemma~\ref{lem:RE}}

After a change of variable $w = \delta / \| \delta \|_\Sigma$, it is equivalent to lower bound $n^{-1} \sn(X_i^\T w)^2$ uniformly over $w \in \partial \BB_\Sigma(1) \cap  \BB_1(l_1)$. For some $R>0$ to be determined,  we apply a truncation argument complemented with a smoothing method as follows.  Define the function
\$
\varphi_R(u) = u^2 \mathbbm{1} (|u|\leq R) + (u -2R)^2  \mathbbm{1} (R<u\leq 2R) + (u+2R)^2  \mathbbm{1} (-2R \leq u < -R ) ,
\$
which is $(2R)$-Lipschitz and satisfies $u^2 \mathbbm{1}(|u|\leq R) \leq \varphi_R(u) \leq u^2  \mathbbm{1}(|u| \leq 2R)$. It follows that
\begin{equation}\label{RE.lbd1}
	\begin{aligned}
		& \frac{1}{n} \sn (X_i^\T w)^2  \geq \frac{1}{n} \sn (X_i^\T w)^2 \mathbbm{1} (|X_i^\T w | \leq 2R)   \\
		& \geq  \frac{1}{n} \sn (1-\EE ) [\varphi_R(X_i^\T w )] + \EE  \{\varphi_R(X_i^\T w )\}  \\
		& \geq \EE \{(X_i^\T w)^2 \mathbbm{1}( |  X_i^\T w | \leq R)\} -  \frac{1}{n} \sn \{ \EE \{\varphi_R(X_i^\T w )\} -  \varphi_R(X_i^\T w ) \}. 
	\end{aligned}
\end{equation}
For the truncated second moment $\EE \{(X_i^\T w)^2 \mathbbm{1}( |  X_i^\T w | \leq R)\}$, 
\#
\EE \{(X_i^\T w)^2 \mathbbm{1}( |  X_i^\T w | \leq R)\} \geq 1 - \frac{1}{R^2} \EE \{(X_i^\T w)^4\} \geq 1 -  \frac{m_4}{R^2}  \geq \frac{3}{4} \label{RE.lbd2}
\#
as long as $R \geq 2 m_4^{1/2}$.  

Next, for every $w \in \partial \BB_\Sigma(1)   \cap  \BB_1(l_1)$, define the function $f_i(x) = \EE \{\varphi_R(X_i^\T w) \}-  \varphi_R(x^\T w) $ for $x\in \RR^p$, satisfying $\EE\{ f_i(X_i)\} = 0$, $\EE \{f_i^2(X_i)\}\leq m_4$, and $\sup_{w \in \partial \BB_\Sigma(1)  \cap  \BB_1(l_1)}\sup_{x\in \RR^p} f_i (x) \leq 1$.  Then, applying Theorem~7.3 in \cite{B2003} to $A_R( l_1) = \sup_{w \in  \partial \BB_\Sigma(1)   \cap   \BB_1(l_1)}  n^{-1} \sn f_i(X_i)$ yields that for any $t \geq 0$,
\#
\PP\bigg[  A_R( l_1)  \geq \EE \{A_R( l_1)\}  + \sqrt{ \frac{2 m_4 t}{n} +   \EE \{ A_R(l_1 ) \} \frac{4 t}{n}  } + \frac{t}{3 n} \bigg] \leq e^{-t}.  \label{RE.concentration}
\#
It remains to bound $\EE \{A_R( l_1)\}$.    Since $\varphi_R(\cdot)$ is $(2R)$-Lipschitz,  $\varphi_R(X^\T w)$ is $(2R)$-Lipschitz continuous in $X^\T w$.  That is, for any sample $X_i$ and parameters $w, w' \in \RR^p$, we have
\$
| \varphi_R(X_i^\T w )  -  \varphi_R(X_i^\T w' )    |   \leq 2R  |X_i^\T w  - X_i^\T w' | .
\$
Hence,  by Lemma~\ref{lem:Rademacher} (Rademacher symmetrization) and  Lemma~\ref{lem:Ledoux} (the contraction inequality for Rademacher averages),  
\$
\EE \{A_R( l_1)\} &   \leq 2 \EE \bigg\{  \sup_{w \in \BB_1(l_1) }   \frac{1}{n} \sn \epsilon_i  \cdot \varphi_R(X_i^\T w)   \bigg\} \\
& \leq  4  R   \cdot  \EE \bigg\{  \sup_{w \in \BB_1(l_1) }   \frac{1}{n} \sn \epsilon_iX_i^\T w  \bigg\}  \leq  4  R l_1   \cdot \EE \bigg\|  \frac{1}{n} \sn \epsilon_i   X_i \bigg\|_\infty  ,
\$
where $\epsilon_1, \ldots, \epsilon_n$ are independent Rademacher random variables.  Arguing as in the proof of Lemma~\ref{lem:first-order.error},  let $\overline X_i = (\overline X_{i,1}, \ldots,\overline X_{i,2p})^\T = (X_i^\T, -X_i^\T)^\T \in \RR^{2p}$,  so that $ \|   \sn \epsilon_i X_i   \|_\infty = \max_{1\leq l\leq 2p}  \sn \epsilon_i\overline X_{i, l}$.  For each $l\in \{1,\ldots,p\}$ and $\lambda \in (0, 3/b_X)$,  we have shown that 
\$
\log \big\{\EE \big(e^{\lambda \sn \epsilon_i X_{i,l} }\big)\big\} \leq \frac{n \sigma_X^2 \lambda^2}{2(1-b_X \lambda/3 )}.  
\$
This together with Theorem~2.5 and Corollary~2.6 in \cite{BLM2013} implies 
\$
\EE  \bigg\| \sn \epsilon_i X_i \bigg\|_\infty \leq  \sigma_X  \sqrt{2 n \log(2p)}  + \frac{b_X}{3} \log(2p).
\$
Taking $R =2 m_4^{1/2}$ in the previous two moment inequalities, we obtain
\$
\EE \{A_R(l_1)\} \leq 8 m_4^{1/2} l_1 \bigg\{ \sigma_X \sqrt{ \frac{2 \log(2p)}{n}} + b_X\frac{\log(2p)}{3 n} \bigg\}.
\$
Combining this with \eqref{RE.lbd1}, \eqref{RE.lbd2} and \eqref{RE.concentration}, we conclude that with probability at least $1-e^{-t}$,
\$
\frac{1}{n} \sn(X_i^\T w)^2 &\geq \frac{3}{4} - \EE \{A_R(l_1)\}  - \sqrt{ \frac{2m_4 t }{n} + 2 \EE \{A_R( l_1)\}  \frac{2t }{n} } - \frac{t}{3 n} \\
&\geq \frac{3}{4} - \EE \{A_R(l_1)\}  -  \sqrt{   \frac{2m_4 t }{n}} -  \sqrt{2 \EE \{A_R( l_1)\}   \frac{2t }{n}} - \frac{t}{3 n}\\
&\geq \frac{3}{4} - \EE \{A_R(l_1)\}  -  \sqrt{ \frac{2m_4  t }{n}} -  \frac{1}{2}\bigg[\frac{1}{4} \cdot 2 \EE \{A_R( l_1)\}  + 4\cdot \frac{2t }{n}\bigg] - \frac{t}{3 n}\\
&\geq \frac{3}{4} -  10 m_4^{1/2}   l_1 \bigg\{ \sigma_X  \sqrt{\frac{2 \log(2p)}{n}} + b_X\frac{\log(2p)}{3 n} \bigg\} - (2  m_4  )^{1/2} \sqrt{\frac{t}{n}} -4.4  \frac{ t}{n}
\$
holds uniformly over $w \in \partial \BB_\Sigma(1)   \cap \BB_1(l_1)$.  This establishes the claim. The same argument can be established for replacing $X_i$ with $X_{i,-j}$, and $\Sigma$ with $\Sigma_{i,-j}$. \qed

\subsection{Proof of Lemma~\ref{lem:qr.diff.bound}}

For each $O_i = (X_i^\T ,  \varepsilon_i)^\T \in \RR^{p+1}$ (with $\varepsilon_i=Y_i-X_i^\T \beta^*$)  and $\delta \in \RR^p$, define the loss difference $d(\delta; O_i) = \rho_\tau( \varepsilon_i  - X_i^\T  \delta)  - \rho_\tau(\varepsilon_i )$ so that $\hat \cD(\delta) = n^{-1} \sn d(\delta ; O_i)$. By the $\bar \tau$-Lipschitz continuity of $\rho_\tau(\cdot)$, it is easy to see that $d(\delta; O_i)$ is $\bar \tau$-Lipschitz continuous in $X_i^\T \delta$, where $\bar \tau = \max(\tau, 1-\tau)$.

Given $r_0, l_1 >0$, define the random quantity $\Delta = \Delta(r_0, l_1) = \sup_{\delta  \in \CC(l_1) \cap \BB_\Sigma(r_0)} n \{ \cD(\delta) -\hat \cD(\delta) \}/( 4   \bar \tau l_1 r_0)$.  For some constant $c>0$ to be specified,  using Chernoff's inequality gives
\#
\PP (  \Delta  \geq y  ) \leq \exp \bigg[ - \sup_{  \lambda \in (0, c) }  \big\{  \lambda y - \log \EE  (e^{\lambda \Delta  } ) \big\}  \bigg],\label{chernoff} 
\#
valid for any $y >0$. 
The key is to bound the exponential moment $\EE (e^{\lambda \Delta } )$.  Note that every $\delta \in \CC(l_1) \cap \BB_\Sigma(r_0)$ satisfies $\| \delta \|_1 \leq l_1 r_0$. By Rademacher symmetrization and the Ledoux-Talagrand contraction inequality for Rademacher processes, we obtain
\$
\EE (  e^{\lambda \Delta   } )& \leq \EE \bigg[\exp \bigg\{ 2\lambda \sup_{\delta\in   \CC(l_1) \cap \BB_\Sigma(r_0) } \frac{1}{4  \bar \tau l_1 r_0 } \sn \epsilon_i  \cdot d(\delta; O_i) \bigg\}\bigg] \\
& \leq \EE \bigg\{\exp \bigg( \frac{\lambda}{  l_1  r_0}  \sup_{\delta\in \CC(l_1) \cap  \BB_\Sigma(r_0) }   \sn \epsilon_i  X_i^\T \delta  \bigg)\bigg\} \leq \EE \bigg\{\exp\bigg( \lambda \bigg\|   \sn \epsilon_i  X_i  \bigg\|_\infty \bigg)\bigg\} ,
\$
where $\epsilon_1 , \ldots,  \epsilon_n$ are independent Rademacher random variables.   The rest of the proof is almost identical to that of Lemma~\ref{lem:first-order.error}.  Write $\overline X_i = (\overline X_{i,1}, \ldots,\overline X_{i,2p})^\T = (X_i^\T, -X_i^\T)^\T \in \RR^{2p}$,  so that $ \|   \sn \epsilon_i X_i   \|_\infty = \max_{1\leq l\leq 2p}  \sn \epsilon_i\overline X_{i, l}$.  Then, for any $0<\lambda < 1/c_X$ with $c_X = b_X/3$, 
\$
\log \EE (e^{\lambda \Delta   } ) &\leq 	  \log~\EE\bigg\{ \exp\bigg(  \lambda  \bigg\| \sn \epsilon_i  X_i  \bigg\|_\infty \bigg)\bigg\}  \\
&\leq    \log \bigg\{  \sum_{l=1}^{2p}  \EE \big(e^{   \lambda \sn \epsilon_i\overline{x}_{i,l} } \big) \bigg\}  \\
&\leq \log(2p)  + \frac{n \sigma_X^2 \lambda^2}{2(1-c_X \lambda) }.
\$
Substituting this exponential moment inequality into \eqref{chernoff} yields that for any $u>0$,
\$
\PP\big\{  \Delta  \geq  c_X u + \sigma_X(2n u)^{1/2} \big\}   \leq   2pe^{-u}.
\$\qed

\subsection{Proof of Lemma~\ref{lem:qr.score.bound}}

Note that for each $l\in \{1,\ldots,p\}$,  $|\mathbbm{1}(\varepsilon_i\leq 0) -\tau| \cdot |X_{i,l}| \leq \bar \tau b_X$ and $\EE[\{ \mathbbm{1}(\varepsilon_i\leq 0) -\tau \}^2 X_{i,l}^2] \leq \tau(1-\tau) \sigma_{ll}$.  Then, by  Lemma~\ref{lem:Berstein} Bernstein's inequality, taking $c = \bar{\tau} b_X/3$ and $v=n\tau(1-\tau)\sigma_X^2$, 
\#
\bigg\| \frac{1}{n} \sn \{ \mathbbm{1}(\varepsilon_i \leq 0) - \tau  \} X_{i,l}\bigg\| \leq  \sqrt{2\tau(1-\tau)} \sigma_X \sqrt{\frac{u}{n}} + \bar \tau b_X \frac{u}{3n} ,
\#
with probability at least $1-2 e^{-u}$. The claimed bound follows from taking union bound over $l=1,\ldots,p$.  
\qed
 
\section{Additional Simulations}\label{append:numerical}
All codes can be found at \href{https://github.com/shushuzh/ES_highD}{https://github.com/shushuzh/ES\_highD}. 

\subsection{Simulations on BIC}
We present numerical results comparing the proposed two-step approach with BIC-based and CV-based tuning approaches in Table~\ref{tab:est_BIC} and Table~\ref{table:inference_BIC} under the same settings as the manuscript. From Table~\ref{tab:est_BIC}, BIC tends to select a larger tuning parameter $\lambda$ than CV, thus resulting in smaller false positives and larger estimation errors for true positives. From Table~\ref{table:inference_BIC}, while the two-step estimator is more biased with BIC, the debiased estimator has a negligible bias for both BIC and CV, and the coverage probabilities are close to nominal levels for both BIC and CV across different quantile levels. 
\begin{table}[h!]
	\fontsize{8}{9}\selectfont
	\caption{
 Numerical comparisons of the two-step with CV or BIC, refitted two-step with CV or BIC, and oracle two-step methods under Model (5.1) with $(s, p) =(10, 1000)$, $n = \lceil 50 s/\tau \rceil$, and covariates $X_i= |z_i|$ where $z_i \sim N\{0,\Sigma= (0.8^{|j-k|})_{1\leq j, k\leq p}\}$. Estimation errors under the relative $\ell_2$-norm,  the false positive rate, averaged over 500 replications, are reported. The true positive rates are all one. The largest standard error is 0.003 for Error (FP), 0.001 for Error (P), 0.001 for FPR with CV, and $2\times 10^{-5}$ for FPR with BIC in the table. }
 \begin{center}
		\begin{tabular}{| c | l | c c c | c c c|}
			\hline
      \multicolumn{2}{|c|}{} & \multicolumn{3}{c|}{ CV} & \multicolumn{3}{c}{BIC} \vline\\ \hline
			$\tau$ & Methods & Error (P) & Error (FP) & FPR ($\times 10^{-4}$) & Error (P) & Error (FP)& FPR($\times 10^{-4}$) \\
			\hline
\multirow[c]{3}{*}{0.05} & two-step & 0.085 & 0.029  & 185 & 0.093 & 0.002  & 2.162 \\
 & two-step+refitted & 0.074 & 0.149  & 185 & 0.072 & 0.012  & 2.162 \\
 & two-step oracle & 0.078 & 0 & NA & 0.078 & 0 & NA \\
\hline
\multirow[c]{3}{*}{0.10} & two-step & 0.088 & 0.031  & 175 & 0.098 & 0.002  & 2.485 \\
 & two-step+refitted & 0.078 & 0.154  & 175 & 0.075 & 0.014  & 2.485 \\
 & two-step oracle & 0.082 & 0  & NA & 0.082 & 0 & NA  \\
\hline
\multirow[c]{3}{*}{0.20} & two-step & 0.096 & 0.033  & 182 & 0.105 & 0.003  & 2.606 \\
 & two-step+refitted & 0.085 & 0.166  & 182 & 0.081 & 0.015  & 2.606 \\
 & two-step oracle & 0.087 & 0 & NA & 0.087 & 0 & NA \\
			\hline	
		\end{tabular}
  \end{center}
 \label{tab:est_BIC}
 \begin{tablenotes}
 \small
 \item
Error (P): relative $\ell_2$-error on the support, i.e., $ \|\hat{\theta}_{S_e^*}-\theta^*_{S_e^*}\|_2/\|\theta^*\|_2$. Error (FP): relative $\ell_2$-error on the false positives, i.e., $ \|\hat{\theta}_{(S_e^*)^c}\|_2/\|\theta^*\|_2$. Here $S_e^* = (1,\ldots,s)$ and $(S_e^*)^c=(s+1,\ldots,p)$. FPR: false positive rate. 
\end{tablenotes}
\end{table}

	\begin{table}[h!]
		 \fontsize{8}{9}\selectfont
		 \centering
		 \caption{Estimation and inference on $\theta_2^*$ under Model (5.1) with $(s, p) = (10, 1000)$,  $n=\lceil 50 s/\tau \rceil$, $\tau \in \{ 0.05, 0.1, 0.2 \}$, and covariates $X_i= |z_i|$ where $z_i \sim  \mathcal{N}(0,\Sigma= (0.8^{|j-k|})_{1\leq j, k\leq p})$. The biases, squared errors of the five estimators, and coverage probabilities of the 95\% confidence interval of the two-step approach with CV and BIC, averaged over 500 replications, are reported. The largest standard error is 0.005 for the bias and mean squared error and 0.01 for the coverage probability in the table.
}		
  \begin{center}
			\begin{tabular}{| c |  c | c|cc | c |cc | c |}
				\hline
    \multicolumn{2}{|c|}{}&\multirow{2}{*}{\makecell[c]{Bias ($\times 10^{-2}$)/\\MSE ($\times 10^{-2}$)}}& \multicolumn{3}{c|}{CV}&\multicolumn{3}{c|}{BIC}\\ \cline{4-9}
    \multicolumn{2}{|c|}{}&&\multicolumn{2}{c|}{Bias ($\times 10^{-2}$)/MSE ($\times 10^{-2}$)}&\multicolumn{1}{c|} {Inference}&\multicolumn{2}{c|}{Bias ($\times 10^{-2}$)/MSE ($\times 10^{-2}$)}&\multicolumn{1}{c|} {Inference}\\
    \hline
				\multirow{2}{*}{$\tau$}&\multirow{2}{*}{$\theta_2^*$} &\multirow{2}{*}{\makecell[c]{two-step\\oracle}} &\multirow{2}{*}{two-step} &\multirow{2}{*}{\makecell[c]{debiased\\two-step}}&\multirow{2}{*}{\makecell[c]{coverage\\ probability}}&\multirow{2}{*}{two-step} &\multirow{2}{*}{\makecell[c]{debiased\\two-step}} &\multirow{2}{*}{\makecell[c]{coverage\\ probability}}\\	
				&&&&&&&&\\
    \hline
    0.05  &        1.312 &        0.0/0.8 & -8.1/1.4 &       1.3/0.7   &    0.94 & -82.0/68.9 & 1.7/0.7 & 0.95\\
0.10  &        1.415 &  0.1/0.9 & -8.8/1.6 &       1.8/0.8 &  0.97& -88.0/79.0 & 2.4/0.8 &  0.96 \\
0.20  &        1.533 &  0.8/1.2 &  -9.5/2.0 &       2.8/1.1 & 0.95& -91.7/86.5 & 3.6/1.1 &  0.95 \\
    \hline
			\end{tabular}
		\end{center}
  \label{table:inference_BIC}
   \begin{tablenotes}
\small
\item
\noindent
Bias: $M^{-1}\sum_{m=1}^M \{\hat \theta^{(m)} - \theta_2^*\}$, where $ \hat \theta^{(m)}$ denotes a generic expected shortfall regression estimator. MSE: $M^{-1}\sum_{m=1}^M  \{ \hat \theta^{(m)} - \theta_2^* \}^2$. Here $M$ is the number of replications, which is set as 500.
\end{tablenotes}
	\end{table}
 
\subsection{Large p}
We have performed additional numerical studies with the same data-generating procedure as in Equation~(5.1) in the manuscript with $p=2n$. Specifically, we consider the case when $p=2n$, $s=10$, $n= \{\lceil 25 s/\tau \rceil\}$, covariates $X_i= |z_i|$ where $z_i \sim  \mathcal{N}(0,\Sigma= (0.8^{|j-k|})_{1\leq j, k\leq p})$. 
The results for $\tau=\{0.05,0.10,0.20\}$, averaged over 500 replications, are presented in Table~\ref{tab:est_largep} and~\ref{table:inference_largep} of this response file. 
We see that the results are very similar to results presented in the main manuscript for $p=1000$.
\begin{table}[!htp]
	\fontsize{8}{9}\selectfont
	\caption{
 Numerical comparisons of the two-step, refitted two-step, and oracle two-step methods under the linear heteroscedastic model (5.1) with $s = 10$, $n= \{\lceil 25 s/\tau \rceil\}$, $p=2n$ and $\tau \in \{0.05, 0.1, 0.2\}$. Estimation errors under the relative $\ell_2$-norm,  the false positive rate, averaged over 500 replications, are reported. The true positive rates are all one. The largest standard error is 0.003 for Error (FP) and 0.001 for Error (P) and FPR in the table. }
 \begin{center}
		\begin{tabular}{| c c c | l | c c c|}
			\hline
 \multicolumn{7}{|c|}{Covariates $X_i= |z_i|$ where $z_i \sim  \mathcal{N}(0,\Sigma= (0.8^{|j-k|})_{1\leq j, k\leq p})$}\\ \hline
			$n$&$p$&$\tau$ & Methods & Error (P) & Error (FP) & FPR  \\
   \hline
\multirow[c]{3}{*}{5000} &\multirow[c]{3}{*}{10000} &\multirow[c]{3}{*}{0.05} & two-step & 0.123 & 0.038  & 0.002 \\
&& & two-step+refitted & 0.100 & 0.271  & 0.002 \\
&& & two-step oracle & 0.111 & 0.0  & NA \\
 \hline
\multirow[c]{3}{*}{2500} &\multirow[c]{3}{*}{5000} &\multirow[c]{3}{*}{0.10} & two-step & 0.127 & 0.043 & 0.005 \\
&& & two-step+refitted & 0.106 & 0.268  & 0.005 \\
&& & two-step oracle & 0.117 & 0.0 & NA \\
 \hline
\multirow[c]{3}{*}{1250} &\multirow[c]{3}{*}{2500} &\multirow[c]{3}{*}{0.20} & two-step & 0.134 & 0.048  & 0.009 \\
&& & two-step+refitted & 0.118 & 0.257 & 0.009 \\
&& & two-step oracle & 0.125 & 0.0 & NA \\
			\hline	
		\end{tabular}
  \end{center}
 \label{tab:est_largep}
 \begin{tablenotes}
 \small
 \item
Error (P): relative $\ell_2$-error on the support, i.e., $ \|\hat{\theta}_{S_e^*}-\theta^*_{S_e^*}\|_2/\|\theta^*\|_2$. Error (FP): relative $\ell_2$-error on the false positives, i.e., $ \|\hat{\theta}_{(S_e^*)^c}\|_2/\|\theta^*\|_2$. Here $S_e^* = (1,\ldots,s)$ and $(S_e^*)^c=(s+1,\ldots,p)$. FPR: false positive rate. 
\end{tablenotes}
\end{table}

	\begin{table}[!htp]
		 \fontsize{8}{9}\selectfont
		 \centering
		 \caption{Estimation and inference on $\theta_2^*$ under the linear heteroscedastic model (5.1) with $s= 10$,  $n=\lceil 25 s/\tau \rceil$, $p=2n$ and $\tau \in \{ 0.05, 0.1, 0.2 \}$. The biases, squared errors of the four methods, and coverage probabilities of the 95\% confidence interval, averaged over 500 replications, are reported. The largest standard error is 0.005 for the bias and mean squared error and 0.01 for the coverage probability in the table.
}		
  \begin{center}
			\begin{tabular}{|cc c   c | ccc | c |}
\hline
 \multicolumn{8}{|c|}{Covariates $X_i= |z_i|$ where $z_i \sim  \mathcal{N}(0,\Sigma= (0.8^{|j-k|})_{1\leq j, k\leq p})$}\\ \hline
\multicolumn{4}{|c|}{} &\multicolumn{3}{c|}{Bias ($\times 10^{-2}$)/MSE ($\times 10^{-2}$)}&\multicolumn{1}{c|} {Inference}\\\hline
				\multirow{2}{*}{$n$}&\multirow{2}{*}{$p$}&\multirow{2}{*}{$\tau$}&\multirow{2}{*}{$\theta_2^*$} &\multirow{2}{*}{two-step} &\multirow{2}{*}{\makecell[c]{debiased\\two-step}}  &\multirow{2}{*}{\makecell[c]{two-step\\oracle}} &\multirow{2}{*}{\makecell[c]{coverage\\ probability}}\\	
				&&&&&&&\\
    \hline
    5000&10000&0.05  &        1.312  & -13.7/3.3 & 3.1/1.4 & 0.8/1.6 &       0.952 \\
2500&5000&0.10  &        1.415 & -14.2/3.6 & 3.4/1.6 & 0.3/1.9&      0.938 \\
1250&2500&0.20  &        1.533 & -14.7/4.2 & 4.3/2.1 & 1.0/2.2 &      0.942 \\
    \hline
			\end{tabular}
		\end{center}
  \label{table:inference_largep}
	\end{table}

\subsection{Sensitivity Analysis}
\label{subsec:Sensitive}

We now investigate whether the proposed method is sensitive to the magnitude of the signal strengths. To this end, we construct the same simulation setting as described in Section 5.1 with different signals by multiplying $\gamma^*$  by $c \in \{0.4,0.6,0.8,1.0\}$.
	Results, averaged over 500 replications, for the case when $\tau=0.2$ are reported in Table~\ref{tab:weak_signal}.
	From Table~\ref{tab:weak_signal}, we see even when the true expected shortfall parameters are as small as $0.333$ and $0.4$, the $\ell_1$-penalized expected shortfall regression has TPR and FPR that are close to one and zero, respectively. 
	While the relative estimation error increases when the signal gets weaker, i.e., for smaller $c$, 
	it is still within a reasonable range compared to the oracle method. 
 
	\begin{table}[h!]
	\fontsize{8}{9}\selectfont
\caption{Sensitivity analysis under the linear heteroscedastic model~(5.1) where the signals vary by $\gamma^* = c \times (\gamma_1^*,\ldots,\gamma_s^*,0,\ldots,0)\in \RR^p, c \in (0.2,0.4,0.6,0.8,1)$, with $(s,p,n)=(10,1000,2500)$. Estimation errors (and standard error) under the relative $\ell_2$-norm, false positive rates, averaged over 500 replications, are reported. 
}
 \begin{center}
		\begin{tabular}{| c | l | c c c c  c c|}
			\hline
			\multicolumn{8}{|c|}{Linear heteroscedastic model, covariates $X_i= |z_i|$ where $z_i \sim N(0,\mathbb{I}_p)$, $\tau=0.2$}\\ \hline
			$c$ & Methods & Error (P) & Error (FP) & TPR & FPR & $\theta_1^*$&$\theta_s^*$\\
			\hline
			0.4 & two-step &  0.416(0.003) &  0.187(0.003) &  0.996(0.001) &  0.043(0.001) &\multirow{3}{*}{0.333} &    	\multirow{3}{*}{0.4} \\
    & two-step+refitted &  0.206(0.002) &  0.821(0.007) &  0.996(0.001) &  0.043(0.001) & &\\
    & two-step oracle &  0.191(0.002) &      0 &      NA &      NA & &\\
    \hline
0.6 & two-step &  0.228(0.002) &  0.103(0.002) &      1.0(0.0) &  0.043(0.001) &  \multirow{3}{*}{0.733} &    	\multirow{3}{*}{0.6} \\
    & two-step+refitted &  0.112(0.001) &  0.453(0.004) &      1.0(0.0) &  0.043(0.001) & &\\
    & two-step oracle &  0.105(0.001) &      0 &      NA &      NA& & \\
    \hline
0.8 & two-step &  0.156(0.001) &   0.07(0.001) &      1.0(0.0) &  0.043(0.001) &   \multirow{3}{*}{1.133} &    	\multirow{3}{*}{0.8} \\
    & two-step+refitted &  0.076(0.001) &  0.309(0.003) &      1.0(0.0) &  0.043(0.001) & &\\
    & two-step oracle &  0.072(0.001) &      0 &      NA &      NA& & \\
    \hline
1.0 & two-step &  0.118(0.001) &  0.054(0.001) &      1.0(0.0) &  0.044(0.001) &   \multirow{3}{*}{1.533} &    	\multirow{3}{*}{1} \\
    & two-step+refitted &  0.058(0.001) &  0.235(0.002) &      1.0(0.0) &  0.044(0.001) & &\\
    & two-step oracle &  0.054(0.001) &      0 &      NA &      NA & &\\
			\hline
			\hline
			\multicolumn{8}{|c|}{Linear heteroscedastic model, covariates $X_i= |z_i|$ where $z_i \sim N(0,\Sigma= (0.8^{|j-k|})_{1\leq j, k\leq p})$, $\tau=0.2$}\\ \hline
			$c$ & Methods & Error (P) & Error (FP) & TPR & FPR & $\theta_1^*$&$\theta_s^*$\\\hline
			0.4 & two-step &  0.335(0.003) &  0.115(0.003) &  0.995(0.001) &  0.018(0.001) &\multirow{3}{*}{0.333} &    	\multirow{3}{*}{0.4} \\
    & two-step+refitted &    0.3(0.004) &  0.578(0.009) &  0.995(0.001) &  0.018(0.001) & & \\
    & two-step oracle &  0.304(0.004) &      0 &      NA &      NA  & &\\
    \hline
0.6 & two-step &  0.184(0.002) &  0.064(0.002) &      1.0(0.0) &  0.018(0.001) &  \multirow{3}{*}{0.733} &    	\multirow{3}{*}{0.6} \\
    & two-step+refitted &  0.164(0.002) &  0.322(0.005) &      1.0(0.0) &  0.018(0.001)  & &\\
    & two-step oracle &  0.167(0.002) &      0 &      NA &      NA  & &\\
    \hline
0.8 & two-step &  0.126(0.001) &  0.043(0.001) &      1.0(0.0) &  0.018(0.001) &   \multirow{3}{*}{1.133} &    	\multirow{3}{*}{0.8} \\
    & two-step+refitted &  0.112(0.001) &  0.219(0.003) &      1.0(0.0) &  0.018(0.001)  & &\\
    & two-step oracle &  0.114(0.001) &      0 &      NA &      NA  & &\\
    \hline
1.0 & two-step &  0.096(0.001) &  0.033(0.001) &      1.0(0.0) &  0.018(0.001) &   \multirow{3}{*}{1.533} &    	\multirow{3}{*}{1} \\
    & two-step+refitted &  0.085(0.001) &  0.166(0.002) &      1.0(0.0) &  0.018(0.001)  & &\\
    & two-step oracle &  0.087(0.001) &      0 &      NA &      NA  & &\\
			\hline		
		\end{tabular}
	  \end{center}
  \label{tab:weak_signal}
  \begin{tablenotes}
  \small
  \item
  Error (P): relative $\ell_2$-error on the support, i.e., $\|\hat{\theta}_{S_e^*}-\theta^*_{S_e^*}\|_2/\|\theta^*\|_2$. Error (FP): relative $\ell_2$-error on the false positives, i.e., $\|\hat{\theta}_{(S_e^*)^{{\rm c}}}\|_2/\|\theta^*\|_2$. Here $S_e^* = (1,\ldots,s)$ and $(S_e^*)^{{\rm c}}=(s+1,\ldots,p)$. 
  TPR: true positive rate. FPR: false positive rate.
  \end{tablenotes}
	\end{table}

	Next, we compare the proposed $\ell_1$-penalized expected shortfall regression to the $\ell_1$-penalized least squares regression, implemented using the \texttt{glmnet} package in \texttt{R} under a linear homogeneous model, 
	\#
	\label{response_iid}
	y_i = X_i^\T \gamma^*+ \varepsilon_i,
	\# 
	where $\gamma^*$ is generated similar to that of the linear heteroscedastic model in Section 5.1.  
	Note that the true quantile regression and expected shortfall coefficients are both equal to $\gamma^*$ at all quantile levels. 
	The goal is to assess whether the proposed estimator, computed at different expected shortfall levels, has similar statistical accuracy as that of the $\ell_1$-penalized least squares regression under the linear homogeneous model.  
	To this end, we implement the proposed method at $\tau=(0.05,0.1,0.2,0.5)$, compared with the lasso regression and the two-step oracle method. Results, averaged over 500 replications, are presented in Table~\ref{tab:iid}. The true positive rates of the three methods are all 1. 
	
	From Table~\ref{tab:iid}, we see that when $\tau=0.5$, the proposed method has similar results as that of the $\ell_1$-penalized least squares regression, suggesting that there is no loss of statistical accuracy.  As we decrease $\tau$ from 0.5 to 0.05, the difference between the relative estimation error of the $\ell_1$-penalized expected shortfall regression estimator and the lasso estimator increases.  This is not surprising since the extremes are usually harder to estimate due to small effective sample size.  In short, the results in Table~\ref{tab:iid} suggest that as long as the expected shortfall level $\tau$ is not too close to zero or one, the proposed estimator yields an estimator that is comparable to that of lasso under the linear homogeneous model.

	\begin{table}[h!]
	\fontsize{8}{9}\selectfont
\caption{Numerical comparisons of the two-step, refitted two-step and oracle two-step methods linear homogeneous model~\eqref{response_iid} (i.e., independently and identically distributed responses) with $(s, p) =(10, 1000)$, $n\in \{\lceil 30 s/\tau \rceil, \lceil 50 s/\tau \rceil\}$ and $\tau \in \{0.05, 0.1, 0.2\}$. Estimation errors under the relative $\ell_2$-norm, true and false positive rates, averaged over 500 replications, are reported. The largest standard error for the results is 0.002. }
		\begin{center}
			\begin{tabular}{| c | l | c c | c c|}
				\hline
				\multicolumn{6}{|c|}{Linear homogeneous model, covariates $X_i= |z_i|$ where $z_i \sim N(0,\mathbb{I}_p)$}\\ \hline
				& & \multicolumn{2}{c|}{ $(n = \lceil 30 s/\tau \rceil, p = 1000)$} & \multicolumn{2}{c}{ $(n = \lceil 50 s/\tau \rceil, p =1000)$} \vline\\ \hline
				$\tau$ & Methods & Error  & FPR & Error  & FPR \\
				\hline
				0.05 & two-step &      0.068 &  0.019 &      0.071 &  0.019 \\
     & lasso &      0.053 &  0.000 &      0.045 &  0.000 \\
     & two-step oracle &      0.012 &    NA &      0.010 &    NA \\
     \hline
0.10 & two-step &      0.088 &  0.028 &      0.059 &  0.023 \\
     & lasso &      0.070 &  0.001 &      0.058 &  0.001 \\
     & two-step oracle &      0.018 &    NA &      0.014 &    NA \\
     \hline
0.20 & two-step &      0.111 &  0.051 &      0.085 &  0.042 \\
     & lasso &      0.091 &  0.003 &      0.074 &  0.002 \\
     & two-step oracle &      0.025 &    NA &      0.019 &    NA \\
     \hline
0.50 & two-step &      0.137 &  0.083 &      0.105 &  0.073 \\
     & lasso &      0.139 &  0.006 &      0.107 &  0.005 \\
     & two-step oracle &      0.040 &    NA &      0.031 &    NA \\
				\hline
				\multicolumn{6}{|c|}{Linear homogeneous model, covariates $X_i= |z_i|$ where $z_i \sim N(0,\Sigma= (0.8^{|j-k|})_{1\leq j, k\leq p})$}\\ \hline
				& & \multicolumn{2}{c|}{ $(n = \lceil 30 s/\tau \rceil, p = 1000)$} & \multicolumn{2}{c}{ $(n = \lceil 50 s/\tau \rceil, p =1000)$} \vline\\ \hline
				$\tau$ & Methods & Error  & FPR & Error  & FPR \\
				\hline
				0.05 & two-step &      0.174 &  0.010 &      0.143 &  0.011 \\
     & lasso &      0.034 &  0.000 &      0.029 &  0.000 \\
     & two-step oracle &      0.018 &    NA &      0.014 &    NA \\
     \hline
0.10 & two-step &      0.070 &  0.008 &      0.059 &  0.009 \\
     & lasso &      0.045 &  0.000 &      0.037 &  0.000 \\
     & two-step oracle &      0.026 &    NA &      0.020 &    NA \\
     \hline
0.20 & two-step &      0.071 &  0.012 &      0.051 &  0.010 \\
     & lasso &      0.058 &  0.000 &      0.047 &  0.000 \\
     & two-step oracle &      0.037 &    NA &      0.028 &    NA \\
     \hline
0.50 & two-step &      0.091 &  0.025 &      0.067 &  0.021 \\
     & lasso &      0.087 &  0.001 &      0.068 &  0.000 \\
     & two-step oracle &      0.059 &    NA &      0.044 &    NA \\
     \hline
			\end{tabular}
	 \end{center}
  	\label{tab:iid}
     \begin{tablenotes}
     \small
     \item
  Error, the relative $\ell_2$-error $ \|\hat{\theta}-\theta^*\|_2/\|\theta^*\|_2$. FPR, false positive rate. 
  \end{tablenotes}
	\end{table}

\bibliographystyle{agsm}
\bibliography{reference}
\end{document}